\documentclass[acmsmall, nonacm]{acmart}

\usepackage{float}
\usepackage{caption}
\usepackage{algorithm}
\usepackage{graphicx}
\usepackage{tabularx} 
\usepackage{threeparttable}
\usepackage{amsmath}
\usepackage{placeins}
\usepackage[normalem]{ulem}
\usepackage{latexsym}
\usepackage{subcaption}
\usepackage{times}
\usepackage{enumerate}
\usepackage{xspace}
\usepackage{calc}
\usepackage{url}
\usepackage{booktabs}
\usepackage[english]{babel}
\usepackage{algorithm, algcompatible}
\usepackage[noend]{algpseudocode}
\usepackage{hhline}
\usepackage{multirow}
\usepackage{makecell} 
\usepackage{hyperref}
\hypersetup{
	citecolor=green
}
\usepackage{wrapfig}
\usepackage{csquotes}
\usepackage{multicol}
\usepackage{tikz}
\usepackage{xparse}
\usepackage{orcidlink}
\usepackage{etoolbox}
\usepackage{amsmath,amsfonts}
\usepackage{graphicx}
\usepackage{textcomp}
\usepackage{xcolor}
\usepackage{listings}
\usepackage{adjustbox}
\usepackage{pifont}

\newcommand{\cmark}{\ding{51}}%
\newcommand{\xmark}{\ding{55}}%

\newcommand{\lightbold}[1]{%
  \textmd{#1\kern-0.05em#1}%
}

\newcommand{\sgn}{\ensuremath {\texttt{SGN}}{\xspace}}

\newcommand{\asgn}{\ensuremath {\texttt{ASGN}}{\xspace}}
\newcommand{\asgnkg}{\ensuremath {\texttt{ASGN.Kg}}{\xspace}}
\newcommand{\asgnsig}{\ensuremath {\texttt{ASGN.ASig}}{\xspace}}
\newcommand{\asgnagg}{\ensuremath {\texttt{ASGN.Agg}}{\xspace}}

\newcommand{\asgnaver}{\ensuremath {\texttt{ASGN.AVer}}{\xspace}}

\newcommand{\oslo}{\ensuremath {\texttt{OSLO}{\xspace}}}

\newcommand{\poslo}{\ensuremath {\texttt{POSLO}{\xspace}}}
\newcommand{\poslop}{\ensuremath {\texttt{POSLO}^+{\xspace}}}
\newcommand{\poslopp}{\ensuremath {\texttt{POSLO}^{++}{\xspace}}}

\newcommand{\posloaver}{\ensuremath {\texttt{POSLO.AVer}}{\xspace}}

\newcommand{\posloaggekeys}{\ensuremath {\texttt{POSLO.Agg-eKeys}{\xspace}}}
\newcommand{\poslopaver}{\ensuremath {\texttt{POSLO.PAVer}}{\xspace}}

\newcommand{\aver}{\ensuremath {\texttt{AVer}}{\xspace}}
\newcommand{\paver}{\ensuremath {\texttt{PAVer}}{\xspace}}

\newcommand{\mmo}{\ensuremath {\texttt{MMO}}{\xspace}}
\newcommand{\mdc}{\ensuremath {\texttt{MDC-2}}{\xspace}}

\newcommand{\poslof}{\ensuremath {\texttt{POSLO-F}{\xspace}}}
\newcommand{\poslofp}{\ensuremath {\texttt{POSLO-F}^+{\xspace}}}
\newcommand{\poslofpp}{\ensuremath {\texttt{POSLO-F}^{++}{\xspace}}}
\newcommand{\poslofkg}{\ensuremath {\texttt{POSLO-F.Kg}}{\xspace}}
\newcommand{\poslofsig}{\ensuremath {\texttt{POSLO-F.Sig}}{\xspace}}

\newcommand{\poslofaver}{\ensuremath {\texttt{POSLO-F.AVer}}{\xspace}}
\newcommand{\poslofdistill}{\ensuremath {\texttt{POSLO-F.Distill}}{\xspace}}
\newcommand{\poslofsebver}{\ensuremath {\texttt{POSLO-F.SeBVer}}{\xspace}}
\newcommand{\poslofagg}{\ensuremath {\texttt{POSLO-F.Agg}}{\xspace}}

\newcommand{\posloc}{\ensuremath {\texttt{POSLO-C}{\xspace}}}
\newcommand{\poslocp}{\ensuremath {\texttt{POSLO-C}^+{\xspace}}}
\newcommand{\poslocpp}{\ensuremath {\texttt{POSLO-C}^{++}{\xspace}}}
\newcommand{\poslockg}{\ensuremath {\texttt{POSLO-C.Kg}}{\xspace}}
\newcommand{\poslocsig}{\ensuremath {\texttt{POSLO-C.Sig}}{\xspace}}

\newcommand{\poslocaver}{\ensuremath {\texttt{POSLO-C.AVer}}{\xspace}}
\newcommand{\poslocdistill}{\ensuremath {\texttt{POSLO-C.Distill}}{\xspace}}
\newcommand{\poslocagg}{\ensuremath {\texttt{POSLO-C.Agg}}{\xspace}}

\newcommand{\poslocsebver}{\ensuremath {\texttt{POSLO-C.SeBVer}}{\xspace}}

\newcommand{\bpv}{\ensuremath {\texttt{BPV}{\xspace}}}
\newcommand{\bpvoff}{\ensuremath {\texttt{BPV.Offline}{\xspace}}}
\newcommand{\bpvon}{\ensuremath {\texttt{BPV.Online}}{\xspace}}

\newcommand{\ds}{\ensuremath { \mathit{ds} }{\xspace}}
\newcommand{\dstop}{\ensuremath { \texttt{Top} }{\xspace}}
\newcommand{\dspush}{\ensuremath { \texttt{Push} }{\xspace}}
\newcommand{\dspop}{\ensuremath { \texttt{Pop} }{\xspace}}
\newcommand{\dsinit}{\ensuremath { \texttt{Init} }{\xspace}}

\newcommand{\ccd}{\ensuremath { \mathit{CCD} }{\xspace}}
\newcommand{\St}{\ensuremath { \mathit{St} }{\xspace}}
\newcommand{\css}{\ensuremath { \texttt{CSS} }{\xspace}}

\newcommand{\oslot}{\ensuremath { \texttt{OSLOT} }{\xspace}}
\newcommand{\poslot}{\ensuremath { \texttt{POSLOT} }{\xspace}}
\newcommand{\sconst}{\ensuremath { \texttt{SC} }{\xspace}}
\newcommand{\sso}{\ensuremath { \texttt{SO} }{\xspace}}
\newcommand{\sret}{\ensuremath { \texttt{SR} }{\xspace}}
\newcommand{\smf}{\ensuremath {\texttt{SMF}{\xspace}}}

\newcommand{\sk}{\ensuremath { \mathit{sk} }{\xspace}}
\newcommand{\pk}{\ensuremath { \mathit{PK} }{\xspace}}

\newcommand{\ro}{\ensuremath {\mathit{RO}(.)}{\xspace}}

\newcommand{\EUCMA}{\ensuremath { \texttt{EU-CMA} }{\xspace}}

\newcommand{\AEUCMA}{\ensuremath { \texttt{A-EU-CMA} }{\xspace}}

\newcommand{\advsgn}{\ensuremath {\mathit{Adv}_{\sgn}^{\EUCMA}(t,q_H,q_s)}{\xspace}}

\newcommand{\advAsgn}{\ensuremath {\mathit{Adv}_{\asgn}^{\AEUCMA}(\mathcal{A})}{\xspace}}

\newcommand{\advsocosa}{\ensuremath {\mathit{Adv}_{\posloc(p,q,\alpha)}^{\AEUCMA}(t, n^{\prime}, n)}{\xspace}}

\newcommand{\advAdl}{\ensuremath {\mathit{Adv}_{\mathbb{G}}^{\dl}(\mathcal{A})}{\xspace}}
\newcommand{\advdl}{\ensuremath {\mathit{Adv}_{G}^{\dl}(t)}{\xspace}}
\newcommand{\advdll}{\ensuremath {\mathit{Adv}_{\mathbb{G}, \alpha}^{\dl}(t')}{\xspace}}

\newcommand{\lh}{\ensuremath {\mathcal{LH}}{\xspace}}
\newcommand{\lm}{\ensuremath {\mathcal{LM}}{\xspace}}

\newcommand{\ls}{\ensuremath {\mathcal{LS}}{\xspace}}

\newcommand{\Areal}{\ensuremath {\overrightarrow{A}_{\mathit{real}}}{\xspace}}
\newcommand{\Asim}{\ensuremath {\overrightarrow{A}_{\mathit{sim}}}{\xspace}}
\newcommand{\Ra}{\ensuremath \stackrel{\$}{\leftarrow}{\xspace}}
\newcommand{\Rq}{\ensuremath \stackrel{\$}{\leftarrow}\mathbb{Z}_{q}^{*}{\xspace}}

\newcommand{\Zq}{\ensuremath \mathbb{Z}_{q}^{*}{\xspace}}

\newcommand{\as}{\ensuremath {\leftarrow}{\xspace}}
\newcommand{\prf}{\ensuremath {\texttt{PRF}}{\xspace}}
\newcommand{\bigo}{\ensuremath{\mathcal{O}}}
\newcommand{\dlp}{\ensuremath { \texttt{DLP} }{\xspace}}
\newcommand{\dl}{\ensuremath {\mathit{DL}}{\xspace}}

\newcommand{\A}{$\mathcal{A}$}
\newcommand{\F}{$\mathcal{F}$}

\newcommand{\RNG}{\ensuremath { \texttt{RNG} }{\xspace}}

\newcommand{\nab}{\ensuremath {\overline{\mathit{E1}}}{\xspace}}
\newcommand{\forge}{\ensuremath {\mathit{E2}}}{\xspace}
\newcommand{\nabb}{\ensuremath {\mathit{\overline{E3}}}{\xspace}}
\newcommand{\suc}{\ensuremath {\mathit{Win}}{\xspace}}

\newcommand{\hsim}{\ensuremath {\mathit{H}\mhyphen\mathit{Sim}}{\xspace}}

\newcommand{\agg}[1]{\ensuremath{ \tilde{#1} } }
\newcommand{\batch}[1]{\ensuremath{ \boldsymbol{#1} } }

\newcommand\tab[1][1cm]{\hspace*{#1}}
\newcommand{\algrule}[1][.2pt]{\par\vskip.5\baselineskip\hrule height #1\par\vskip.5\baselineskip}
\newcommand{\specialcell}[2][c]{
	\begin{tabular}[#1]{@{}c@{}}#2\end{tabular}}
\newcommand{\floor}[1]{\lfloor #1 \rfloor}

\usepackage{mathtools, stmaryrd}
\usepackage{xparse} \DeclarePairedDelimiterX{\Iintv}[1]{\llbracket}{\rrbracket}{\iintvargs{#1}}
\NewDocumentCommand{\iintvargs}{>{\SplitArgument{1}{,}}m}
{\iintvargsaux#1} %
\NewDocumentCommand{\iintvargsaux}{mm} {#1\mkern1.5mu,\mkern1.5mu#2}
\newcommand{\romannum}[1]{\uppercase\expandafter{\romannumeral #1\relax}}
\mathchardef\mhyphen="2D 

\newcolumntype{P}[1]{>{\centering\arraybackslash}p{#1}}
\newcolumntype{M}[1]{>{\centering\arraybackslash}m{#1}}

\newcommand{\xor}{\oplus}             

\AtBeginDocument{%
}

\setcopyright{acmcopyright}

\begin{document}

    \title{Lightweight and High-Throughput Secure Logging for Internet of Things and Cold Cloud Continuum}

\author{Saif E. Nouma}
\email{saifeddinenouma@usf.edu}
\author{Attila A. Yavuz}
\email{attilaayavuz@usf.edu}
\affiliation{%
		\institution{University of South Florida}
		\streetaddress{3720 Spectrum Blvd, Interdisciplinary Research Building (IDR)-400}
		\city{Tampa}
		\state{Florida}
		\country{USA}
		\postcode{33612}
	}
    


\begin{abstract}
The growing deployment of resource-limited Internet of Things (IoT) devices and their expanding attack surfaces demand efficient and scalable security mechanisms. System logs are vital for the trust and auditability of IoT, and offloading their maintenance to a Cold Storage-as-a-Service (Cold-STaaS) enhances cost-effectiveness and reliability. However, existing cryptographic logging solutions either burden low-end IoT devices with heavy computation or create verification delays and storage inefficiencies at Cold-STaaS. There is a pressing need for cryptographic primitives that balance security, performance, and scalability across IoT–Cold-STaaS continuum.

In this work, we present {\em Parallel Optimal Signatures for Secure Logging} (\poslo), a novel digital signature framework that, to our knowledge, is the first to offer constant-size signatures and public keys, near-optimal signing efficiency, and tunable fine-to-coarse-grained verification for log auditing. \poslo~achieves these properties through efficient randomness management, flexible aggregation, and multiple algorithmic instantiations. It also introduces a GPU-accelerated batch verification framework that exploits homomorphic signature aggregation to deliver ultra-fast performance. For example, \poslo~can verify $2^{31}$ log entries per second on a mid-range consumer GPU (NVIDIA GTX 3060) while being significantly more compact than state-of-the-art. \poslo~also preserves signer-side efficiency, offering substantial battery savings for IoT devices, and is well-suited for the IoT–Cold-STaaS ecosystem.


\end{abstract}


	\maketitle
	\section{Introduction} \label{sec:Introduction}

Internet of Things (IoT) refers to a large-scale ecosystem of heterogeneous, Internet-connected devices \cite{mosenia2016comprehensive}. Cyber-Physical Systems (CPS) harness IoT devices (e.g., sensors) to monitor the physical environment and construct Digital Twins (DT) for autonomous decision-making and real-time actuation \cite{minerva2020digital}. 
%
Despite their proliferation, IoT devices remain intrinsically vulnerable due to stringent constraints in computation, bandwidth, and memory. These limitations impede the deployment of advanced security mechanisms. Moreover, their exposure to open and untrusted networks further amplifies their susceptibility to diverse attacks (e.g., tampering~\cite{sasi2024comprehensive}). 

System auditing is a critical security measure for early detection of malware and intrusions \cite{ahmad2022hardlog, li2021threat, DSSE:Yavuz:patent:ForwardSecureLog:2015}. It involves maintaining system logs that record security-relevant events (e.g., user activity, errors, breaches), serving as a foundational element for post-incident investigation, attack reconstruction, and forensic analysis. However, modern cyberattacks employ anti-forensics techniques, such as log deletion or manipulation, to hide evidence \cite{mitrelog}, thereby hindering investigators and system administrators from tracing the origin of security incidents. As a result, the importance of ensuring the trustworthiness of logs is vital for both authorities\footnote{\url{https://bidenwhitehouse.archives.gov/briefing-room/presidential-actions/2021/05/12/executive-order-on-improving-the-nations-cybersecurity/}} and practitioners \cite{Logging_Seekable2,Yavuz:2012:TISSEC:FIBAF, chen2024last}.

IoT devices (e.g., smartwatches, pacemakers) often lack the storage capacity to retain log streams locally. Their vulnerabilities (e.g., cyber-physical attacks) further increase the risk of log tampering. A common approach is to securely offload log data to cloud storage for secure archival and forensic analysis \cite{liao2024semantic, nouma2023practical}. While Storage-as-a-Service (STaaS)\footnote{\url{https://www.intel.com/content/www/us/en/cloud-computing/storage-as-a-service.html}} offers scalable infrastructure, retaining append-only log files on cloud servers is prohibitively expensive. A {\em cold storage solution} \cite{chi2020coldstore} is a type of cost-effective data warehouse designed for large-scale archives. Cold-STaaS, therefore, becomes a suitable choice to store rarely accessed yet valuable system logs. Secure offloading requires integrity protection, authentication, and confidentiality to prevent tampering and unauthorized access \cite{wang2011privacy}. In this work, we focus on achieving data authentication and integrity. To this end, an ideal secure log-authentication scheme for IoT-STaaS must, at a minimum, offer the following properties:
\vspace{2pt}

\textsc{1) Scalability, Public Verifiability, and Non-Repudiation.} {\em (i)} The cryptographic solution should be scalable to large IoT networks. {\em (ii)}  It should allow any authorized entity to publicly verify and attest the trustworthiness of information (e.g., metadata, logs) when requested by external parties. {\em (iii)} It should provide the non-repudiation property, in which the signer cannot later deny that they signed the message. This is an 
is essential feature for digital forensics and legal dispute resolution (e.g., financial, health). These features are usually offered by digital signatures \cite{SecureLogAttacks:2017:Hartung,Yavuz:2012:TISSEC:FIBAF}.
\vspace{2pt}

\textsc{2) Logger Efficiency.} Cryptographic mechanisms must efficiently manage the limited resources (e.g., battery, memory) of low-end IoT loggers, which are expected to operate unattended for extended periods.
{\em (i) Efficient Authentication}: Authentication must impose minimal computational overhead to preserve energy. 
{\em (ii) Compact Signatures}: Signatures should be compact to reduce transmission and storage overhead.  
{\em (iii) Optimized Memory}: A lightweight cryptographic code with minimal memory footprint is useful to extend the life of the device (e.g., 8-bit microcontroller) \cite{rohde2008fast}. 
\vspace{2pt}

\textsc{3) Verification Efficiency and Minimal Cloud Storage.} Cold-STaaS platforms manage big data that requires efficient verification and cryptographic compression for secure and scalable archival. 
{\em (i) Real-time Security Audits}: Regular integrity checks are essential to mitigate undetected data tampering and for early breach detection \cite{ahmad2022hardlog}.  
{\em (ii) Regulatory Compliance}: An efficient verification complies with strict integrity requirements (e.g., General Data Protection Regulation (GDPR) \cite{shah2019analyzing}).   
\vspace{2pt}

\textsc{4) Flexible Verification Granularity.} There is a performance and precision trade-off for secure log verification. 
Authenticating the entire log stream with a single authentication tag offers minimal storage and fast batch verification. However, having a single altered log entry (e.g., attack, error) voids the authentication of the entire log stream. Conversely, per-entry signatures enables the highest precision (i.e., the maximum granularity), but with a high storage overhead. Hence, the authentication scheme should permit for both logger and cold storage to adjust the storage granularity and verification precision based on the application requirements~\cite{SecureLogDiMaACMTrans09,Logging_VerifiableExcerpt}. 
\vspace{2pt}

\textsc{5) Configurability for Different Security and Performance Demands.} An effective authentication primitive must support tunable parameters to align with various security requirements and resource constraints across diverse IoT environments. These tunable parameters include, for example, cryptographic primitives (e.g., standard cryptographic or lighweight non-cryptographic hash function \cite{chen2021does}). As such, it enables system designers to tailor latency and memory usage based on the device constraints and adversarial models, enabling practical deployment.
\vspace{1pt}
 
Overall, it is a highly challenging task to devise a digital signature scheme that meets the stringent performance and security requirements of both IoT devices and STaaS simultaneously.  The current state-of-the-art techniques prioritize the needs of either the logger or verifier side while omitting performance and security features for the other side. In the following, we outline the research gap in existing secure logging schemes with a focus on digital signature schemes.

\subsection{Related Work and Research Gap} \label{subsec:Introresearch_gap}
We first discuss the most closely related works to our solutions with a focus on digital signature-based secure logging approaches. We then discuss other relevant and complementary works.
\vspace{2pt}

\underline{Related Work in our Scope}:  The proposed signature framework, {\em Parallel Optimal Signatures for Secure Logging} (\poslo), leverages aggregate digital signature schemes, specialized seed management, and GPU-accelerated batch verification.  The relevant research is outlined below.

\vspace*{1mm}

{\em \text{Aggregate Signatures.} }
\poslo~follows prominent Aggregate Signature (AS)-based secure logging schemes (e.g., \cite{SecureLogDiMaACMTrans09,Yavuz:2012:TISSEC:FIBAF, Logging_VerifiableExcerpt, SecureLogAttacks:2017:Hartung, FAS_Asymptotic}), wherein the logger computes a compact aggregate signature over log entries to enable post hoc attestation. Digital signatures provide data integrity, authentication, public verifiability, and non-repudiation through Public Key Infrastructures (PKI), making them ideal for scalable authentication in IoT and Cold-STaaS. 

Conventional digital signatures (e.g., Ed25519 \cite{Ed25519}) 
incur expensive operations (e.g., modular exponentiation, Elliptic Curve (EC) scalar multiplication), which are costly for resource-limited IoTs.  Moreover, they lack signature aggregation, resulting in $\mathcal{O}(n)$ signature overhead for $n$ log entries, which poses a heavy storage burden on cold storage servers.  Finally, most fail to support batch verification, an important property for efficient large-scale log authentication.

AS schemes \cite{Yavuz:TDSC:OutsourcedDB,BLS:2004:Boneh:JournalofCrypto} enable the aggregation of multiple distinct signatures into a single compact tag, with some schemes also supporting batch verification \cite{ferrara2009practical}.  Therefore, they are instrumental tools for building cryptographic forensic schemes \cite{SecureLogDiMaACMTrans09,Logging_Seekable2,Yavuz:2012:TISSEC:FIBAF, Logging_VerifiableExcerpt}.  Condensed-RSA (C-RSA)~\cite{Yavuz:TDSC:OutsourcedDB}  and BLS~\cite{BLS:2004:Boneh:JournalofCrypto} are two essential AS schemes but with a costly computation in both signing and verification. BLS requires highly expensive pairings and EC scalar multiplication with a heavy special hash function on the verifier and signer sides, respectively. Conversely, C-RSA relies on expensive modular exponentiation during signing with large key sizes. As demonstrated in our evaluations (see Table \ref{tab:OverallPerfComp}), these schemes are not suitable for our envisioned IoT-STaaS applications. 

Forward-secure and Aggregate Signatures (FAS) \cite{Yavuz:2012:TISSEC:FIBAF,FAS_Asymptotic} offer both key-compromise resiliency and signature aggregation. Despite their merits, most FASs introduce high computational and storage overhead either at the signer and/or verifier side. BAF signatures~\cite{Yavuz:2012:TISSEC:FIBAF} are tailored for signer-efficient signatures, but at the cost of a linear public key size.  Our experiments demonstrate that large keys introduce substantial storage overhead at Cold-STaaS. Moreover, they cannot offer storage at different granularities due to fixed public key sizes. Hence, they are not suitable for cold storage applications.

Recent AS schemes with extended properties for IoTs (e.g., \cite{li2020permissioned, verma2021scbs, vallent2021efficient}) are either based on BLS \cite{BLS:2004:Boneh:JournalofCrypto} or Schnorr~\cite{costello2016schnorrq}, inheriting their expensive operations overhead (e.g., pairing, EC scalar multiplication) at the signer. Our empirical analysis confirms that these operations impose high overhead for ultra low-end IoT devices. Moreover, a critical gap in the literature is the lack of performance benchmarks on low-end devices (e.g., 8-bit ATMega2560). In our evaluations, we focus on Ed25519~\cite{Ed25519}, SchnorrQ~\cite{costello2016schnorrq}, and BLS \cite{BLS:2004:Boneh:JournalofCrypto} to represent the signer overhead of these signature primitives, with the note that they do not offer a (full) signature aggregation property. \vspace*{1mm}

{\em \text{Hardware-Accelerated Signatures.}} \poslo~parallel batch verification follows a prominent line of research \cite{kim2024parallel, feng2022accelerating, pan2016efficient, hu2023high, dong2018sdpf, lee2019hybrid} that accelerates digital signature primitives by exploiting the massive parallelism and computational throughput of modern GPU architectures. For example, Dong et al. \cite{dong2018sdpf} improve the RSA signing and verification throughputs, but RSA remains prohibitively expensive for low-end IoT signers, rendering it impractical for IoT-STaaS. Other works (e.g., \cite{kim2024parallel, feng2022accelerating}) demonstrate high-throughput GPU implementations of conventional ECDSA and NIST post-quantum signature standards, but they focus solely on optimizing independent signature operations without tackling structural efficiencies of mutable AS schemes and batch verification algorithms.

\vspace*{1mm}
\underline{Complementary Related Work}:   
\poslo~is a special class of ASs, and therefore does not offer data confidentiality. \poslo~can be complemented by privacy services: (i) data encryption on the logger \cite{goyal2006attribute}, (ii) private auditing on the STaaS side, and (iii) privacy enhancement tools~\cite{wang2010secure}. 

There is a line of work focusing on  Proof of Data Possession (PDP) \cite{ateniese2008scalable} and Proof of Retrievability (PoR)~\cite{anthoine2021dynamic} on the outsourced user data. 
Some efforts also address privacy-preserving public auditing \cite{wang2011privacy}. 
However, these approaches differ from our system model and primary performance objectives. Specifically, PoR/PDP schemes allow IoT devices to offload log files to STaaS providers without generating data signatures or performing authentication checks. Instead, integrity verification is usually initiated by administrators (or STaaS) via interactive audit protocols, whereas AS-based schemes are generally non-interactive. PoR/PDP schemes offer fast audit time that is achieved by Homomorphic Linear Authenticators (HLA) \cite{wang2011privacy}. These enable an external entity to audit the data without retrieving the entire set. However, it comes at the cost of a very high computational overhead on IoT devices since most deployed HLAs (i.e., BLS, RSA) suffer from expensive signing (see Table \ref{tab:OverallPerfComp} and Fig. \ref{fig:energy-consumption}). In a different line,  Li et al. \cite{li2017iot} proposed a public auditing scheme with data sampling.
\vspace{2pt}

Herein, {\em our goal is to achieve optimal signing and small cryptographic payload for IoT devices, while offering high verification efficiency and compact storage at STaaS. By doing so, we permit low-end IoT to actively compute signatures, thereby ensuring public verifiability and non-repudiation}. 

\subsection{Our Contribution} \label{subsec:contribution}
In this work, {\em we create Parallel Optimal Signatures for secure Logging (\poslo), a novel AS-based secure logging framework. To the best of our knowledge, \poslo~is the first to achieve simultaneously small constant-size tags and public key sizes while enabling near-optimal signing and high-throughput parallel batch verification at multiple granularities. These features make \poslo~ideal for IoT-STaaS, where lightweight signing on constrained IoTs and scalable verification on Cold-STaaS are essential}. 

\subsubsection{Main Idea} \sloppy Our key observation is that EC-based signatures (e.g., Ed25519 \cite{Ed25519}, SchnorrQ \cite{costello2016schnorrq}) provide compact signature sizes and superior signing efficiency compared to RSA~\cite{Yavuz:TDSC:OutsourcedDB, liu2010efficient} and pairing-based schemes~\cite{BLS:2004:Boneh:JournalofCrypto}. However, they still incur the cost of a full EC scalar multiplication for each signature generation. Seed-based signing methods~\cite{Yavuz:2012:TISSEC:FIBAF, Yavuz:CNS:2019} improve signing efficiency by employing commitment separation and precomputation during key generation. This gain comes at the expense of significant verifier-side storage and high verification costs. For instance, FI-BAF~\cite{Yavuz:2012:TISSEC:FIBAF} demands 2TB of storage and 264 hours of verification time on commodity hardware for $2^{35}$ log entries, each 32 bytes in size. As detailed in Section~\ref{subsec:Introresearch_gap} and Section~\ref{sec:performance_analysis}, existing AS schemes fall short of jointly minimizing signer and verifier overheads, and do not sufficiently address the challenge of efficient verification at scale for large-volume logs.


\poslo~addresses these bottlenecks through novel design strategies:
{\em (i) Efficient Seed Management}: A tree-based randomness for seed management in constrained signers reduces signer-side seed storage from $\mathcal{O}(n)$ to intermediate $\mathcal{O}(\log_2{n})$ and final $\mathcal{O}(1)$ overhead.  
{\em (ii) Flexible Tag Aggregation}: Our scheme can aggregate additive or multiplicative signature components at arbitrary granularity. This allows for signature aggregation either at the signer per epoch or at the verifier. 
{\em (iii) Tailored Variants}: We propose two \poslo~variants: Coarse-grained signer-optimal \poslo~(\posloc) optimized for minimal computational, bandwidth, and memory overhead; and Fine-grained public-key \poslo~(\poslof) provides fine-grained auditing with a constant-size public key and efficient signing.  
{\em (iii) High-Throughput Parallel Batch Verification}: \poslo~introduces, to the best of our knowledge, the first GPU-accelerated verification (\poslopaver), with orders-of-magnitude speedup compared to AS-based secure logging schemes. 

\subsubsection{Improvements over Preliminary Version.} This article is the extended version of initial \oslo~signatures appeared in~\cite{nouma2023practical} (IEEE/ACM IoTDI'23). Our current article makes substantial contributions over initial \oslo~\cite{nouma2023practical} with new  algorithmic approaches and experimental optimizations: \\ \vspace{-2mm}

\textbf{1) Enhanced Seed Management.} In \poslo, we introduce a new seed management algorithm that replaces \oslo's hash-table-based structure with a stack-based tree structure \cite{Logging_Seekable2}. This reduces the signing overhead and memory footprint on resource-constrained IoT devices.

\textbf{2) New Instantiations ($\poslop$ and $\poslopp$).} We instantiate the message processing and enhanced seed manager of \poslo~with symmetric/arithmetic primitives \cite{Yavuz:ESCAR:SecuritySafety} beyond cryptographic hash functions: (i) $\poslop$: is an AES-based instantiation, optimized for energy efficiency and parallelism on low-end IoT \cite{rohde2008fast} and Cold-STaaS \cite{tezcan2021optimization}. (ii) $\poslopp$: combines AES and modular arithmetic to offer superior performance for small inputs but with a reduced security level.


\textbf{3) Parallel Batch Verification (\poslopaver) and Expanded Performance Analysis.} Most notably, we created \poslopaver, the first GPU-accelerated batch verification system for AS schemes, optimized for \poslo~scheme. \poslopaver~achieves high-throughput verification while maintaining lightweight signing.  It leverages \poslo's commitment separation and the additive homomorphism of its signature aggregation to offload hashing operations to GPUs and perform parallel reduction for computing sub-aggregate ephemeral keys. Our extended performance analysis includes a detailed evaluation of signer-side efficiency with reduced energy consumption and verifier-side scalability with several orders of magnitude speedup from GPU-parallel batch processing.  

\begin{figure*}[t]
	\centering
	\includegraphics[width=0.85\textwidth]{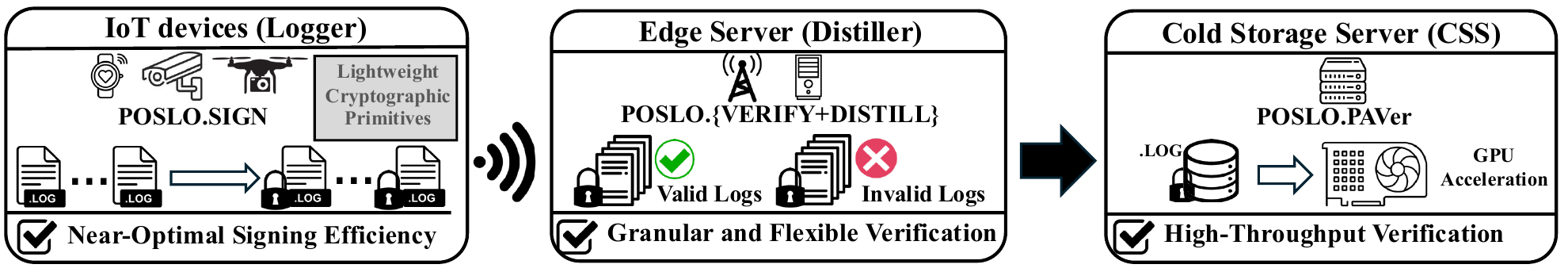}
	\vspace{-2mm}
	\caption{ A high-level illustration of \poslo~system model}
	\label{fig:system-model}
\end{figure*}

These novel features combine to significantly improve the efficiency of \poslo, surpassing not only its previous version \oslo~but also current state-of-the-art secure logging schemes. In Figure \ref{fig:system-model} and Table \ref{tab:OverallPerfComp},  we present a high-level architectural and comparative evaluation of \poslo~against state-of-the-art cryptographic secure logging schemes (details are discussed in Section \ref{sec:performance_analysis}).

\subsubsection{Desirable Properties} The security and efficiency properties of \poslo~are as follows: \\ \vspace{-2mm}

$\bullet$ \textbf{Highly Efficient Verification and Minimal Cold Cryptographic Storage.} Table~\ref{tab:OverallPerfComp} provides a comparative analysis of verification time and cryptographic storage requirements for a 1TB log dataset ($2^{35}$ entries, 32 bytes each). (i) \textit{Verification Time:} Our most efficient instantiation, the AES-based $\poslop$, achieves the lowest verification latency in the Cold-STaaS setting, outperforming C-RSA and BLS by factors of $4.8\times$ and $70\times$, respectively, on commodity CPUs. Its GPU-accelerated version, \poslop.\paver, completes verification of 1TB of log data in just 24.8 seconds, achieving a throughput of $2^{31}$ log entries per second, and significantly outpaces the GPU-accelerated baseline SchnorrQ by several orders of magnitude.
(ii) \textit{Minimal Cold Storage:} \poslo~reduces cryptographic storage to only 0.06~KB, representing a maximum compression compared to EC-based schemes such as Ed25519 and FI-BAF, which require several TBs of storage.

\vspace{1pt}

$\bullet$ \textbf{Granular and Adaptive Verification Architectures.} \poslo~supports flexible verification architectures tailored for constrained signers and cold storage environments.
(i) \textit{Coarse-Grained Signer-Optimal \posloc:} This variant authenticates each log entry and aggregates them into a single epoch-level umbrella signature. \posloc~produces the most compact signature and enables the most efficient verification, outperforming BLS by several orders of magnitude for an epoch of size $n_1 = 256$. Although it initially requires $\mathcal{O}(n_1)$ public key storage at the verifier, it ultimately compresses to a constant-size $\mathcal{O}(1)$ public key in Cold-STaaS.
(ii) \textit{Fine-Grained Public-Key \poslof:} This variant transmits individual signatures and enables on-the-fly aggregation at the verifier (distiller), allowing maximum granularity while maintaining a constant-size public key.
(iii) \textit{Configurable Distillation:} \poslo~introduces a configurable distillation process, enabling entries to be verified and aggregated with flexible granularity. Distillation can be performed by an intermediate verifier (e.g., edge cloud) or directly at the cold storage server.

\vspace{1pt}

$\bullet$ \textbf{Near-Optimal Logging Efficiency.} \poslo~schemes are highly efficient for signing operations, making them particularly well suited for secure logging in constrained IoT environments.
(i) \textit{\posloc:} By eliminating EC scalar multiplications, \posloc~achieves one to several orders of magnitude faster signing compared to the most compact traditional and aggregate schemes, namely SchnorrQ and BLS. It is $2\times$ faster than FI-BAF and significantly more compact in Cold-STaaS, with a $34\times$ verification speedup at the distiller.
(ii) \textit{\poslof:} It offers fine-grained auditability and a constant-size public key. It uses a larger private key for precomputation, which can be optionally replaced with EC scalar multiplication to reduce its size.

\vspace{1pt}

$\bullet$ \textbf{Full-Fledged Implementation.} We implement the \poslo~schemes on both a low-end IoT device and commodity hardware, benchmarking them against existing alternatives. Our experiments confirm that the theoretical advantages of \poslo~yield practical performance benefits. The implementation is open-sourced for public testing and further adaptation:
~~~~  \fbox{\url{https://github.com/SaifNOUMA/POSLO}}

\begin{table*}[t]
	\centering
	\caption{Performance of \poslo~and its counterparts on embedded IoT and cold storage servers }
	\label{tab:OverallPerfComp}
        \footnotesize
	\begin{minipage}{\textwidth}
		\resizebox{\textwidth}{!}{
			\begin{tabular}{|l|@{}c@{}|@{}c@{}|@{}c@{}||@{}c@{}|@{}c@{}||r|@{}c@{}|@{}c@{}||@{}c@{}||@{}c@{}|@{}c@{}|@{}c@{}|@{}c@{}|c|}
				\hline
				\multirow{3}{*}{\textbf{Scheme}}
				& \multicolumn{3}{c||}{ \specialcell[]{\textbf{Logger (Signer)} \\ IoT Device: AtMega2560 (8-bit)} }
				& \multicolumn{2}{c||}{ \specialcell[]{\textbf{Edge Cloud (Distiller)} \\ Commodity Hardware } } 
				& \multicolumn{4}{c||}{ \specialcell[]{\textbf{Cold Storage Server} \\ Commodity Hardware (Desktop)} }
				
				& \multirow{3}{*}{\specialcell[]{\textbf{Dynamic} \\ \textbf{Granularity}}}
				& \multirow{3}{*}{\specialcell[]{\textbf{Granul.} \\ \textbf{Level} \\ (C: Coarse \\ F: Fine) }}
				& \multirow{3}{*}{\specialcell[]{\textbf{Initial/Final} \\ \textbf{Public Key}}} 
				\\
				\cline{2-10}
				& \specialcell[]{\textbf{Signing (sec)}}
				& \specialcell[]{\specialcell[]{\textbf{ Crypto. }}} 
				& \multirow{2}{*}{\specialcell[]{\textbf{Priv Key}}}
				& \multirow{2}{*}{\specialcell[]{\textbf{Ver time} \\  \textbf{(ms)} }}				
				& \multirow{2}{*}{\specialcell[]{\textbf{Distill \& Agg} \\ \textbf{($\mu$s)}}} 
				& \multicolumn{2}{c|}{\specialcell[]{\textbf{Cold Cryptographic Data}}}
				& \multicolumn{2}{c||}{\specialcell[]{\textbf{Verification time}}}
				&
				& 
				& 
				\\
				\cline{7-10}
				& \specialcell[]{\textbf{(per item)}}
				& \specialcell[]{\textbf{ Payload } \\ \textbf{(KB)} }
				& \specialcell[]{\textbf{ Size (KB)}}
				&
				&
				& \specialcell[]{\textbf{Entire Sig/PK Set} \\ \textbf{(for $2^{35}$ entries)} }
				& \specialcell[]{ \textbf{One Sig} \\  \textbf{(KB)} }
				& \specialcell[]{\textbf{AVer} \\ \textbf{(hours)} }
				& \specialcell[]{\textbf{PAVer} \\ \textbf{(seconds)} }
				&
				&
				&
				\\ \hline
				Ed25519~\cite{Ed25519} 
				& $0.869$
				& $8$
				& $0.03$
				& $91.38$
				& -
				& $1$ TB
				& $0.03$
				& $3,406.89$
				& $ $ -
				& $\times$
				& F
				& $\mathcal{O}(1)$ / $\mathcal{O}(1)$
				\\ \hline
				SchnorrQ~\cite{costello2016schnorrq} 
				& $0.323$
				& $8$
				& $0.03$
				& $28.02$
				& -
				& $1$ TB
				& $0.03$
				& $1,044.66$
				& $6 \text{h} 26 \text{m}$
				& $\times$
				& F
				& $\mathcal{O}(1)$ / $\mathcal{O}(1)$
				\\ \hline
				FI-BAF~\cite{Yavuz:2012:TISSEC:FIBAF} 
				& $0.004$
				& $0.05$
				& $0.10$
				& $56.04$
				& $0.02$
				& $2$ TB
				& $0.77$
				& $2,089.32$
				& -
				& $\times$
				& C
				& $\mathcal{O}(n)$ / $\mathcal{O}(n)$
				\\ \hline
				C-RSA \cite{Yavuz:TDSC:OutsourcedDB}
				& $35.828$
				& $0.25$
				& $0.51$
				& $1.48$
				& $5.27$
				& $0.77$ KB
				& $0.25$
				& $55.18$
				& -
				& $\checkmark$
				& C/F
				& $\mathcal{O}(1)$ / $\mathcal{O}(1)$
				\\ \hline
				BLS \cite{BLS:2004:Boneh:JournalofCrypto} 
				& $4.08$
				& $0.05$
				& $0.03$
				& $77.31$
				& $0.02 $
				& $0.1$ KB
				& $0.05$
				& $2,882.33$
				& -
				& $\checkmark$
				& C/F
				& $\mathcal{O}(1)$ / $\mathcal{O}(1)$
				\\ \hline \hline
				SOCOSLO \cite{nouma2023practical}
				& $0.005$
				& $0.03$
				& $0.06$
				& $1.27$
				& $1.45$
				& \multirow{2}{*}{$0.06$ KB}
				& \multirow{2}{*}{$0.05$}
				& \multirow{2}{*}{$45.35$} 
				& \multirow{2}{*}{-}
				& $\checkmark$
				& \textbf{C}
				& $\mathcal{O}(n / n_1)$/$\mathcal{O}(1)$
				\\  \cline{1-6} \cline{11-13}
                    FIPOSLO \cite{nouma2023practical}
				& $0.016$
				& $8$
				& $65.6$
				& $28.38$
				& $0.37$
				& 
				& 
				& 
				& 
				& \textbf{$\checkmark$}
				& \textbf{F}
				& $\mathcal{O}(1)$/$\mathcal{O}(1)$
				\\  \hline

				$\textbf{\poslofp}$
				& $\boldsymbol{0.002}$
				& $\boldsymbol{0.03}$
				& $\boldsymbol{0.06}$
				& $\boldsymbol{1.92}$
				& $\boldsymbol{1.45}$
				& \multirow{2}{*}{$\boldsymbol{0.06}$ KB}
				& \multirow{2}{*}{$\boldsymbol{0.05}$}
				& \multirow{2}{*}{\textbf{$\boldsymbol{71.58}$}} 
				& \multirow{2}{*}{$\boldsymbol{24.8}$}
				& $\checkmark$
				& \textbf{C}
				& $\mathcal{O}(n / n_1)$/$\mathcal{O}(1)$
				\\  \cline{1-6} \cline{11-13}
				$\textbf{\poslofp}$
				& $\boldsymbol{0.014}$
				& $\boldsymbol{8}$
				& $\boldsymbol{65.6}$
				& $\boldsymbol{28.38}$
				& $\boldsymbol{0.37}$
				& 
				& 
				& 
				& 
				& \textbf{$\checkmark$}
				& \textbf{F}
				& $\mathcal{O}(1)$/$\mathcal{O}(1)$
				\\  \hline
				
			\end{tabular}
		}
		
		\vspace{2pt}
		{
			\scriptsize
			The experiment settings, hardware/software configurations, and cryptographic parameters are given in Section \ref{sec:performance_analysis}. 
			We chose our counterparts to cover the primary signature schemes deployed for secure logging in IoT environments (see Section \ref{sec:performance_analysis} for selection rationale).
			The total number of entries and the epoch size are $n=2^{35}$ and $n_2=2^8$, respectively. The input message is of size $32$ bytes. The overall data is of size 1TB.
			At the logger (signer), the signature is measured per epoch, and signing time (in seconds) is given for a single entry. 
			At the distiller, verification time (in ms) is for all entries within an epoch. 
			At the cold storage server, the cryptographic storage is the total size of signatures and public keys needed to verify $n$ entries. The verification time (\aver~on x86/64 and \paver~on GTX 3060) is the total runtime for batch verifying $n$ items. 
		}
	\end{minipage}	
    \vspace{-4pt}
\end{table*}

\section{Preliminaries} \label{preliminaries}

\textbf{ Notations and Algorithmic Definitions.} ~$||$ and $|x|$ denote concatenation and the bit length of variable $x$, respectively.
$x \Ra \mathcal{S}$ means that the variable $x$ is randomly selected from the finite set $\mathcal{S}$ using a uniform distribution.
$|\mathcal{S}|$ denotes the cardinality of set $\mathcal{S}$. 
$\{0,1\}^*$ denotes a set of binary strings of any finite length.
For $q$ a prime power, $\mathbb{Z}_q$ denotes a finite field of order $q$ and $\Zq$ denotes $\mathbb{Z}_q \textbackslash \{0\}$.
$\boldsymbol{x}=\{x_i\}_{i=1}^{n}$ denotes the set of items $(x_1, x_2, \ldots, x_n)$. $\log x$ denotes $\log_2 x $.
$H : \{0,1\}^* \rightarrow \{0,1\}^{h}$ is a cryptographic hash function~\cite{CryptoHandBook}, where $h$ is the output size. 
$\prf_{0,1}: \{0,1\}^\kappa \rightarrow \{0,1\}^\kappa$ are two pseudorandom functions. 
$n$ denotes the maximum number of items to be signed in a given signature scheme.  Our schemes operate in epochs in which $n_2$ items are processed, with a total of $n_1$ epochs available such that $n=n_1 \cdot n_2$. 
The variable $\agg{x}$ denotes an aggregate value. 
The variable $\agg{x_i}$ denotes the aggregate value for an epoch $i$. 
$m_i^j \in \boldsymbol{m_i}$ means that $m_i^j$ belongs to a vector of $n_2$ items $\boldsymbol{m_i}=\{m_i^j\}_{j=1}^{n_2}$. 
$\vec{\boldsymbol{m}}= \{\boldsymbol{m_i}\}_{i\in \boldsymbol{I}}$ denotes a super vector where each $\boldsymbol{m_i}$ contains $n_2$ items and $\boldsymbol{I}$ are epoch indexes of $\vec{\boldsymbol{m}}$. 

\begin{definition}
	\label{def:mmo}
	Matyas-Meyer-Oseas (MMO) \cite{CryptoHandBook} consists of a generic single-length cryptographic hash algorithm $\mmo$ that generates an $\ell$-bit digest given a pre-defined $\ell$-bit block cipher ($E$) and its initialization vector $h_0$. \mmo~is defined as follows:
	
	\begin{enumerate}[-]
            \item $h \as \mmo(m)$: Given a $k$-bit string $m$, it splits $m$ into $t=\floor{\frac{k+1}{\ell}}$ $\ell$-bit blocks $\{m_i\}_{i=1}^t$ (padding the last block). Then, it computes $h_i=E_{h_{i-1}}(m_i) \xor m_i , \forall i=1,\ldots, t$. Finally, it outputs $h \as h_t$. 
	\end{enumerate}
\end{definition}

\begin{definition}
	\label{def:mdc}
	MDC-2~\cite{CryptoHandBook} consists of a generic double-length cryptographic hash algorithm $\mdc$ that generates a $2\ell$-bit digest using a pre-defined $\ell$-bit block cipher ($E$) and two initialization vectors $\{h_0, h_0'\}$. \mdc~is defined as follows:
	
	\begin{enumerate}[-]
            \item $h \as \mdc(m)$: Given a $k$-bit string $m$, it splits $m$ into $t=\floor{\frac{k+1}{\ell}}$ $\ell$-bit blocks  $\{m_i\}_{i=1}^t$ (padding the last block). Then, it computes $h_i=E_{h_{i-1}}(m_i) \xor m_i, \text{ and } h_i'=E_{h_{i-1}'}(m_i) \xor x_i, \forall i=1,\ldots, t$, interprets $h_i \as h_i^1 \| h_i^2, h_i' \as {h_i^1}' \| {h_i^2}'$, and updates $h_i \as h_i^1 \| {h_i^2}', h_i' \as {h_i^1}' \| {h_i^2}$. Finally, it outputs $h \as h_t \| h_t'$. 
	\end{enumerate}
\end{definition}

\begin{definition}
	A single-signer multiple-time aggregate signature scheme \asgn~consists of four algorithms \texttt{(Kg}, \texttt{ASig}, \texttt{Agg}, \texttt{AVer)} defined as follows:
	
	\begin{enumerate}[-]
		\item $ (I,\sk, \pk) \as \asgnkg(1^{\kappa}, n ) $: Given the security parameter $\kappa$ and maximum number of generated signatures $n$, it returns a private/public key pair $(\sk,\pk)$ with a public system-wide parameter $I$.
		\item $ \sigma_i \as \asgnsig(\sk, m_i) $: Given $\sk$ and a message $m_i$, it returns a signature $\sigma_i$ as output.
		\item $ \agg{\sigma} \as \asgnagg(\sigma_1, \ldots, \sigma_\ell) $: Given $\ell$ signatures $\{\sigma_i\}_{i=1}^\ell$, it returns an aggregate signature $\agg{\sigma}$.
		\item $ b \as \asgnaver(\pk, \{m_i\}_{i=1}^\ell, \agg{\sigma}) $: Given $\pk$, a vector of messages $\{m_i\}_{i=1}^\ell$ and their corresponding aggregated signature $\agg{\sigma}$, it outputs $b=1$ if $\agg{\sigma}$ is valid or $b=0$ otherwise.
		
	\end{enumerate}
\end{definition}

\poslo~schemes rely on the intractability of \textit{Discrete Logarithm Problem (DLP)} \cite{CryptoHandBook}. Remark that, while our notations are based on DLP, our implementation is based on {\em Elliptic Curves (EC)} for efficiency, and {\em the definitions also hold under ECDLP~\cite{costello2016schnorrq}}. 

\begin{definition}
	\label{def.dlp}
	Let $q$ and $p$ two prime numbers where $p>q$, $\mathbb{G}$ be a cyclic group of order $q$, $\alpha$ be a generator of $\mathbb{G}$, and \dlp~attacker \A~be an algorithm that returns an integer in $\mathbb{Z}_q$. 
	We consider the following experiment:
	
	Experiment $Expt_{\mathbb{G}, \alpha}^{DL}(\mathcal{A})$:
	\newline 
	\tab $y \Ra \mathbb{Z}_q^*$, $Y \as \alpha^{y} \mod p$,
	$y' \as \mathcal{A}(Y)$
	\newline 
	\tab If $\alpha^{y'} \mod p = Y$ then return 1, else return 0
	\newline
	The DL-advantage of \A~in this experiment is defined as:
	$\advAdl = Pr[ Expt_{\mathbb{G}, \alpha}^{DL}(\mathcal{A}) = 1]$.
	\newline
	The DL-advantage of $(\mathbb{G}, \alpha)$ in this experiment is defined as follows:
	$\advdl = \underset{\mathcal{A}}{\max}\{\advAdl\}$, where the maximum is over all \A~having time complexity $t$.
\end{definition}

\poslof~uses Boyko-Peinado-Venkatesan (\bpv) generator \cite{boyko1998speeding}.
It reduces the computational cost of expensive operations (e.g., EC scalar multiplication) via a pre-computation technique.

\begin{definition} \label{def:bpv}
    \bpv~generator consists of two algorithms $(\bpvoff, \bpvon)$:
    
    \begin{enumerate}[-]
	\item $\underline{ (\Gamma, v, k) \as \bpvoff(1^{\kappa} , p , q , \alpha) }$:
	It chooses \bpv~parameters $(v,k)$ as the size of the pre-computed table and number of randomly selected elements, respectively.
	Then, it generates the pre-computed table $\Gamma = \{r_i, R_i\}_{i =1}^v$.
	\item $\underline{ (r , R) \as \bpvon(\Gamma) }$: Generate a random set $S \in \{ 1 , \ldots , v \}$ of size $|S|=k$ and compute a one-time commitment pair $(r \as \sum_{i \in S}{r_i \mod q} ~,~R \as \prod_{i \in S}{R_i} \bmod p)$.
\end{enumerate}
\end{definition}

\vspace{2pt}
\noindent \textbf{Parallel Computing on GPU.} 
A Graphical Processing Unit (GPU) is an electronic circuit that aims to accelerate computer computations by leveraging its parallel structure. It consists of a set of Stream Multiprocessors (SMs). Each SM contains a set of relatively constrained cores (e.g., 1320 MHz base clock frequency). The GPU provides different memory types 
{\em (i) global memory ($\approx$GBs)}: can be accessed by all threads in SMs. It is an off-chip memory and represents the data transfer medium between system memory and GPU with high access throughput ($\approx$100 GB/s via PCI-Express link). 
{\em (ii) shared memory ($\approx$KBs)}: is an on-chip memory shared among cores in a single SM and provide a much faster memory access compared to (i). 
{\em (iii) register file}: is an on-chip memory that resides in each core. It has the fastest memory access to hold the frequently accessed and local data.

\vspace{2pt}
\noindent \textbf{CUDA.} It is a program computing platform and application programming interface developed by NVIDIA \cite{kirk2007nvidia}. It allows programmers to define and execute operations on GPUs. 
A CUDA program launches a {\em kernel} function that executes multiple {\em threads} in parallel. The fundamental execution unit is a {\em warp}, consisting of 32 threads, and multiple warps form a {\em block}. Blocks are organized into a {\em grid}, containing all the threads. 
Before executing the kernel, (i) the programmer must specify the number of both threads and blocks (ii) the input data is copied from system memory to global (device) memory. After kernel execution, the output data is copied from global memory to system memory to resume CPU processes. 
Note that CUDA follows Single Instruction Multiple Threads (SIMT) model. Running different tasks on threads causes warp divergence, forcing sequential execution. 


	\section{Models} \label{sec:models}

\subsection{System Model}
Our system model follows a well-established AS-based secure logging framework (e.g., \cite{SecureLogDiMaACMTrans09, Yavuz:2012:TISSEC:FIBAF, SecureLogAttacks:2017:Hartung, Logging_VerifiableExcerpt}), in which the logger (i.e., IoT device) generates authentication tags for its log entries to enable future public verification.
We consider an {\em IoT–Cloud continuum}, where numerous IoT devices produce log streams and report them to an edge or core cloud for processing, verification, and archival. As illustrated in Figure \ref{fig:system-model}, our model consists of three main entities:

\textit{(i) Logger (Signer):} This component represents resource-constrained end-user IoT devices (e.g., medical sensors) that collect sensitive data such as health or personal information. These devices periodically send the data along with their corresponding log entries to a nearby edge server. They are limited in computational power, memory, energy, and communication bandwidth.

\textit{(ii) Distiller:} This is an authorized entity responsible for verifying log entries using associated public keys. For example, in a smart-building scenario, IoT sensors may periodically transmit sensing reports to a local edge cloud. Before transferring logs to cold storage, the edge cloud performs a distillation process that generates {\em Cold Cryptographic Data (\ccd)}, a curated dataset containing valid log batches with compressed, adjustable-granularity cryptographic tags. Invalid (flagged) entry-signature pairs are stored individually. In most real-world applications, such invalid entries are rare, so the storage of \ccd~is dominated by valid entries. After distillation, the edge cloud uploads \ccd~to cold storage servers for long-term archival and audit readiness.

\textit{(iii) Cold Storage Server (\css):} This server provides a Cold-STaaS platform within the IoT–STaaS continuum. As discussed in Section~\ref{sec:Introduction}, these systems require periodic audits to demonstrate the trustworthiness of archived digital content~\cite{ahmad2022hardlog, shah2019analyzing}. Verifiers regularly check the authenticity and integrity of log data maintained in \css. For simplicity, we assume verifiers are integrated within \css~in our system model, and \css~is equipped with a GPU hardware card. 

\subsection{Threat and Security Model} 

We follow the threat model of cryptographic audit log techniques originally introduced by Schneier et al. in \cite{SecureLogBruceACMTrans99} and then improved in various subsequent cryptographic works~\cite{Yavuz:2012:TISSEC:FIBAF,Logging_VerifiableExcerpt}.  In this model, the adversary \A~is an active attacker that aims to forge and/or tamper with audit logs to implicate other users. The state-of-the-art cryptographic secure logging schemes rely on digital signatures to thwart such attacks with public verifiability and non-repudiation.  As stated in Section \ref{sec:Introduction}, we focus on signer-efficient (EC-based) AS-based approaches due to their compactness and fast batch verification. 

We follow the {\em Aggregate Existential Unforgeability Under Chosen Message Attack} (\AEUCMA)~\cite{BLS:2004:Boneh:JournalofCrypto} security model that captures our threat model. \AEUCMA~considers the homomorphic properties of aggregate signatures and can offer desirable features such as log order preservation (if enforced)  and truncation detection for signature batches \cite{Yavuz:2012:TISSEC:FIBAF}.  \poslo~schemes are single-signer aggregate signatures, and therefore we do not consider inter-signer aggregations.  

\begin{definition}
	\label{Def:AIEUCMA}
	\AEUCMA~experiment for \asgn~is as follows:
	
	Experiment $\mathit{Expt}_{\asgn}^{\AEUCMA}(\mathcal{A})$
	\begin{enumerate}[]
		\setlength{\parskip}{0pt}
		\setlength{\parsep}{0pt}
		
		\item $(I,sk,PK)\leftarrow		\asgnkg(1^{\kappa},n)$,
		
		\item $(\boldsymbol{m^{*}},\sigma^{*})\leftarrow  \mathcal{A}^{\ro,~\asgnsig_{sk}(.)}(PK)$,
		
		\item If $\asgnaver(PK,\boldsymbol{m^{*}},\sigma^{*})=1~\text{and}~\boldsymbol{m^*} \not\subset \{ \boldsymbol{m_j} \}_{j=1}^{n_1}$, return $1$ otherwise return $0$. 
	\end{enumerate}
	\noindent
	
	The \AEUCMA~of \A~is defined as
	\begin{eqnarray*}
		\advAsgn=Pr[\mathit{Expt}_{\asgn}^{\AEUCMA}(\mathcal{A})=1].
	\end{eqnarray*}
	
	The  \AEUCMA~advantage of \asgn~is defined as
	\begin{eqnarray*}
		\advsgn=\max_{\mathcal{A}}\{Adv_{\asgn}^{\AEUCMA}(\mathcal{A})\},
	\end{eqnarray*}
	where the maximum is over $\mathcal{A}$ having time complexity
	$t$, with at most $n'$ queries to the random oracle \ro~and $n$ queries to $\asgnsig(.)$.
\end{definition}

The oracles reflect how \poslo~works as an \asgn~scheme. The signing oracle $\asgnsig(.)$~returns an aggregate signature $\agg{\sigma}$ on a batch of messages $\vec{\boldsymbol{m}}=(\boldsymbol{m_1},\ldots,\boldsymbol{m_{n_1}})$ computed under $sk$. $\asgnagg(.)$~aggregates the individual (or batch) signatures of these messages. $\asgnagg(.)$~can be performed during the signing or before verification (e.g., in the distillation). It can aggregate additive or multiplicative components $\delta_i \in \sigma_i$. \ro~is a random oracle from which \A~can request the hash of any message of her choice up to $n'$ messages. In our proofs (see Section \ref{sec:SecurityAnalysis}), the cryptographic hash function $H$ is modeled as a random oracle~\cite{CryptoHandBook}	via \ro.


\section{Proposed Schemes} \label{sec:proposed_schemes}
Our goal is to design novel cryptographic secure logging schemes that satisfy the stringent requirements of low-end IoT devices by enabling efficient signing and compact signatures, while also ensuring fast verification and optimal storage at the Cold-StaaS. Specifically, we aim to achieve: \textit{(i)} Near-optimal signer efficiency without relying on costly operations such as EC scalar multiplication, \textit{(ii)} Compact aggregate tag storage and communication overhead, \textit{(iii)} $\mathcal{O}(1)$ final cryptographic storage at cold storage, i.e., constant-size public key and signature for maximal compression, \textit{(iv)} Fast batch verification over large message sets, and \textit{(v)} Flexible aggregation support at any desired granularity, either at the signer or verifier side.

EC-based signatures (e.g., Ed25519~\cite{Ed25519}, SchnorrQ~\cite{costello2016schnorrq}) provide compact signature sizes and improved signing efficiency over RSA~\cite{Yavuz:TDSC:OutsourcedDB, liu2010efficient} and pairing-based schemes~\cite{BLS:2004:Boneh:JournalofCrypto}. However, they still require at least one EC scalar multiplication per signing operation, which limits their applicability in constrained environments. Prior efforts to address this include: (i) \textit{Precomputation}, which shifts signing costs to key generation through commitment precomputation. While this reduces signing time, it incurs linear storage at the signer, making it impractical for resource-limited devices. (ii) \textit{Seed-based signing}, which replaces precomputed public commitments with one-time seeds~\cite{Yavuz:2012:TISSEC:FIBAF, Yavuz:CNS:2019}. This transforms Schnorr signatures~\cite{schnorr1991efficient} into one-time aggregate signatures, moving the generation and storage of expensive commitments $(R \as \alpha^r \bmod q, r \Rq)$ to key generation and the verifier, respectively. The signing process decouples the message $m$ from the commitment by substituting $H(m || R)$ with $H(m || x)$, where $x$ is a one-time randomness. Since $x$ cannot be revealed before signing and is non-aggregatable, this leads to $\mathcal{O}(n)$ storage at Cold-STaaS (e.g., 2TB for $2^{35}$ log entries of 32 bytes each) and expensive batch verification (hundreds of hours to verify 1TB), as detailed in Section~\ref{sec:performance_analysis}. In Sections~\ref{subsec:Introresearch_gap} and~\ref{sec:performance_analysis}, we analyze AS-based signatures and their limitations in detail. Overall, existing AS-based secure logging solutions either favor signer efficiency at the cost of excessive verifier-side overhead or vice versa, failing to provide a scalable and efficient path for verifying large volumes of log data in Cold-STaaS environments.

In \poslo, we address the persistent {\em signer-versus-verifier bottleneck} through a set of novel and synergistic design strategies:
{\em (i) Efficient Seed Management:} We introduce a tree-based randomness structure that reduces signer-side seed storage from $\mathcal{O}(n)$ to intermediate $\mathcal{O}(\log_2{n})$ and ultimately to $\mathcal{O}(1)$ in the final form. This design eliminates the need for linear signer storage while preserving deterministic and secure EC-based signing.
{\em (ii) Flexible Tag Aggregation:} \poslo~supports adaptive aggregation of additive and multiplicative signature components with configurable granularity. This allows aggregation either at the signer per epoch or on-demand by the verifier.
{\em (iii) Tailored Variants:} We design two core \poslo~variants to serve distinct application needs: the {\em Coarse-grained signer-optimal} \posloc, which minimizes signer-side overhead and maximizes compression, and the {Fine-grained public-key} \poslof, which enables precise auditing with constant-size public key and efficient signing.
{\em (iv) High-Throughput Parallel Batch Verification:} We introduce \poslo.\paver, the first GPU-accelerated batch verification framework (to our knowledge) for mutable AS schemes. \poslo~eliminates sequential bottlenecks such as repeated hashing, enabling order-of-magnitude speedups over traditional digital signature schemes.
{\em (v) Multiple Instantiations}: We provide two distinct instantiations: \poslop, and \poslopp, each offering unique performance-security trade-offs. \poslo+~utilizes MMO and MDC-2 constructions using AES as a block cipher to replace SHA-256 in the original \poslo~for the key derivation $\prf$ and message hashing $H$, respectively. \poslopp~replaces MDC-2  with modular addition \cite{chen2021does}, enabling reduced overhead in low-end signers.  



We first describe our seed management architecture and associated data structures that addresses the signer storage challenge in Section~\ref{subsec:opas_seed_management}. We then present the \posloc~and \poslof~schemes, detailing their efficiency in signing, storage compression, and batch verification at configurable granularity. Lastly, we introduce \poslo.\paver, our parallelized batch verification engine that achieves significant throughput gains, facilitating scalable and efficient log auditing at Cold-STaaS.

\subsection{POSLO Data Types and Seed Management} 
\label{subsec:opas_seed_management}

\begin{figure*}[ht!]
    \begin{subfigure}[b]{0.5\textwidth}
        
	\centering
	\begin{minipage}{0.95\textwidth}
		\centering
		\noindent \fbox{\parbox{\columnwidth} {
				\scriptsize
				
				\begin{algorithmic}[1]
					\Statex $\underline{ x_{d_1}[i_1] \as \sconst(x_{d_0}[i_0], d_{1}, i_1) }$: require $d_1 < d_0$ and $i_0 \leq  i_1 \cdot 2^{d_1-d_0} < i_0+1 $
					\vspace{3pt}
					
					\State $x_p \as x_{d_0}[i_0]$ , $i \as i_1 - i_0 \cdot 2^{d_0-d_1}$
					
					\For{$j = d_0-d_1-1 , \ldots, 0$}
					\State  $x_p= 
					\begin{cases}
						\prf_0(x_p),& \text{if } \floor{i / 2^j} \equiv 0  ~\bmod~2  \\
						\prf_1(x_p),& \text{if } \floor{i / 2^j}  \equiv 1  ~\bmod~2
					\end{cases}$
					\EndFor
					
					\State \Return $x_{d_1}[i_1] \as x_p$
				\end{algorithmic}
				\algrule
				
				\begin{algorithmic}[1]
					\Statex $\underline{ ({\ds}_i, x_0[i]) \as \sso(\ds_{i-1}, x_{D}[0], i) }$: require $0 \leq i < 2^D $
					\vspace{3pt}
					
					\State $d_1 \as 0$ , $i_1 \as i$ , $\ds_i \as \ds_{i-1}$
					\State \textbf{while} Depth($x \as \dstop(\ds_{i}) ) = d_1$ \textbf{do}
					\State \indent $ x \as \dspop(\ds_{i}) $ , $ d_1 \as d_1+1$ , $i_1 \as \floor{i_1/2} $
					\State $\dspush(\ds_i, x_{d_1}[i_1]) $ , where $ x_{d_1}[i_1] \as \sconst(x_D[0], d_1, i_1) $
					
					\State \Return $({\ds}_i , x_0[i] \as \sconst(x_{d_1}[i_1] , 0 , i ) )$
				\end{algorithmic}
				\algrule
				
				\begin{algorithmic}[1]
					\Statex $\underline{ x_0[i'] \as \sret(\ds, i') }$:
					\vspace{3pt}
					
					\State $x_d[i] \as \dspop(\ds)$
					\State \textbf{if} $i' \geq (i+1) \cdot 2^{d} $ \textbf{then return} $\perp$
					\State \textbf{while} $ i' < i \cdot 2^d $ \textbf{do}
					\State \indent $x_d[i] \as \dspop(\ds) $
					
					\State \Return $x_0[i'] \as \sconst( x_d[i] , 0 , i' ) $
				\end{algorithmic}
		}}
		
	\end{minipage}
    
	\caption{Seed Management Functions (\smf)}
	\label{alg:sso}
    \end{subfigure}
    \hfill
    \begin{subfigure}[b]{0.49\textwidth}
	\centering
	
        \includegraphics[scale=0.23]{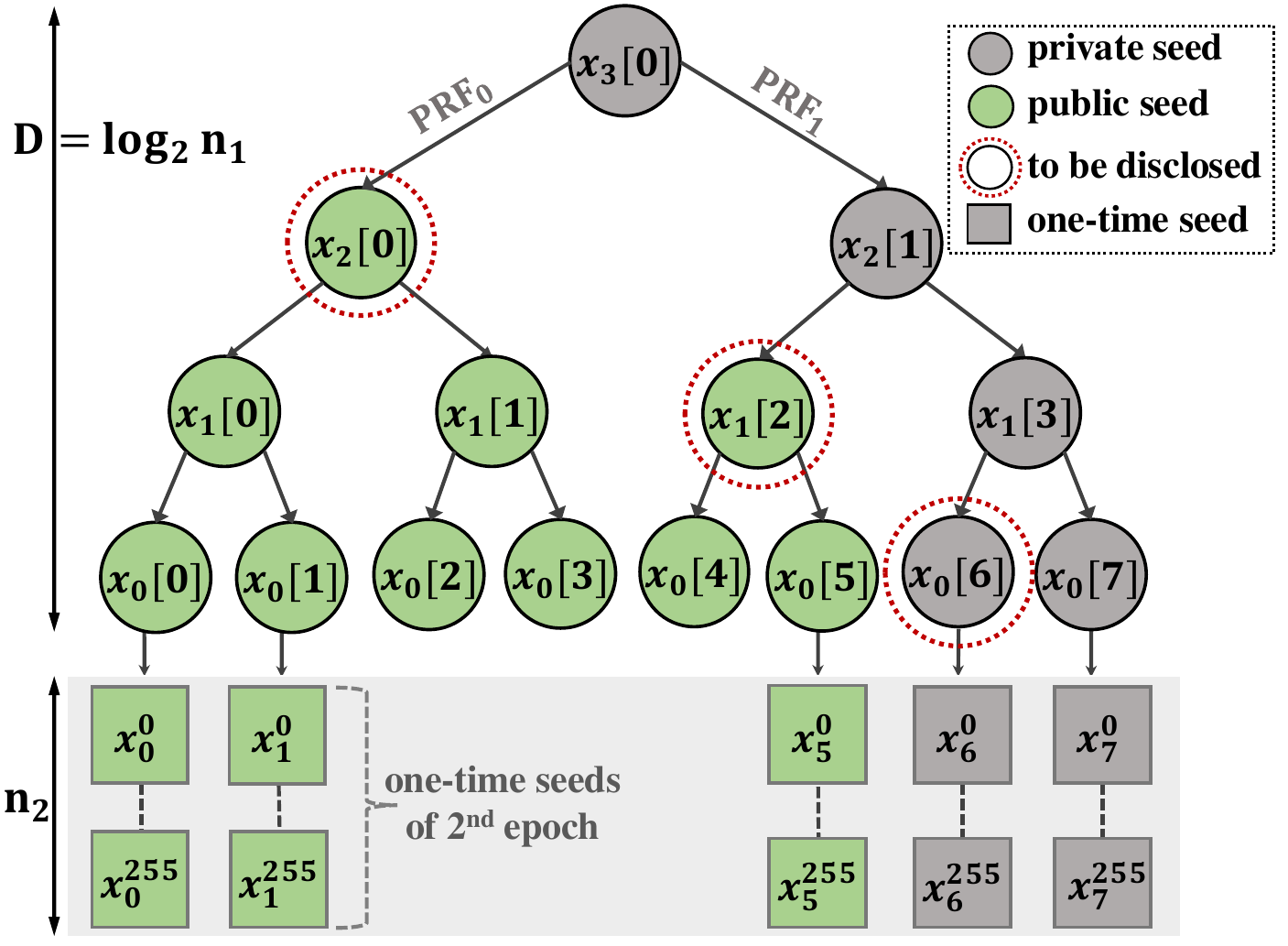}
	\includegraphics[scale=0.33]{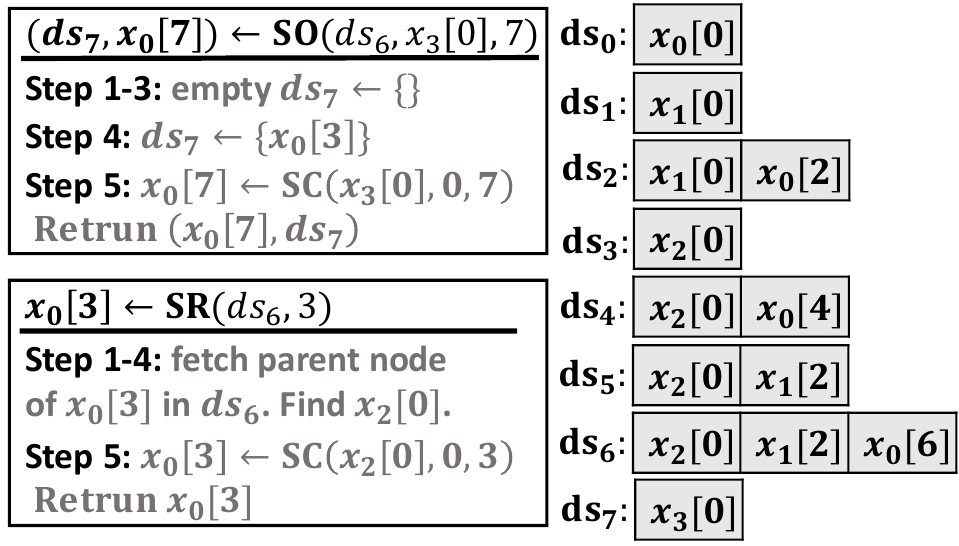}
	\caption{Example of \smf~execution ($n_1=2^3,n_2=2^8$)}
	\label{fig:seed-management}
	
    \end{subfigure}

	\vspace*{-2mm}
\end{figure*}

\noindent \textbf{\poslo~Data Types:}
\poslo~Tree-based structure (\poslot) is designed for seed storage and management, in which the leaves represent one-time seeds $x$, and the left and right children of each node are computed using $\prf_{0,1}$, respectively. Let $n_1$ and $D=\log{n_1}$ be the total number of leaves and tree height, respectively. The \poslot~root node, $x_D[0]$, serves as the source of randomness in order to derive $2^D$ one-time seeds $x_0[i], i = 0, \ldots 2^D-1$. The inner nodes of \poslot~$x_d[i]$, $0 \leq i < 2^{D-d}$, where $d = 0, \ldots, D-1 $, are  computed as follows:  

\begin{equation}
x_d[i]= 
\begin{cases}
	\prf_0\left(x_{d+1}\left[ \left\lfloor \frac{i}{2} \right\rfloor \right]\right), & \text{if } i \equiv 0 \pmod{2} \\
	\prf_1\left(x_{d+1}\left[ \left\lfloor \frac{i}{2} \right\rfloor \right]\right), & \text{if } i \equiv 1 \pmod{2}
\end{cases}
\end{equation}

{\em Disclosed Seeds (\ds)} is a stack data structure with the following operations: $\dsinit(\ds)$ initializes the stack \ds~to an empty stack. $x \as \dstop(\ds)$ returns the top element of the stack. $\dspush(\ds,x)$ pushes $x$ onto the top of \ds. $x \as \dspop(\ds)$ removes and return the top element of \ds. 

\ds~maintains the disclosed one-time seeds $\{x_0[i]\}_{i=0}^{n_1-1}$ in a $\bigo(\log(n_1))$ compact storage by replacing leaves with their parent nodes whenever it is possible.

%
%

%
%
%

\textbf{Seed Manager}: The Seed Management Functions (\smf) are described in Figure \ref{alg:sso}: 
{\em (i)  Seed Computation (\sconst)} computes the node $x_{d_1}[i_1]$  from the source node $x_{d_0}[i_0]$  by traversing the \poslot~tree. 
{\em (ii) Seed storage Optimizer} (\sso) progressively discloses ancestor nodes as the signer completes epochs. For a given leaf index $i$, root $x_D[0]$, and prior $\ds_{i-1}$, it outputs $\ds_i$ that represents disclosed nodes during the epoch $i$.
{\em (iii) Seed Retrieval (\sret)} takes a \ds~instance and leaf index $i$ as inputs. It returns the seed $x_0[i]$ if \ds~contains an ancestor node for leaf index $i$.

	
	

An instance of \poslot~is provided in Figure \ref{fig:seed-management}, where $(n_1=2^3, n_2=2^8)$. It shows the \poslot~status after completing the $6^{\text{th}}$ epoch and the execution of \smf~algorithms. The seeds, to be disclosed, are highlighted. They can be determined by running \sso~algorithm. The \sso~output is:
$(\ds_6, x_0[6]) \as \sso(\ds_5, x_3[0], 6)$
$\text{ where } \ds_5 = \{ x_2[0] , x_1[2]\}$.  The output $\ds_6$ is equal $\{ x_2[0] , x_1[2] , x_0[6]\}$.

The advantage of \poslot~seed management is apparent over the linear disclosure of one-time commitments in Schnorr-like schemes. It reduces both $\mathcal{O}(n)$ signer transmission and verifier storage into $\mathcal{O}(\log{n_1})$.
Upon finishing all epochs, the signer discloses the \poslot~root $x_D[0]$, enabling $\mathcal{O}(1)$  storage at verifiers.

%
%

%

\subsection{Coarse-grained signer-optimal POSLO (POSLO-C)} \label{subsec:socosa}

\posloc~offers a near-optimal signing efficiency in terms of both computational and storage overhead. It offloads an aggregate tag upon signing an epoch of individual log entries. Unlike previous EC-based signature designs, \posloc~pre-stores a $\mathcal{O}(n_1)$ sublinear number of public commitments $(R)$ at the verifier side and compacts them after receiving the authenticated logs from IoT devices. In the following, we give a detailed description of \posloc~algorithms.

\subsubsection{POSLO-C Digital Signature Algorithms} We give the aggregate signature functions of \posloc~in Fig. \ref{alg:sopas}. 


\begin{figure*}[ht!]
	\centering
	
	\begin{subfigure}[b]{0.5\textwidth}
		\centering
		\noindent \fbox{\parbox{0.99\textwidth} {
				\scriptsize
				
				\begin{algorithmic}[1]
					\Statex $\underline{(I, \sk, \pk)\as \poslockg(1^{\kappa}, n)}$:
					\vspace{3pt}
					\State Choose $(n_1,n_2)$ s.t. $n=n_1 \cdot n_2$ and $n_1=2^D$ where $D \in \mathbb{N}^*$.
					\State Generate large primes $q$ and $p > q$ such that $(p-1)$ divides $q$. Select a generator $\alpha$ of the subgroup $\mathbb{G}$ of order $q$ in $\mathbb{Z}_q^*$. 
					\State $y \Ra \mathbb{Z}_q^*$ , $Y \as \alpha^y \bmod p$
					\State $x_D[0] \Ra \{0,1\}^{\kappa}$ , $r \Ra \mathbb{Z}_q^*$ , $\dsinit(\ds_0)$
					
					\For{$i =0, \ldots, n_1-1$}
					\State $\agg{r_i} \as \sum_{j=0}^{n_2-1}{r_i^j} \bmod q$, where $r_i^j \as \prf_0(r \mathbin\Vert i \mathbin\Vert j)$
					\State $\agg{R_i} \as \alpha^{\agg{r_i}} \bmod p$
					\EndFor
					\State $\sk \as (y, r, x_D[0])$ , $ \pk \as (Y, \batch{R})$, where $\batch{R} \as \{\agg{R_i}\}_{i=0}^{n_1-1}$
					\State The public parameter $I \as (p,q,\alpha, n_1, n_2, \St:=(i=1, \ds_0))$ 
					
					\State \Return ($I$, $\sk$, \pk)
				\end{algorithmic}
				
				\algrule
				\begin{algorithmic}[1]
					\Statex $\underline{\agg{\delta} \as \poslocagg(\{\delta_j \in \sigma_j \}_{j=1}^{\ell})}$: 
					\vspace{3pt}
					\If{$\delta_1 \in \mathbb{Z}_q^* $} $\agg{\delta} \as \sum_{j=1}^{\ell} \delta_j  \bmod q$ \textbf{else}  $\agg{\delta} \as \prod_{j=1}^{\ell} \delta_j \bmod p$
					\EndIf
					\State \Return $\agg{\delta}$
				\end{algorithmic}
				
				\algrule
				\begin{algorithmic}[1]
					\Statex $\underline{\agg{\sigma_i} \as \poslocsig(\sk, \batch{m_i})}$: 
					\vspace{3pt}
					\State \textbf{if} $i > n_1$ and $\batch{m_i} \neq \{ m_i^j \}_{j=0}^{n_2-1}$ \textbf{then return $\perp$} 
					\State $ (\ds_i , x_{0}[i] ) \as \sso(\ds_{i-1}, x_D[i], i)$ 
					\For{$j=0, \ldots, n_2-1$}
					\State  $ e_i^j \as H(m_i^j \mathbin\Vert x_i^j) \bmod q $, where $ x_i^j \as \prf_0(x_{D}[i] \mathbin\Vert j)$
					\State $s_i^j \as r_i^j - e_i^j \cdot y \bmod q$, where $ r_i^j \as \prf_0(r \mathbin\Vert i \mathbin\Vert j) \bmod q $
					\EndFor
					
					\State $ \agg{s_i} \as \poslocagg( \{ s_i^j \}_{j=0}^{n_2-1} )$ , $\St \as (i+1 , \ds_i)$
					\State \Return $\agg{\sigma_i} \as ( \agg{s_i},~\ds_i ) $
					
				\end{algorithmic}
				
				\algrule
				\begin{algorithmic}[1]
					\Statex $\underline{b \as \poslocaver(\pk, \vec{\batch{m}}, \agg{\sigma})}$: 
					$\vec{\batch{m}}$ = $\{ \batch{m_i} \}_{i \in \batch{I}}$
					\vspace{3pt}

                        \State \textbf{if} $|\batch{m_i}| \nequiv 0 \bmod n_2, \forall i \in \batch{I} $ \textbf{then return $\perp$} 
					\If{$\agg{R} \notin \agg{\sigma}$} $\agg{R} \as \poslocagg(\{\agg{R_i}\in \pk\}_{i \in \batch{I}})$ 
					\EndIf
					
					\State $\agg{e} \as 0$
					\For{$i \in \batch{I}$}
					\State $x_{0}[i] \as \sret(\ds, i)$
					\For{$j=0, \ldots, n_2-1$}
                        \State $ x_i^{j} \as \prf_0(x_{0}[i] \mathbin\Vert j)$
					\State $\agg{e} \as \agg{e} + H(m_{i}^j \mathbin\Vert x_i^{j}) \bmod q$
					\EndFor
					\EndFor
					\State \textbf{if} $\agg{R} = Y^{\agg{e}} \cdot \alpha^{\agg{s}} \bmod p$ \textbf{then} \Return $b=1$ \textbf{else} \Return $b=0$
				\end{algorithmic}
		}}
		\caption{Digital signature algorithms}
		\label{alg:sopas}
	\end{subfigure}
	\hfill
	\begin{subfigure}[b]{0.48\textwidth}
		\centering
		
		\noindent \fbox{\parbox{\textwidth} {
				\scriptsize
				
				\begin{algorithmic}[1]
					\Statex $\underline{ \ccd_i \as \poslocdistill(\pk, \ccd_{i-1}, \batch{m_i}, \agg{\sigma_i}, n^u )}$: require $n_1 \equiv 0 \bmod n^u $
					\vspace{1pt}
					
					\noindent \textbf{Init} $\agg{s}^u \as 0$ , $\agg{R}^u \as 1$ , $\agg{s} \as 0$ , $\agg{R} \as 1$
					
					\State $b_i \as \poslocaver(\pk, \batch{m_i}, \agg{\sigma_i})$
					\If{$b_i = 1$} 
					\State $\agg{s}^v_i \as \poslocagg(\agg{s}^v_{i-1},\agg{s_i}\in \agg{\sigma}_i)$
					\vspace{1pt}
					\State $\agg{R}^v_i \as \poslocagg(\agg{R}^v_{i-1},\agg{R_i}\in \agg{\sigma}_i)$
					\State $\agg{\sigma}^v_i \as ( \agg{s}^v_i , \agg{R}^v_i )$ \Comment{valid signature}
					\State $\ccd_i^v \as \ccd_{i-1}^v \cup \{ \agg{\sigma}^v_i \} $
					
					\State $\agg{s}^u \as \poslocagg(\agg{s}^u, \agg{s_i})$
					\State $\agg{R}^u \as \poslocagg(\agg{R}^u, \agg{R_i})$
					\If{$i \equiv 0 \mod \floor{n1/n^u} $} 
					\State $\agg{\sigma}^u_i \as ( \agg{s}^u , \agg{R}^u )$ \Comment{umbrella signature}
					\State $\ccd_i^u \as \ccd_{i-1}^u \cup \{\agg{\sigma}^u_i ~, \floor{\frac{i}{ \floor{n_1/n^u} } } \}$
					\State reset $\agg{s}^u = 0 ~,~ \agg{R}^u  = 1$
					\EndIf
					
					\Else~
					\State $\agg{\sigma_i}^i \as ( \agg{s_i} , \agg{R_i} )$, where $\agg{s_i} \in \agg{\sigma_i}$ and $\agg{R_i} \in \pk$ \Comment{invalid signature}
					\State $\ccd_i^i \as \ccd_{i-1}^i \cup \{ \agg{\sigma_i}^i , i \}$
					\EndIf
					
					\State remove $\agg{R_i}$ from $\batch{R} \in \pk$ 
					\State $\ccd_i \as (\ccd_i^v, \ccd_i^u, \ccd_i^i, {\ds}_i \in \agg{\sigma_i})$
					\State \Return $\ccd_i$
				\end{algorithmic}
				\algrule
				
				\begin{algorithmic}[1]
					\Statex $\underline{ \batch{b} \as \poslocsebver(\pk, \vec{\batch{m}}, \ccd, \mu)}$: require $|\vec{\batch{m}}| \equiv 0 \bmod n_2 $
					\vspace{3pt}
					
					
					\State \textbf{switch} $(\mu)$
					\State ~~~~~~\textbf{case} \text{\enquote{V}}:
					\State ~~~~~~~~~~~~$\vec{\batch{m}} \as \{ \batch{m_i} \}_{i \in \batch{I} \textbackslash \ccd^i}$
					\State ~~~~~~~~~~~~$b^v \as \poslocaver(\pk, \vec{\batch{m}}, \agg{\sigma}^v \in \ccd^v)$
					\State ~~~~~~~~~~~~$\batch{b} = \{ b^v \}$
					
					\State ~~~~~~\textbf{case} \text{\enquote{U}}:
					\State ~~~~~~~~~~~~ \textbf{for} $(\agg{\sigma_\ell}, i_\ell) \in \ccd^u$  \textbf{do}
					\State ~~~~~~~~~~~~~~~~~~$\vec{\batch{m}}' \as \{ \batch{m_i} \}_{i \in \{ i_\ell \cdot \floor{n^u / n} , \ldots, (i_\ell+1) \cdot \floor{n^u / n}-1 \} - \ccd^i}$
					\State ~~~~~~~~~~~~~~~~~~$b_\ell^u \as \poslocaver(\pk, \vec{\batch{m}}' , \agg{\sigma_\ell})$
					\State ~~~~~~~~~~~~$\batch{b} \as \{b_\ell^u\}_{\ell\in |\ccd^u|}$
					
					\State ~~~~~~\textbf{case} \text{\enquote{I}}:
					\State ~~~~~~~~~~~~ \textbf{for} $(\agg{\sigma_\ell}, i_\ell) \in \ccd^i$ \textbf{do}
					\State ~~~~~~~~~~~~~~~~~~$b_{\ell}^i \as \poslocaver(\pk, \batch{m_{i_{\ell}}} , \agg{\sigma_{\ell}})$
					\State ~~~~~~~~~~~~ $\batch{b} \as \{b_{\ell}^i\}_{\ell\in |\ccd^i|}$
					
					\State \Return $\batch{b}$
					
				\end{algorithmic}
		}}			
		\caption{Distillation and selective batch verification}
		\label{alg:sopasa}
	\end{subfigure}
	
	\vspace*{-2mm}
	\caption{Coarse-grained signer-optimal \poslo~(\posloc)}
	\label{fig:sopasall}
	
\end{figure*}

In $\poslockg(.)$, for a given $n$, we first select the number of epochs ($n_1$) and items to be signed per epoch ($n_2$) (Step 1). We generate EC-based parameters $(p,q,\alpha)$ and private/public key pair $(y,Y)$ (Steps 2-3). We then generate the initial ephemeral randomness $r$ and the root of \poslot~tree $x_D[0]$ (Step 4). These values will be used to generate ephemeral public commitments ($R$) and one-time randomness ($x$) for a given epoch state $\St$. \posloc~is coarse-grained, thus we combine the commitments for each epoch as in Steps 5-7, which results in initial $\mathcal{O}(n_1)$ and final $\mathcal{O}(1)$ storage at the verifier via aggregation. 
The private/public keys and public parameters are as in Steps 8-9. 

$\poslocagg(.)$~is a keyless signature aggregate function with dual signature combination mode. That is, given a signature element $s \in \sigma$ or $R \in \sigma$, it performs an additive or multiplicative aggregation, respectively. This generic construction enables the aggregation of different keys during the signing and/or batch verification algorithms.  

$\poslocsig(.)$~is an aggregate signature generation that signs each entry and sequentially aggregates into a single umbrella signature (i.e., the tag authenticates all items in the epoch). The signer first checks if the message set complies with the epoch size $n_2$ (Step 1). The seed $x_0[i]$ is then computed via \sso~once per epoch $i$ (Step 2) and is used to derive one-time seeds $x_i^j$ (Step 4). The aggregate signature $\agg{s_i}$ is computed with only a few hashing and modular additions plus a modular multiplication (Steps 3-6). This makes \posloc~the most signer efficient \poslo~scheme. At the end of epoch $i$, the signer updates its internal state and outputs the aggregate signature $\agg{\sigma_i}$~(Steps 6-7). 

$\poslocaver(.)$~accepts as input the public key \pk, a set of messages $\vec{\batch{m}}$, and an aggregate signature $\agg{\sigma}$. The verifier checks if messages comply with the epoch size (Step 1), and then identifies the format of the aggregate signature to choose a component $R$ (Step 2). $\poslocaver(.)$~can be invoked by the edge cloud or \css~as the final verifier. This difference dictates if the aggregate commitment $R$ is included in the initial public key \pk~or the aggregate signature $\sigma$. Below, we will elaborate further that the $\poslocdistill(.)$~function can be used to verify the entries and then compact them according to a granularity parameter $\rho$. Hence, if verification is done during the distillation, the verifier already has $\agg{R_i} \in \batch{R}$ as part of \pk~and this value is used during verification (Step 9). Otherwise, if the verification is run by the \css, then $\agg{R}$ can be found as a part of the signature in \ccd. The verifier retrieves the seeds in the given epoch via \sret~(Step 5) and then computes the aggregate hash component $\agg{e}$ (Steps 6-8). 
Finally, the aggregate signature is verified (Step 9). 
Figure \ref{fig:seed-management} depicts the mechanism for seed retrieval. It consists of the verifier's view after finishing $6^{\text{th}}$ epoch. It illustrates the request to retrieve the seed of the $3^{\text{rd}}$ epoch (i.e., $x_0[3] \as \sret(\ds_6, 3)$). 

\subsubsection{POSLO-C Distillation and Selective Batch Verification}  
The verification involves two entities from our system model (as in Section \ref{sec:models}): \textit{(i) Distiller} and \textit{(ii) Cold Storage Server} (\css~).

The \css~maintains the cryptographic cold data (\ccd~), which is updated by the distiller.  
Figure \ref{alg:sopasa} formally describes the distillation and batch verification algorithms.  
Initially, both entities initialize the \ccd~as empty sets of signatures.  
$\poslocdistill(.)$ updates the \ccd~structure by aggregating valid signatures and adding them to $\ccd^v$ (Steps 2-6).  
The \css~stores valid signatures according to the granularity level $\rho$, resulting in:  
(i) an overall valid tag $\ccd^v$,  
(ii) a set of umbrella valid signatures $\ccd^u$ (Steps 7-12), and  
(iii) individual invalid signatures $\ccd^i$ (Step 15).  

By storing sub-aggregates (i.e., umbrella aggregates) in $\ccd^v$, umbrella signatures enable localization of invalid log data chunks when the batch verification of the overall tag fails.  

$\poslocsebver(.)$ is a selective batch verification algorithm that operates in three modes:  
{\em (i)} Mode \enquote{V} verifies $\ccd^v$, which consists of one aggregate signature for all valid messages.  
{\em (ii)} Mode \enquote{U} checks partial umbrella signatures if the overall verification (mode ``V'') fails.  
The storage overhead for \ccd~can be adjusted according to the granularity parameter $\rho$.  
{\em (iii)} Mode \enquote{I} verifies the invalid set by checking each entry individually.  

The generic batch verification $\poslocaver(.)$ supports both the edge server and \css~in the distillation and verification processes, respectively. 


\subsection{Fine-grained public-key POSLO~(POSLO-F)}
\label{sec:fiposa}

Our fine-grained variant \poslof, as shown in Figure~\ref{alg:fiposa}, utilizes \bpv~\cite{boyko1998speeding} (see Definition \ref{def:bpv}) to pre-store a constant size of one-time commitments at the signer. Prior works \cite{Yavuz:CNS:2019} have shown that this storage overhead is tolerable for some low-end IoT devices. The pre-computation is important for immediate and fine-grained verification at the edge server and allows \css~to authenticate log entries individually, enabling precise investigation and optimal recovery. 

\poslof~provides several performance advantages on the distiller side. It enables immediate verification of each message within an epoch by attaching the seed $x_i^j$ to the signature, as shown in $\poslofsig(.)$~(Step 6). Unlike \posloc, \poslof~allows $\mathcal{O}(1)$ public-key storage at the distiller.  
The signer generates a commitment $R_t$ using \bpv~(Step 3) and includes it in the signature (Step 6). This eliminates the initial $\mathcal{O}(n_1)$ public key storage (as in \posloc) and allows the highest granularity by enabling individual signature verification.

The distillation and selective batch verification functionalities are similar to those of \posloc~with minor differences. The edge server aggregates each signature separately. The invalid set $\ccd^i$ contains individual signatures (at the highest granularity), requiring \css~to verify each invalid entry separately.  
Thus, \poslof~offers better verification precision than \posloc, albeit with slightly slower verification and higher signer communication overhead.

\begin{figure}[ht!]
	\centering
	\begin{minipage}{0.65\textwidth}
		\centering
		\noindent \fbox{\parbox{0.9\columnwidth} {
				\scriptsize
				
				\begin{algorithmic}[1]
					\Statex $\underline{(I, \sk, \pk)\as \poslofkg(1^{\kappa}, n)}$: Steps 1-4 are identical to $\poslockg(.)$, the rest is as follows:
					\vspace{3pt}
					
					\State $(\Gamma, v, k) \as \bpvoff(1^{\kappa}, p, q, \alpha)$
					\State $\sk \as (y, r, x_D[0], \Gamma)$ , $ \pk \as Y$
					\State The public parameters $I \as (p,q,\alpha, v, k, n_1, n_2, \St:= (t=0 , \ds_0) )$
					\State \Return ($I$, $\sk$, \pk)
				\end{algorithmic}
				\algrule
				
				\begin{algorithmic}[1]
					\Statex $\underline{\sigma_t \as \poslofsig(\sk, m_t)}$: Given $t \le n$
					\vspace{3pt}
					\State $i \as \floor{\frac{t}{n_2}}$ ; $j \as i \bmod n_2$
					\If{$j=1$}
					$(\ds_i , x_{0}[i] ) \as \sso(\ds_{i-1}, x_D[0])$
					\EndIf
					\State $ (r_t, R_t) \as \bpvon(\Gamma, v, k) $
					\State $ e_t \as H(m_t \mathbin\Vert x_t) \bmod q $ , where  $ x_t \as \prf_0(x_{0}[i] \mathbin\Vert j) $
					\State $s_t \as r_t - e_t \cdot y \bmod q$ , $\St := ( t+1 , \ds_i) $
					
					\State \textbf{if} $j=n_2$ \textbf{then return} $\sigma_t \as (s_t, R_t, \ds_i)$ \textbf{else return} $\sigma_t \as (s_t, R_t, x_t)$
				\end{algorithmic}
				\algrule
				
				\begin{algorithmic}[1]
					\Statex $\underline{b \as \poslofaver(\pk, \vec{\batch{m}}, \sigma)}$:
					\If{$|\vec{\batch{m} }| = 1$}   $e \as H(m \mathbin\Vert x) \bmod q$, where $\vec{\batch{m}}=\{m\}$ and $x \in \sigma$
					\Else~execute \poslocaver~Steps 2-8
					\EndIf
					
					\If{$R = Y^{e} \cdot \alpha^{s} \bmod p$} \Return $b = 1$ \textbf{else} \Return $b=0$
					\EndIf
					
				\end{algorithmic}
		}}
		
	\end{minipage}
	
	\vspace{-2mm}
	\caption{Fine-grained public-key \poslo~(\poslof)}
	\label{alg:fiposa}
	\vspace{-2mm}
\end{figure}

\subsection{POSLO Parallel Batch Verification (\poslopaver)} \label{POSLO.PAVer}

\poslo~signature verification, as well as other batch verification algorithms (e.g., \cite{ferrara2009practical}), offers significantly increased computational efficiency compared to traditional per-message verification. This improvement stems primarily from reducing the number of expensive operations, such as EC scalar multiplications in ECDLP-based signatures. Nonetheless, batch verification can remain costly for large datasets due to sequential hashing operations needed to derive one-time ephemeral keys $e_i^j$ (see Step 8 in \poslocaver, Fig.~\ref{alg:sopas}). For example, without hardware acceleration, verifying 1\,GB of log data (with 32-byte entries) using \posloaver~takes approximately 30 seconds on commodity hardware. In contrast, the BLS aggregate signature scheme requires roughly 2.55 hours due to costly map-to-point hashing. However, as the size of dataset grows, the cumulative overhead of hashing becomes a bottleneck for real-time log verification, even for efficient primitives like \poslo.

To address this gap, we propose {\em Parallel batch verification} (\poslopaver), which exploits two key properties: (i) the independence of per-signature hashing (e.g., in \poslocaver), and (ii) the additive homomorphism of the aggregation function (i.e., modular addition) used to compute the aggregate ephemeral key $\agg{e}$. These properties allow the hashing and aggregation steps to be parallelized or shuffled without compromising correctness. This observation motivates the design of \poslopaver~for efficient batch verification at the \css. While the edge server (i.e., the distiller) can also leverage \poslo~verification, the benefits of parallelization are most pronounced at the \css, which archives the entire log stream.

\begin{figure*}[t]
	\centering
	\begin{subfigure}[b]{0.38\textwidth}
		\centering
		\includegraphics[scale=0.365]{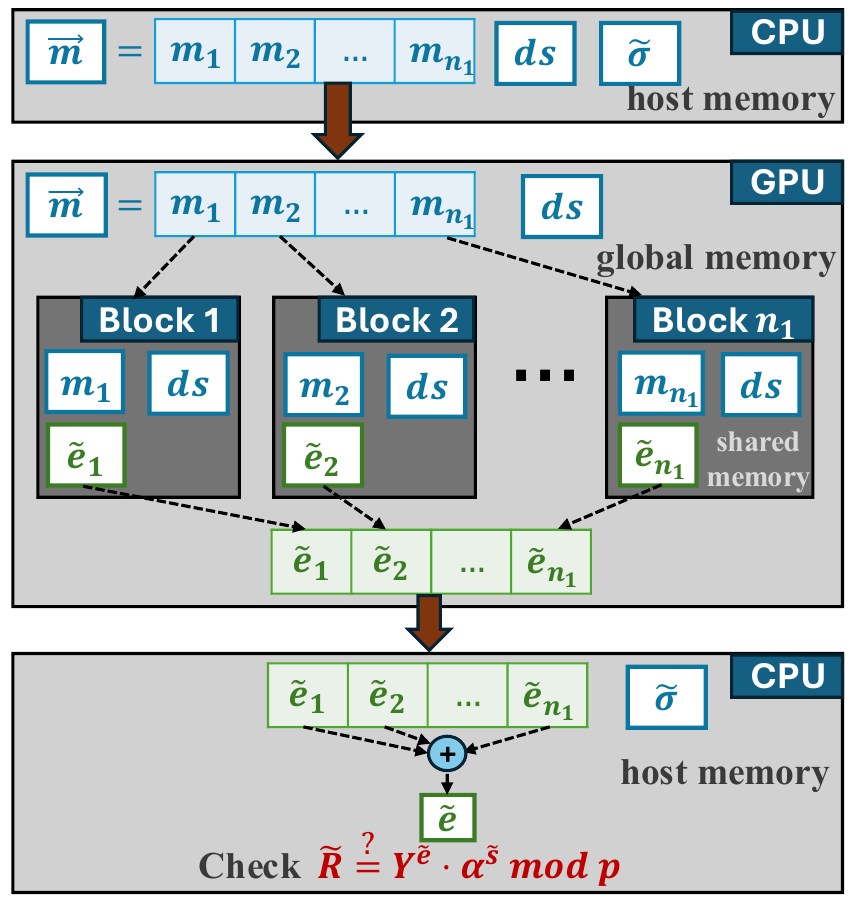}
		\caption{CPU-GPU communication workflow}
		\label{fig:gpu-cpu-flow}
	\end{subfigure}
	\hfill
	\begin{subfigure}[b]{0.61\textwidth}
		\centering
		\includegraphics[scale=0.29]{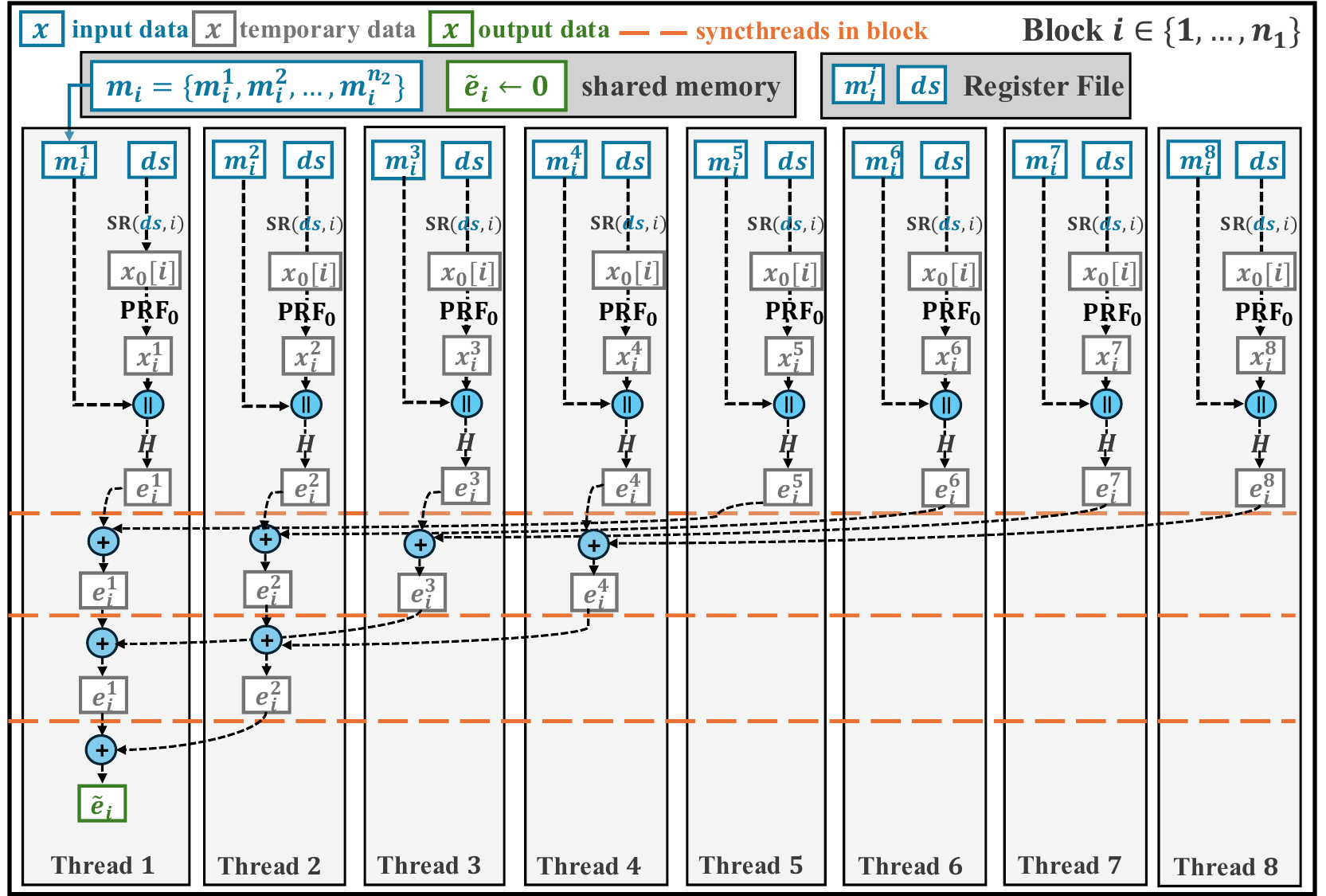}
		\caption{CUDA kernel execution within a block ($n_2=8$)}
		\label{fig:cuda-kernel}
	\end{subfigure}
	
	\vspace*{-2mm}
	\caption{High-level illustration of the parallel batch verification algorithm (\poslopaver)}
	\label{fig:poslo}
	
\end{figure*}

\begin{figure}[ht!]
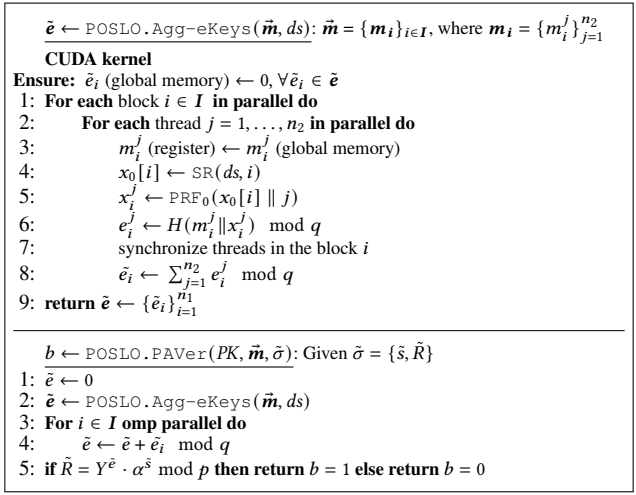

	\centering
	\begin{minipage}{0.65\textwidth}
		\centering
		\noindent \fbox{\parbox{0.9\columnwidth} {
				\scriptsize
				
				\begin{algorithmic}[1]
					\Statex $\underline{ \batch{\agg{e}} \as \posloaggekeys(\vec{\boldsymbol{m}}, \ds)}$:
					$\vec{\batch{m}}$ = $\{ \batch{m_i} \}_{i \in \batch{I}}$, where $\batch{m_i}=\{m_i^j\}_{j=1}^{n_2}$
					
					\vspace{2pt}
					\noindent \textbf{CUDA kernel}
					
					\Ensure $\agg{e}_i~\text{(global memory)} \as 0, \forall \agg{e}_i\in \batch{\agg{e}}$ 
					
					\State \textbf{For each} block $i \in \batch{I}$ \textbf{ in parallel do}
					
					\State \indent \textbf{For each} thread $j = 1,\ldots,n_2$ \textbf{in parallel do}
					\State \indent \indent $m_i^j~\text{(register)} \as m_i^j~\text{(global memory)} $
                        \State \indent \indent $x_0[i] \as \sret(\ds, i) $
					
					\State \indent \indent $x_i^j \as \prf_0(x_0[i]~\|~j)$
					\State \indent \indent $e_i^j \as H(m_i^j \| x_i^j) \mod q$
					\State \indent \indent synchronize threads in the block $i$
					\State \indent \indent $\tilde{e_i} \as \sum_{j=1}^{n_2}{e_i^j} \mod q$
					
					\State \Return $ \batch{\agg{e}} \as \{ \agg{e}_i \}_{i=1}^{n_1} $
					
				\end{algorithmic}
				
				\algrule
				
				\begin{algorithmic}[1]
					\Statex $\underline{b \as \poslopaver(\pk, \vec{\boldsymbol{m}}, \agg{\sigma})}$: Given $\agg{\sigma} = \{\agg{s}, \agg{R}\}$
					
					\State $\agg{e} \as 0$
					\State $ \batch{ \agg{e} } \as \posloaggekeys(\vec{\batch{m}},\ds) $
					\State \textbf{For} $i \in \batch{I}$ \textbf{omp parallel do}
					\State \indent $\agg{e} \as \agg{e} + \agg{e_i} \mod q $
					\State \textbf{if} $\agg{R} = Y^{\agg{e}} \cdot \alpha^{\agg{s}} \bmod p$ \textbf{then} \Return $b=1$ \textbf{else} \Return $b=0$
					
				\end{algorithmic}
		}}
		
	\end{minipage}
	
		\vspace{-2mm}
	\caption{Parallel \poslo~signature verification algorithm (\poslopaver)}
	\label{alg:polsoagg}
\end{figure}

\poslopaver~is a CUDA-based parallel verification algorithm designed to exploit the inherent parallelism in mutable aggregate signatures, focusing on \poslo~batch verification. The architecture of \poslopaver~is shown in Figure~\ref{fig:poslo}, with a formal description in Figure~\ref{alg:polsoagg}.

Verification begins by initializing the aggregate ephemeral key $\agg{e}$, then executing the CUDA kernel \posloaggekeys, which computes the individual ephemeral keys $e_i^j$ in parallel and aggregates them into per-epoch subaggregates $\{\agg{e}_i\}_{i \in \boldsymbol{I}}$ using a tree-based parallel reduction (Step 8). CUDA blocks are mapped to batches in $\boldsymbol{\vec{m}}=\{\boldsymbol{m_i}\}_{i\in \boldsymbol{I}}$, and the threads per block correspond to the number of items to be processed per epoch $n_1$. Figure~\ref{fig:cuda-kernel} shows the kernel execution for a single block with $n_2 = 8$. Each batch $\boldsymbol{m_i}$ is first loaded from global memory into shared memory, and each log entry $m_i^j$ is placed into the register space of thread $j$ in block $i \in \boldsymbol{I}$. Each thread $j\in \{1,\ldots, n_2\}$ computes the seed $x_0[i]$ using the \sret~algorithm (Step 4), and then computes its own ephemeral key $e_i^j$ (Steps 5-6), and a synchronization is performed to ensure readiness for aggregation. A tree-based parallel reduction is performed within each block (see Figure~\ref{fig:poslo}) to compute an aggregate $\agg{e}_i$, which is then written back to global memory for final aggregation on the CPU.

The final aggregation of $\agg{e}_i$ values into $\agg{e}$ can be done using another CUDA kernel, OpenMP~(OMP)~\cite{chandra2001parallel}, or sequentially on the CPU. Because aggregation involves $\mathcal{O}(n_1)$ modular addition and $\agg{e}_i \in \mathbb{Z}_q$ for all $i \in \boldsymbol{I}$, this step is lightweight even when done sequentially.

\poslopaver~is applied to the valid set $\ccd^v$~, which contains the overall aggregate tag. It can also be applied to the umbrella set $\ccd^u$~ by re-aggregating $\{\agg{e}_i\}_{i \in \boldsymbol{I}}$ into coarser aggregates according to the selected granularity parameter $\rho$.

Overall, \poslopaver~significantly reduces CPU load by offloading expensive hashing and \prf~operations to the GPU. Section~\ref{sec:performance_analysis} details the speedups obtained with various instantiations of $H$ and \prf, and compares them with both baseline and state-of-the-art digital signature schemes.

	\section{Security Analysis} \label{sec:SecurityAnalysis}

We formally prove that \posloc~and \poslof~schemes are  \AEUCMA-secure AS schemes in the Random Oracle Model (ROM) in \cite{CryptoHandBook} Theorem 5.1 and Lemma 5.1. We ignore terms that are negligible in terms of $\kappa$. We prove that \poslop~and \poslopp~instantiations are as secure as \poslo~in Corollary 5.3. The full security proofs are given in Appendix A.

\begin{theorem} \label{the:Theorem1}
	$\advsocosa \le \advdll$, where  $t'=O(t)+ O(n \cdot (\kappa^{3}+RNG))$. 
\end{theorem}

\vspace{1pt}

\begin{lemma}
	\poslof~is as secure as \posloc. 
    
\end{lemma}

\vspace{1pt}

\begin{corollary}
    \poslop and \poslopp~instantiations are as secure as \poslo.
\end{corollary}

	\section{Performance Analysis} \label{sec:performance_analysis}
This section presents a detailed performance comparison of \poslo~schemes and their counterparts.

\subsection{Evaluation metrics}
Our evaluation considers the following metrics:
\textit{(i)} signer's execution time and energy consumption,
\textit{(ii)} private/public key sizes,
\textit{(iii)} signature size,
\textit{(iv)} individual or sub-aggregate signature verification and batch verification execution times at \css, and
\textit{(v)} cryptographic cold storage at \css.

We select the main counterparts that represent key families of AS schemes:
\textit{(i)} Factorization-based: C-RSA~\cite{Yavuz:TDSC:OutsourcedDB} is an AS scheme offering near-optimal signature verification.
\textit{(ii)} ECDLP-based: SchnorrQ~\cite{costello2016schnorrq} is one of the fastest EC-based signature schemes (compared to ECDSA or Ed25519~\cite{Ed25519}), with high efficiency on embedded devices. FI-BAF~\cite{Yavuz:2012:TISSEC:FIBAF} is a signer-optimal FAS scheme, and is our closest logger-efficient comparator.
\textit{(iii)} Pairing-based: BLS~\cite{BLS:2004:Boneh:JournalofCrypto} is a multi-user AS scheme based on bilinear maps, offering the most compact storage among the alternatives. Moreover, BLS is widely adopted in recent AS schemes with enhanced properties (e.g.,~\cite{li2020permissioned, verma2021scbs}) in IoT networks, making it a natural benchmark to highlight the efficiency advantage of \poslo.

In the remainder of this section, we present the parameters and hardware/software setup used in our evaluation. We then provide an in-depth performance analysis on the resource-constrained logger (signer), edge cloud (distiller), and cold storage server (final log repository and auditor).

\subsection{Instantiations and Configurations}
\label{subsec:parameter_selection}

We set $\kappa$ to $128$-bit security. We used FourQ curve \cite{costello2016schnorrq} and set $|q| = 256$ bits for the EC-based schemes. 
The BPV parameters are $(v,k)=(1024, 16)$. 
The composite modulus in C-RSA is $|n|=2048$.

\begin{table}[ht!]
	\centering
	\caption{ Instantiations of \poslo~schemes using different \prf~and $H$ functions}
	\label{tab:instantations}
	\footnotesize
	\vspace{-2mm}
	\begin{minipage}{0.82\textwidth}
		\resizebox{\textwidth}{!}{
			\begin{tabular}{|l|c|c|c|c|c|c|c|}
				
				\hline
				\textbf{Scheme} 
				& \textbf{$\boldsymbol{\prf_{0,1}}$} 
				& \textbf{$\boldsymbol{H}$} 
				& \specialcell[]{\textbf{Standard} \\ \textbf{compliance} }
				& \specialcell[]{\textbf{Hardware} \\ \textbf{Support}}
                    & \specialcell[]{\textbf{High} \\ \textbf{Efficiency}}
                    & \specialcell[]{\textbf{Large} \\ \textbf{Input Sizes}}
				\\
				\hline

                    $\poslo$ & SHA-256 & SHA-256 & \cmark & \xmark & \xmark & \cmark \\
                    \hline
                    
				\poslop & MMO-AES-128 & MDC-AES-128 & \xmark & \cmark & \cmark & \cmark \\
                    \hline

                    \poslopp & MMO-AES-128 & $Add_q$ & \xmark & \cmark & \cmark & \xmark \\
                    \hline
            
			\end{tabular}
		}
		{
			\scriptsize
			Modular addition ($Add_q$) cannot evaluate inputs larger than the modulus $q$.
		}		
		\vspace{-5pt}
	\end{minipage}
	
\end{table}

\subsubsection{Instantiations of PRF$_{0,1}$}
We derive the pseudo-random functions $\prf_{0,1}$ from: (i) Standard cryptographic hash function SHA-256. (ii) MMO construction with AES-128 as a block cipher ($E$).
\[
\prf_j(x) = F(x \|j), \forall F \in \{\text{MMO}\mhyphen\text{AES}\mhyphen\text{128}, \text{SHA-256}\}, j=0,1, x \in \{0,1\}^\kappa
\]
    


%
$\prf_{0,1}$ are used in \poslo~signature algorithms to derive one-time keys (i.e., $x_i^j$ and $r_i^j$) and in \smf~algorithms to derive the left and right child nodes, respectively. Each node $\{x_d[i]\}_{d,i}$ in the \poslot~tree has a bit-length of $\kappa$. 
For $\kappa=128$, $|x_d[i] \| j|$ is 129 bits, where $j\in\{0,1\}$, requiring two AES evaluations per \prf~call. However, if $\kappa=127$, the concatenated input $|x_d[i] \| j|$ becomes exactly 128 bits, matching the AES block size and allowing a single AES call, reducing the computational overhead by half.  This optimization makes the $\kappa=127$ configuration the lightest and fastest variant, with slightly relaxed security guarantees. 
%


\subsubsection{Instantiations of $H$}
We use multiple instantiations of $H$ in \poslo~signature schemes:

\begin{enumerate}[(i)]
    \item {\em Standard cryptographic hash function.} SHA-256 is used to ensure standard compliance. This mode incurs more computational overhead since cryptographic hash functions are sequential and do not provide high parallelism and efficiency.  
    \item {\em MDC-2 construction with AES-128.} Given a $128$-bit input, MDC-2 \cite{CryptoHandBook} outputs a $256$-bit digest using two iterations of the MMO construction (see Definition \ref{def:mdc}). When $E$ is instantiated with AES-128, MDC-2 delivers efficient performance on both constrained and commodity devices, benefiting from highly optimized hardware and software AES implementations.
    \item  {\em Modular addition.} For small input messages (i.e., $|m|<q$), Chen et al. \cite{chen2021does} demonstrate that modular addition can be securely and efficiently used in Schnorr-based digital signatures. Given a message $m$, private key $(y,r)$, and one-time seed $x$, the \poslo~signature becomes $\sigma\as(s,x)$ where $s \as r+(m+x) \cdot y \mod q$. This variant enables ultra-lightweight signature generation that is especially well-suited for resource-constrained devices. 
\end{enumerate}
 
%

\subsubsection{Instantiations of POSLO signatures} 
Table \ref{tab:instantations} summarizes our \poslo~signature instantiations, each defined by a particular combination of \prf~and $H$ functions. The \poslo~variant ensures standard compliance with cryptographic standards by exclusively using SHA-256. \poslop~and \poslopp~variants prioritize efficiency through the reliance on AES-based constructions. These benefit from the hardware acceleration available across platforms, including AES-NI \cite{hofemeier2012introduction} and GPU-accelerated AES \cite{tezcan2021optimization, hajihassani2019fast}. We demonstrate the high efficiency of $\poslop$ in subsequent sections on both resource-constrained IoT devices and commodity platforms.

\subsubsection{Hardware and Software Configurations} 

We evaluate \poslo~on the following testbeds:

\noindent {\em (i) Signing/Verification on x86/64.} A desktop with an Intel i9-11900K@3.5GHz and 64 GB of RAM. 
\vspace{1pt}

\noindent {\em (ii) Signing on 8-bit.} A low-end 8-bit AVR ATmega2560 microcontroller, due to its low energy consumption and extensive use in practice. It is equipped with 256KB flash memory, 8KB SRAM, and 4KB EEPROM, with a clock frequency of 16MHz. 
\vspace{1pt}

\noindent {\em (iii) Verification on GPUs.} A commodity desktop with an Intel i9-11900K@3.5GHz processor and 64 GB of RAM. It also includes an NVIDIA GTX 3060 GPU, which provides CUDA 3584 cores, 12GB GDDR6-based memory, and 360GB/s memory bandwidth. In our benchmarks, we include the memory communication overhead between the CPU’s main memory and the GPU’s global memory. \vspace{1pt}

\begin{table*}[t]
	\centering
	\captionsetup{justification=centering}
	\caption{Private/public key and signature sizes, and signature generation/verification costs of \poslo~and its counterparts}
	\label{tab:analytical_table_sig.ver}
	\footnotesize
	\vspace{-2mm}
	\begin{minipage}{\textwidth}
		\resizebox{\textwidth}{!}{
			\begin{tabular}{|l||@{}c@{}|c|@{}c@{}||@{}c@{}|@{}c@{}|@{}c@{}|}
				\hline
				\multirow{2}{*}{\textbf{Scheme}}
				& \multicolumn{3}{c||}{\textbf{Logger (Signer)}}
				& \multicolumn{3}{c|}{\textbf{Edge Server (Distiller)}}
				\\
				\cline{2-7}
				& \specialcell[]{\textbf{Signature Generation}}
				& \specialcell[]{$\boldsymbol{|\sk|}$}
				& \specialcell[]{$\boldsymbol{|\sigma|}$}
				& \specialcell[]{$\boldsymbol{|\pk|}$}
				& \specialcell[]{\textbf{Signature Verification ($\boldsymbol{\times n_2}$) }} 
				& \specialcell[]{\textbf{Distill \& Agg ($\boldsymbol{\times \tau_S \cdot n_2}$)}} \\
				\hline
				SchnorrQ~\cite{costello2016schnorrq}
				& \specialcell[]{$2H + Add_q + Mul_q + EMul $} 
				& $|q|$ 
				& $2|q|$ 
				& $|q|$ 
				& \specialcell[]{$ H + 1.3 \cdot EMul $}  
				& N/A \\ \hline
				FI-BAF \cite{Yavuz:2012:TISSEC:FIBAF}
				& $3H + 2Add_q + Mul_q $ 
				& $2 \cdot (|q| + \kappa)$ 
				& $|q|+\kappa$ 
				& $2 n \cdot (|q|+\kappa)$
				& \specialcell[]{$   2\cdot( H +  Add_q) + 2.3 \cdot EMul $}  
				& \specialcell[]{$ Add_q $} \\ \hline 
				C-RSA \cite{Yavuz:TDSC:OutsourcedDB} 
				& $H + Exp_{|n_{RSA}|}^{|d_{RSA}|}$ 
				& $ 2|n_{RSA}| $ 
				& $|n_{RSA}|$ 
				& $2|n_{RSA}|$ 
				& \specialcell[]{$ H + Exp_{|n_{RSA}|}^{|e_{RSA}|} $}  
				& \specialcell[]{$ Mul_{n_{RSA}} $} \\ \hline 
				BLS \cite{BLS:2004:Boneh:JournalofCrypto} 
				& \specialcell[]{$MtP + EMul'$} 
				& $|q|$ 
				& $|q|$ 
				& $2|q|$ 
				& \specialcell[]{$ MtP + Pr $ } 
				&  \specialcell[]{$ Mul_q $} \\ \hline \hline 
				\posloc
				& \specialcell[]{$ 3 \cdot PRF + H + 2 Add_q + Mul_q $} 
				& $|q|+2\kappa$ 
				& $|q|$ 
				& $ n_1 \cdot |q|$ 
				& \specialcell[]{$ PRF + H + Add_q + (PRF + 1.3 \cdot EMul) / n_2 $ }  
				& \specialcell[]{$ (Add_q + EAdd) / n_2 $} \\ \hline 
				\poslof
				& \specialcell[]{$ 2 \cdot PRF + H + Add_q + Mul_q $\\$+ k \cdot (Add_q+EAdd)$} 
				& \specialcell[]{$ 2 \cdot v \cdot |q| + \kappa $ } 
				& $2|q|+ \kappa$ 
				& $|q|$ 
				& \specialcell[]{$H + 1.3 \cdot EMul $ } 
				& $ Add_q + EAdd $ 
				\\ \hline 
			\end{tabular}
		}	
		
		\vspace{2pt}
		{
			\scriptsize
			$Add_q$ and $Mul_q$ denote modular addition and multiplication, respectively, with modulus $q$. 
			$EMul$, $EMul'$ are EC scalar multiplication on Four$\mathbb{Q}$ and pairing-based curves, respectively. 
			We used double-point scalar multiplication (e.g., $1.3 EMul$ instead of $2 EMul$ for Four$\mathbb{Q}$). $Pr$ is a pairing operation. $Exp_{|y|}^{|x|}$ denotes modular exponentiation with exponent $x$ and modulus $y$.
			$\tau_S$ denotes the success verification rate.
		}
		
	\end{minipage}
		\vspace{-5pt}
\end{table*}

\subsection{Performance Analysis on Signer (Logger)}
\label{subsec:perf_analysis_signer}

\subsubsection{Analytical Evaluation}
\label{subsubsec:analytical_evaluation_signer}
Table \ref{tab:analytical_table_sig.ver} illustrates the efficiency of \posloc~signature generation, which only requires 2 \prf~and one $H$ calls (on average), two and one modular additions and multiplication, respectively. This makes it as lightweight as its most signer-efficient counterpart FI-BAF, but with a vastly superior performance at \css.   \posloc~is significantly more signer efficient than all other alternatives in terms of runtime, with a highly compact signature and small key sizes. \poslof~is the second most signer efficient alternative that requires constant number (e.g., 16) of $EAdd$ operations. It relies on a pre-computed \bpv~table, which increases its private key size in exchange for better signing efficiency. Note that the use of \bpv~can be avoided by accepting a single $EMul$, which makes \poslof~signing cost equal to that of SchnorrQ. We remind that \poslof~accepts extra signing/verification cost over \posloc~in exchange for finer granularity. 

\vspace{2pt}
\noindent {\em Seed Management Overhead Analysis.} 
%
The amortized seed management computational overhead of \poslo~signing algorithms across $n$ messages is on average one \prf~call based on the derivation and disclosure of seeds by \sconst~and \sso~algorithms, respectively. 
The average amortized cost is $( 1 + \frac{\log{n_1}}{n_2}) \cdot \prf$, which corresponds to less than two \prf~calls, therefore we conservatively accept it as two $\prf$ calls.

The storage overhead of \poslot~stack structure (i.e., \ds) is $\log{(n_1)} \cdot \kappa$ bytes. We used an array-based implementation of the stack (i.e., \ds)~\cite{hopcroft1983data} since the maximum number of elements in \ds~is known and limited to $D = \log{(n_1)}$ (height of \poslot) nodes. 
At the end of the last epoch (i.e., $i=n_1$), the signer discloses the \poslot~root $x_D[0]$, enabling the \css~to verify any previous message-signature pair with $\mathcal{O}(1)$ final storage. 



\subsubsection{Experimental Evaluation on 8-bit AVR}
\label{subsubsec:signer_avr}

\begin{figure*}[ht!]
	\centering
	\begin{subfigure}[b]{0.49\textwidth}
		\centering
		\includegraphics[scale=0.37]{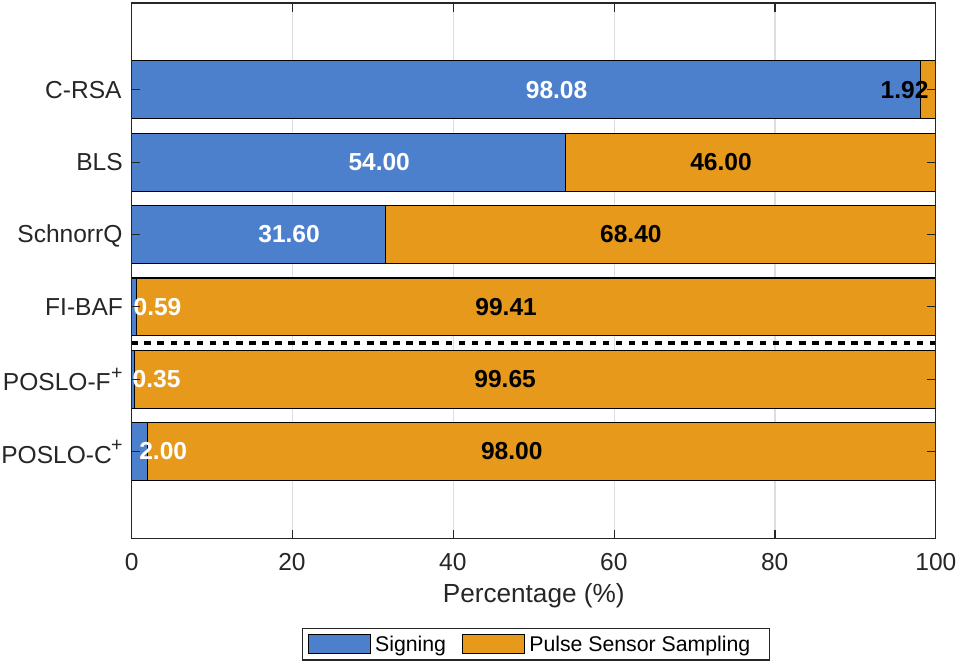}
		\label{fig:pulse_sensor}
	\end{subfigure}
	\hfill
	\begin{subfigure}[b]{0.49\textwidth}
		\centering
		\includegraphics[scale=0.37]{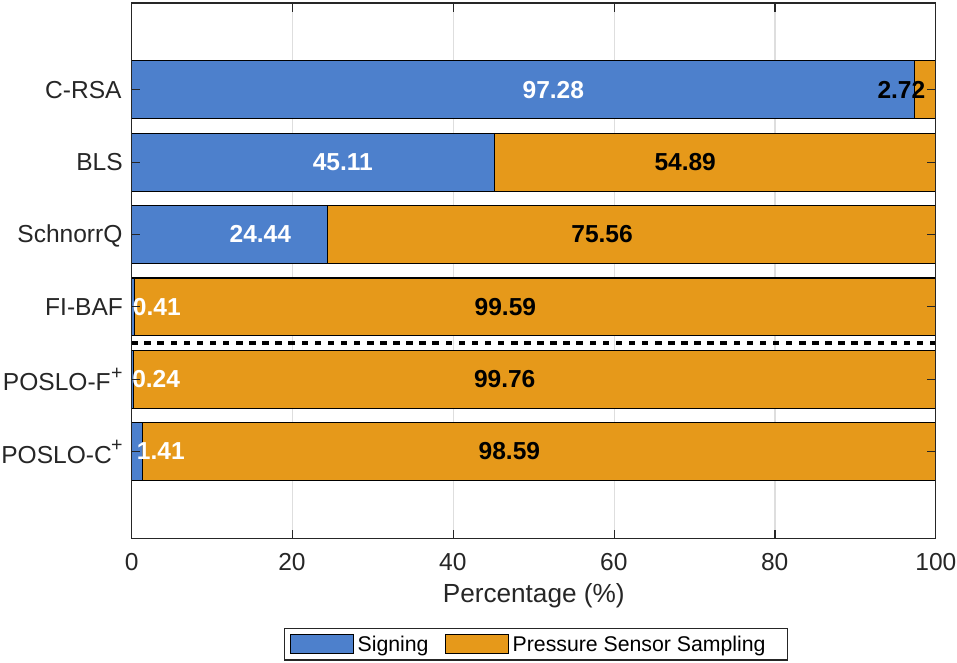}
		\label{fig:pressure_sensor}
	\end{subfigure}
	
	\vspace*{-2mm}
	\caption{Energy consumption of \poslo~schemes and their counterparts at the logger side}
	\label{fig:energy-consumption}
	\vspace*{-2mm}
\end{figure*}

\sloppy Figure \ref{fig:energy-consumption} showcases the energy usage of \poslo~schemes and their counterparts compared to that of sensors typically found in IoT devices. Specifically, we compared the energy usage of a single signature generation with that of sampling via pulse\footnote{\url{https://pulsesensor.com/}} and pressure\footnote{\url{https://cdn-shop.adafruit.com/datasheets/1900_BMP183.pdf}} sensors ($10s$ per sampling time with $1ms$ reading time).

Our most efficient instantiations using AES-based \prf~and hashing functions, $\poslocp$~and $\poslofp$, have remarkably low energy usage with $0.35\%$ and $2\%$, respectively, compared to that of the pulse sensor. For $\poslocp$, this translates into $90\times$ and $154\times$ lower energy usage than the most efficient standard SchnorrQ and verifier compact  BLS, respectively. $\poslocp$~is $1.65\times$ lower energy than FI-BAF, but with substantial gains in cold storage, which will be discussed in further detail in the following. $\poslofp$~is the third most energy efficient alternative, while offering fine granularity and higher verification efficiency.

{\em Traffic Variation and Bandwidth Usage.} Table \ref{tab:OverallPerfComp} depicts the signer cryptographic payload, by enabling full aggregation (per epoch). The signer-efficient variant, \posloc, has equal bandwidth overhead compared to both of the short signature scheme BLS and the most signer-efficient counterpart FI-BAF. \posloc~is the most suitable during a low-frequency upload since it has a lightweight signature generation with a compact signature size. For a high-frequency upload and/or more available battery lifetime, \poslof~offer higher precision by uploading individual signatures to a nearby edge cloud, to be verified and distilled separately. Table \ref{tab:cryptographic overhead} depicts the variation of the signer's cryptographic payload under different epoch sizes. Recall that the epoch size represents the number of individual tags to be aggregated. That is, the low-end devices can increase the epoch size when low bandwidth or battery is observed. \poslof~have equal cryptographic payload compared to SchnorrQ, while having $3\times$ faster signature generation time. Similarly, \posloc~have equal bandwidth overhead compared to the most signer-efficient counterpart FI-BAF but with constant and flexible storage at the distiller and \css~sides. \posloc~is considered the best scheme to offer both low bandwidth overhead and fast signature generation on the signer side.

The sign-aggregate-forward approach can be adopted in a hop-by-hop setting, wherein each IoT device signs a set of log entries, aggregates the individual signatures, and forwards the resulting tag to the next IoT device.
Another possible design is to employ a clustering approach \cite{grissa2019trustsas} wherein IoT devices elect a cluster leader to communicate authenticated log entries to the distiller. The leader adjusts the cryptographic payload based on network conditions. For instance, for a set of $2^{10}$ loggers and $2^8$ of epoch size, the bandwidth overhead for a maximum compression across multiple signers is $16.03$KB, which is $3\times$ and $171\times$ smaller than single-signer aggregate and non-aggregate approaches.

\subsection{Performance Analysis on Edge Cloud (Distiller)}
\label{subsec:perf_analysis_distiller}

\subsubsection{Analytical Evaluation}
\label{subsubsec:distiller_signer}
The edge cloud (or distiller) performs the signature verification on the received log entries. While \posloc~enjoys an efficient batch verification with one single expensive EC scalar multiplication per epoch of size $n_2$, \poslof~validates each individual log entry-signature pair individually and ensure a fine-grained auditing on the cloud storage server. Therefore, \poslo~schemes offer different granularity levels depending on the application context. We further add a set of {\em umbrella} signatures for the valid set to avoid binary epoch verification and bit-flipping events. The verification of \poslo~schemes is more computationally efficient compared to the pairing-based BLS by replacing expensive map-to-point and pairing operations with hashing and EC scalar multiplication, respectively. Moreover, the distillation process for \poslo~schemes only requires modular addition and EC point addition, which are computationally efficient.

\begin{table}[t]
	\centering
	\caption{ Bandwidth overhead  and signature generation time of \poslo~and its counterparts at signer}
	\label{tab:cryptographic overhead}
	\footnotesize
	\vspace{-2mm}
	\begin{minipage}{0.8\textwidth}
		\resizebox{\textwidth}{!}{
			\begin{tabular}{|l|c|c|c|c|c|c|c|}
				
				\hline
				\multirow{2}{*}{ \textbf{Scheme} } 
				& \multirow{2}{*}{ \specialcell[]{\textbf{Analytical} \\  \textbf{complexity} } }
				& \multicolumn{5}{c|}{\textbf{Cryptographic payload (KB)}}
				& \textbf{Signing (s)}
				\\
				\cline{3-7}
				
				& 
				& 16
				& 32
				& 64
				& 128
				& 256
				& \textbf{(per item)}
				\\ \hline \hline
				SchnorrQ~\cite{costello2016schnorrq}
				& $ 2 \cdot n_2 \cdot |q| $
				& 0.5
				& 1
				& 2
				& 4
				& 8
				& 0.323
				\\  \hline
				FI-BAF \cite{Yavuz:2012:TISSEC:FIBAF} 
				& $ |q| + \kappa $
				& 0.05
				& 0.05
				& 0.05
				& 0.05
				& 0.05
				& 0.004
				\\ \hline
				C-RSA \cite{Yavuz:TDSC:OutsourcedDB} 
				& $ |n_{RSA}| $
				& 0.25
				& 0.25
				& 0.25
				& 0.25
				& 0.25
				& 35.828
				\\ \hline
				BLS \cite{BLS:2004:Boneh:JournalofCrypto} 
				& $ |q| $
				& 0.05
				& 0.05
				& 0.05
				& 0.05
				& 0.05
				& 4.080
				\\ \hline \hline
				$\text{\posloc}$
				& \multirow{3}{*}{\specialcell[]{$|q| + 2\cdot \kappa$}}
				& \multirow{3}{*}{\specialcell[]{0.05}}
				& \multirow{3}{*}{\specialcell[]{0.05}}
				& \multirow{3}{*}{\specialcell[]{0.05}}
				& \multirow{3}{*}{\specialcell[]{0.05}}
				& \multirow{3}{*}{\specialcell[]{0.05}}
				& 0.005
				\\\cline{1-1}  \cline{8-8}
                    $\text{\poslocp}$ & & & & & & & 0.002 \\ \cline{1-1} \cline{8-8}
                    $\text{\poslocpp}$ & & & & & & & 0.002 \\ \hline
                    
				$\text{\poslof}$
				& \multirow{3}{*}{$ n_2 \cdot (|q|+\kappa) $}
				& \multirow{3}{*}{0.5}
				& \multirow{3}{*}{1}
				& \multirow{3}{*}{2}
				& \multirow{3}{*}{4}
				& \multirow{3}{*}{8}
				& 0.016
                    \\ \cline{1-1} \cline{8-8}
                    $\text{\poslofp}$ & & & & & & & 0.014 \\ \cline{1-1} \cline{8-8}
                    $\text{\poslofpp}$ & & & & & & & 0.014 \\ \hline
			\end{tabular}
		}
		{
			\scriptsize
			The cryptographic payload is under various epochs (i.e., $n_2$) to illustrate the impact on signer bandwidth.
		}		
	\end{minipage}
	
\end{table}

\subsubsection{Experimental Evaluation on x86/64}
\label{subsubsec:distiller_x86}
The distiller storage overhead is more cumbersome than the cold storage server, especially when the hardware is not a resourceful device (e.g., hotspot). The latter receives sets of authenticated logs, from a large number of IoT devices, to be verified and aggregated following the pre-determined policy. Therefore, the authentication mechanism must have a low-cost verification algorithm and a flexible aggregation capability. By introducing the valid and invalid cryptographic sets, along with the umbrella valid signatures, the distiller can adjust the signature sizes depending on its resource capabilities and/or the network conditions at the cost of auditing precision. 
In table \ref{tab:OverallPerfComp}, the verification time per epoch is performed by the edge cloud. \posloc~verification is $89\times$ and $26\times$ much faster than the short signature BLS and most signer-efficient counterpart FI-BAF, respectively. While \poslof~offer a finer granularity, it still outperforms BLS and FI-BAF by a factor of $5.4$ and $1.6$, respectively.

\subsection{Performance Analysis on Cold Storage Server}
\label{subsec:perf_analysis_ccs}

\subsubsection{Analytical Evaluation}
\label{subsubsec:ccs_anaytical}
Table \ref{tab:analytical_table_auditor} shows the overall performance of \poslo~schemes and their counterparts at the server side. The aggregate signatures offer batch verification for all valid entries (i.e., $\tau=\tau_S$ and $\lambda=1$) and umbrella signatures. The batch verification of all valid entries requires only one $EMul$, while that of umbrella tags requires $n^u \cdot EMul$ where $n^u$ is the number of umbrella signatures. In terms of storage, the final public key and aggregate signature sizes of \poslo~schemes are as efficient as the most compact alternative BLS, but with a faster running time since they do not require expensive pairing $(Pr)$ operation. \poslo~schemes have more efficient verification compared to the most signer-efficient counterpart, FI-BAF, which suffer from a linear public key size at \css. 

\begin{table}[ht!]
	\centering
	\caption{Storage and computation costs of \poslo~and their counterparts at cold storage server (\css)}
	\label{tab:analytical_table_auditor}
	\footnotesize
	\vspace{-2mm}
	\begin{minipage}{0.81\textwidth}
		\resizebox{\textwidth}{!}{
			\begin{tabular}{|l|@{}c@{}|@{}c@{}|@{}c@{}|@{}c@{}|@{}c@{}|@{}c@{}|}
				\hline
				\multirow{2}{*}{\textbf{Scheme}}
				& \multicolumn{2}{c|}{\specialcell[]{\textbf{Cold Cryptographic Data ($\boldsymbol{\ccd}$)}}}
				& \multirow{2}{*}{\specialcell[]{\textbf{Verification Overhead}}} \\
				\cline{2-3}
				& \textbf{|$\boldsymbol{\pk}$|}
				& \textbf{|$\boldsymbol{\sigma}$|}
				& 
				\\ \hline
				SchnorrQ~\cite{costello2016schnorrq} 
				& \specialcell[]{$ |q| $}
				& $ 2 \cdot \tau \cdot n \cdot |q| $
				& \specialcell[]{$\tau \cdot n \cdot (H + 1.3 \cdot EMul) $}
				\\  \hline
				FI-BAF \cite{Yavuz:2012:TISSEC:FIBAF} 
				& \specialcell[]{$ 2 \cdot \tau \cdot n \cdot (|q|+\kappa) $}
				& \specialcell[]{$ \lambda \cdot (|q| + \kappa) $}
				& \specialcell[]{$ \tau \cdot n \cdot (2H + 2Add_q + 1.3 \cdot EMul)$ \\ $+ \lambda \cdot 1.3 \cdot EMul $}
				\\ \hline
				C-RSA \cite{Yavuz:TDSC:OutsourcedDB} 
				& \specialcell[]{$ 2|n_{RSA}| $}
				& \specialcell[]{$ \lambda \cdot |n_{RSA}| $}
				& \specialcell[]{$ \tau \cdot n \cdot (H + Mul_{n_{RSA}}) + \lambda \cdot Exp_{|n_{RSA}|}^{|e_{RSA}|}$}
				\\ \hline
				BLS \cite{BLS:2004:Boneh:JournalofCrypto} 
				& \specialcell[]{$ 2|q| $}
				& $\lambda \cdot |q|$ 
				& \specialcell[]{$\tau \cdot n \cdot (MtP + Mul_q) +  \lambda \cdot Pr$}
				\\ \hline \hline
				\posloc
				& \specialcell[]{$ 2|q| $}
				& \specialcell[]{$ \lambda \cdot |q|$}
				& \multirow{2}{*}{\specialcell[]{$ \tau \cdot n \cdot (2\cdot PRF / n_2 + H + Add_q) $ \\ $+ \lambda \cdot 1.3 \cdot EMul $}}
				\\ \cline{1-3}
				\poslof
				& \specialcell[]{$ |q| $}
				& \specialcell[]{$ \lambda \cdot 2|q| $}
				& 
				\\ \hline 
			\end{tabular}
		}
		
		\vspace{2pt}
		{
			\scriptsize
			The notations are as in Table \ref{tab:analytical_table_sig.ver}. 
			The \ccd~storage and ver. overhead of valid set ($\ccd^v$), umbrella valid set ($\ccd^u$), and invalid set ($\ccd^i$) are displayed when ($\tau=\tau_S,\lambda=1$), ($\tau=\tau_S,\lambda=n^u$), and ($\tau=\tau_F,\lambda=\tau_F\cdot n$), respectively.
			%
		}
		\vspace{-3pt}
	\end{minipage}
\end{table}

\subsubsection{Experimental Evaluation on x86/64}
\label{subsubsec:ccs_x86}
\begin{figure*}[t]
	\centering
	
	\begin{subfigure}[b]{0.32\textwidth}
			\centering
			\includegraphics[width=\textwidth]{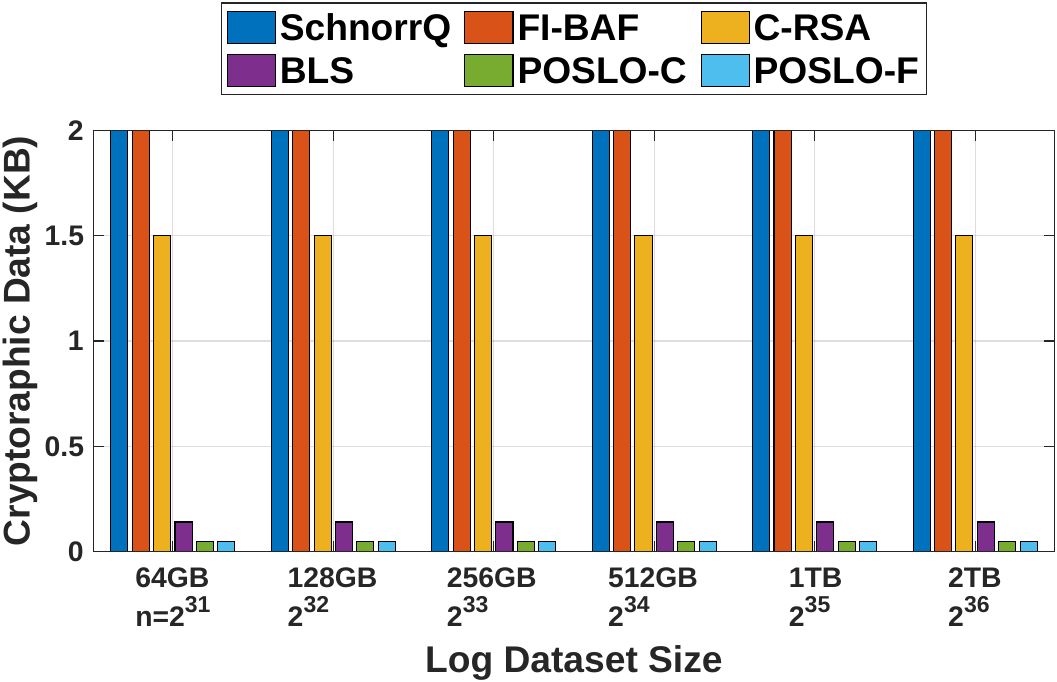}
		\end{subfigure}
	\hfill
	\begin{subfigure}[b]{0.32\textwidth}
			\centering
			\includegraphics[width=\textwidth]{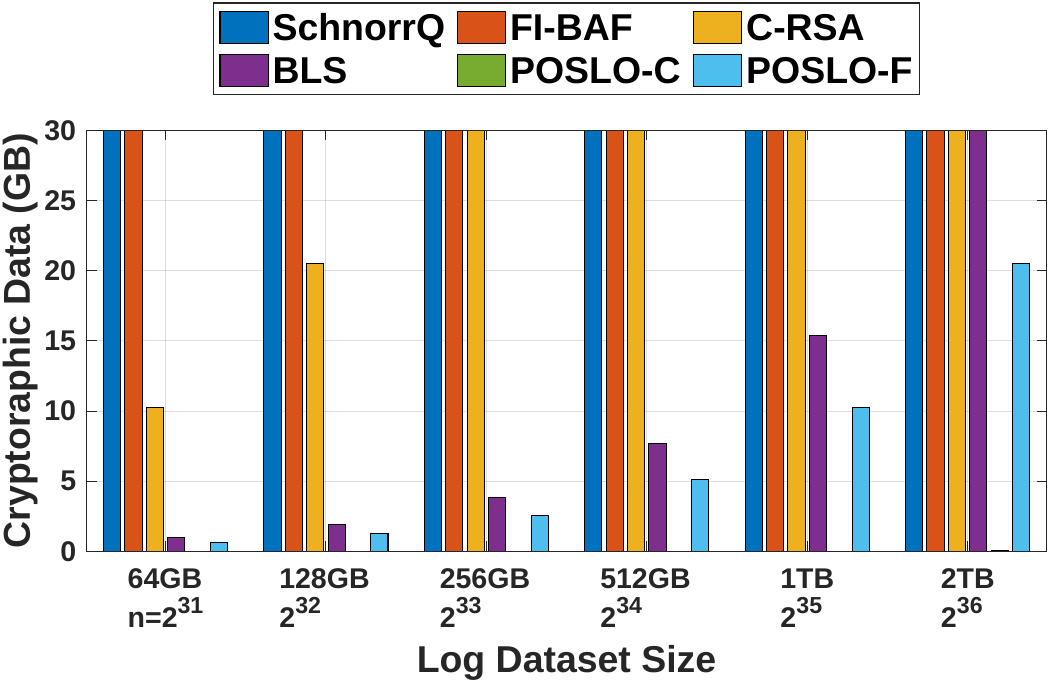}
		\end{subfigure} 
	\hfill
	\begin{subfigure}[b]{0.32\textwidth} 
			\centering
			\includegraphics[width=\textwidth]{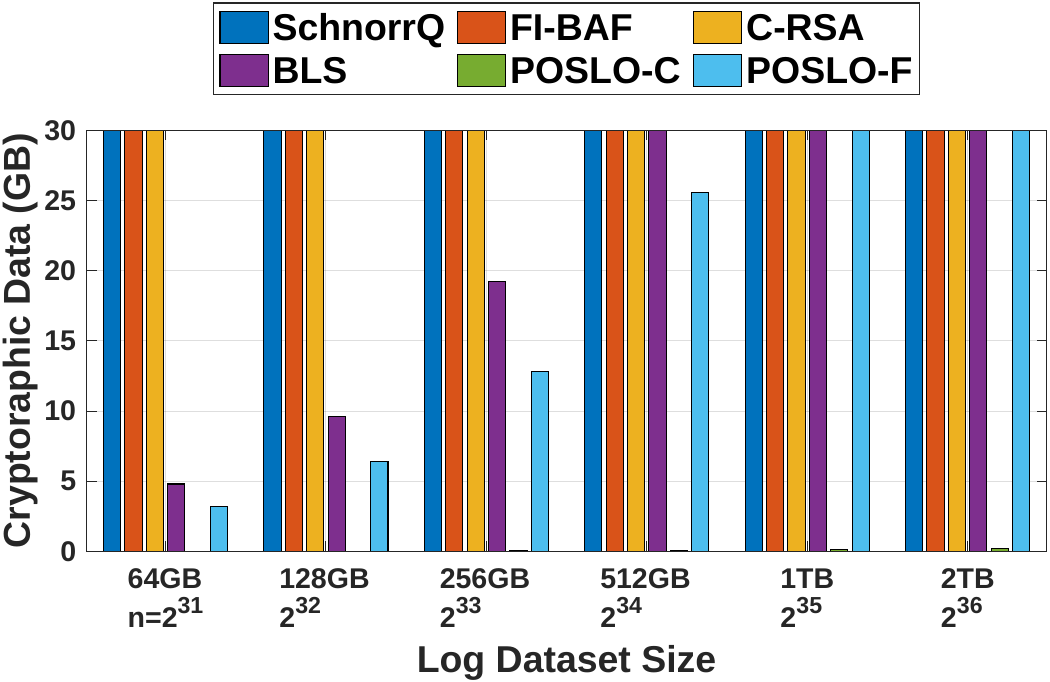}
		\end{subfigure}

    \begin{subfigure}[b]{0.32\textwidth}
			\centering
			\includegraphics[width=\textwidth]{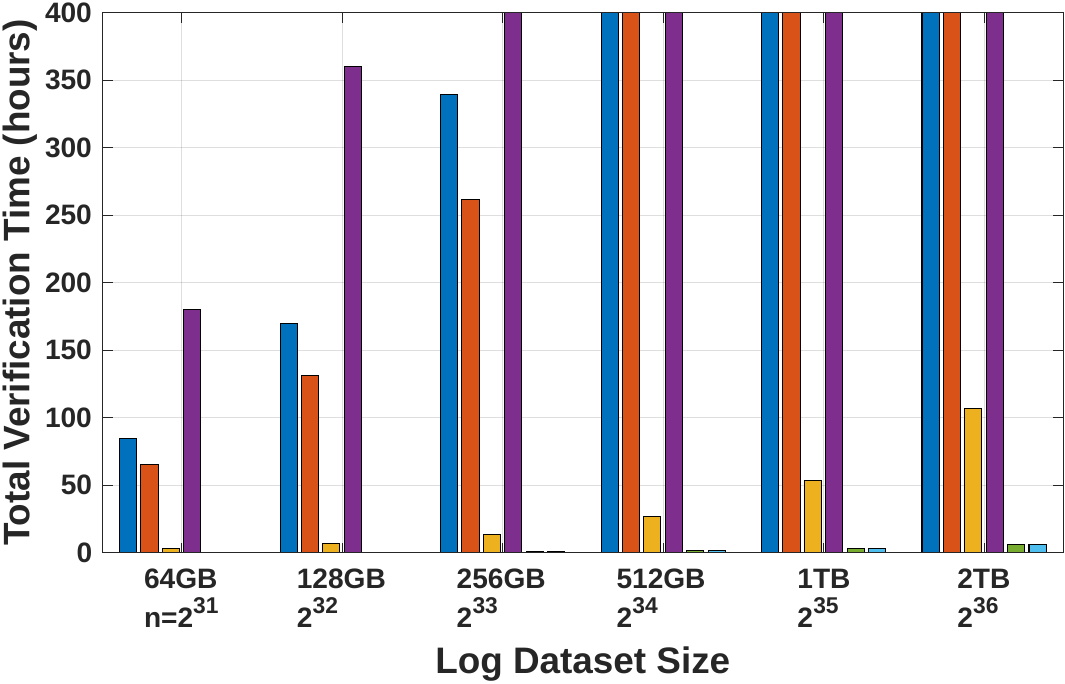}
                \caption{$\tau_F=0\%$}
		\end{subfigure}
	\hfill
	\begin{subfigure}[b]{0.32\textwidth}
			\centering
			\includegraphics[width=\textwidth]{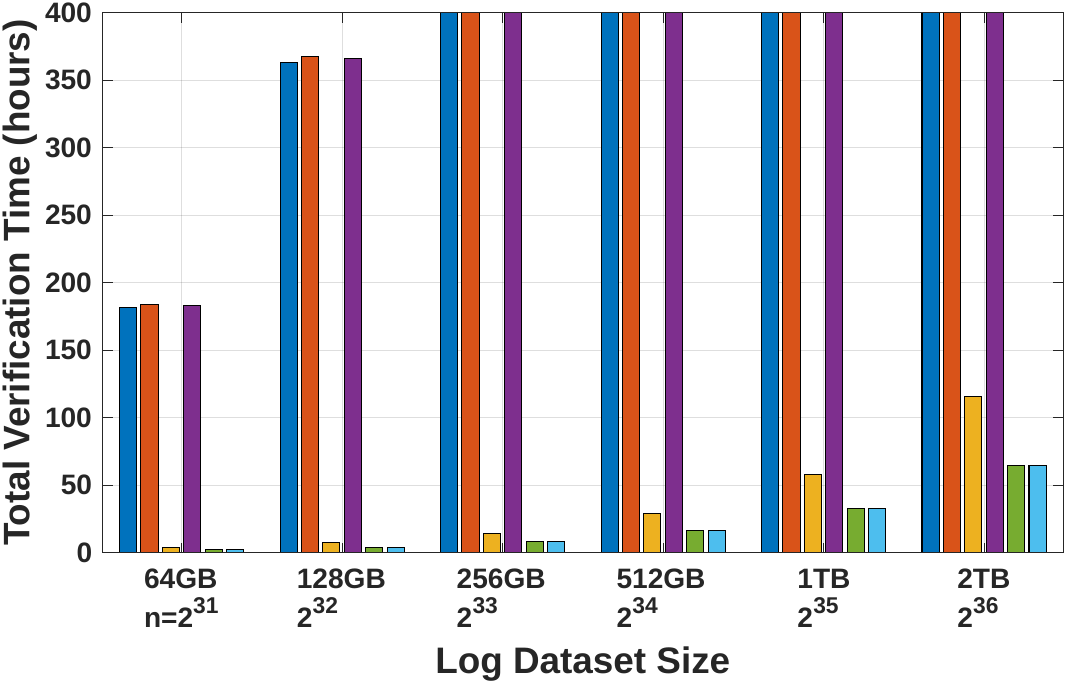}
                \caption{$\tau_F=1\%$}
		\end{subfigure} 
	\hfill
	\begin{subfigure}[b]{0.32\textwidth} 
			\centering
			\includegraphics[width=\textwidth]{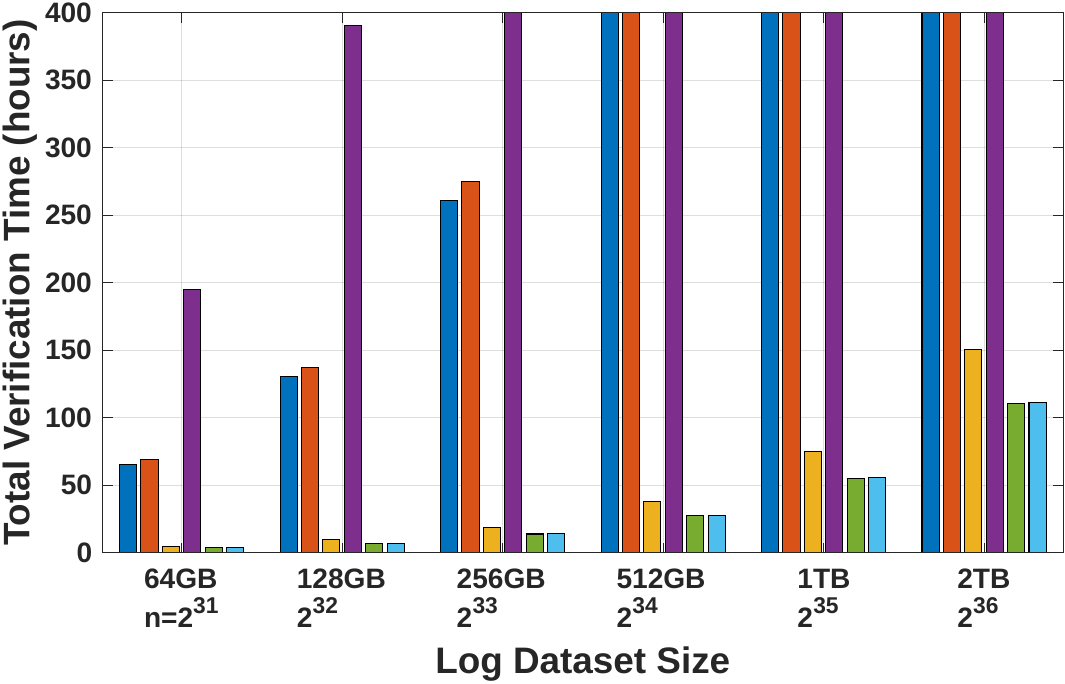}
            \caption{$\tau_F=5\%$}
		\end{subfigure}

	\vspace*{-3mm}
	\caption{Storage and verification time (on x86/64) comparison of \poslo~schemes and their counterparts at the cold storage server (\css)~under different failure rates ($|m|$=$256$-bit, $n_2$=$2^8$)}
	\label{fig:perf-comparison}
	
\end{figure*}

Figure \ref{fig:perf-comparison} presents the verification time and the storage overhead for different log entry set sizes (with each entry being 32 bytes) and failure rates $\tau_F$. Recall that $\tau_F$ represents the proportion of entries marked as \enquote{invalid} during distillation. As discussed in Section \ref{sec:models}, in the vast majority of real-world applications, flagged events (\enquote{invalid} logs) constitute only a small fraction of the overall log. Therefore, it is preferable to avoid compressing invalid tags, allowing them to be attested individually. 

In the case of full signature aggregation  (i.e., $\tau_F=0$), we refer the reader to Table \ref{tab:OverallPerfComp} that summarizes the verification time and storage advantages of our schemes. In Fig.~\ref{fig:perf-comparison},  we further investigate the efficiency of compared schemes for varying failure rate $(\tau_F)$ and umbrella signature $(n^u)$. 

Specifically, we vary  $\tau_F = 0, 1, 5\%$ to observe storage overhead and verification time in Fig. \ref{fig:perf-comparison}-($1^\text{st}$ row) and Fig. \ref{fig:perf-comparison}-($2^\text{nd}$ row), respectively. We increase the size of log entries from 64 GB ($2^{31}$ entries) to 2 TB ($2^{36}$ entries). 
We crop the y-axis to ensure better visibility of schemes with lower overhead and to prevent schemes having linear storage (e.g., SchnorrQ) and/or computational (e.g., Fi-BAF) from overshadowing the comparison.
In our experiments, for large logs, we processed them in small batches and included repeated disk I/O time in our results. 
We experimented with $\rho$ from $10^{-7}\%$ to $1\%$ and we observed that it has a minimal impact on performance in these margins. Further increase mainly impacts storage with only a slight increase in verification time.

	


	

{\em Disk I/O and Cold Storage Cost.} Consider a large IoT network where several low-end IoT devices are offloading their authenticated log entries to a remote edge cloud, and ultimately to the cold storage server (\css). The overall storage at both the edge clouds and \css~becomes exponentially large and costly. Recall that log files are infrequently accessed data, and therefore it is preferred to store them at a cold line solution (e.g., Google cloud\footnote{\url{https://cloud.google.com/storage}}), which is relatively low-priced (i.e., \$49.15/year for each TB). However, we argue that \poslo~is able to offer the best trade-off between low-cost compact server storage, low disk I/O, and fast verification. According to Table \ref{tab:OverallPerfComp}, \posloc's cryptographic storage overhead is only $0.05$KB for $1$TB of log entries, whereas it is $2$TB for the most signer-efficient counterpart FI-BAF. 
As a result, \posloc~have lower disk I/O time and cheaper storage cost since both metrics are directly proportional to the storage overhead. 
Additionally,  \poslo~optimizes the disk memory access time by only loading the overall aggregate tag to verify the set of log files. In case the verification fails, the partially condensed signatures are loaded to locate the flagged log. 
The storage cost at the distiller is more expensive than that of the cold storage server. As the distiller represents the medium between IoT devices and \css, its stored data is frequently accessed since it receives the authenticated log entries, and distills them after performing the verification. Afterward, it offloads the logs along with the associated cryptographic payload upon finishing a pre-defined set of epochs. This fits the standard storage for data stored within only brief periods of time. Based on the Google cloud solution, the storage cost of one Terabyte is equal to $\$245.76$/year. Similarly, the disk I/O becomes a key metric at the distiller side.

\subsubsection{Experimental Evaluation on GPUs}
\label{subsubsec:ccs_gpu}

In the following, we describe the implementation details and optimization techniques used to achieve an improved speedup compared to CPU version. 
\vspace{2pt}

\noindent {\em Finite Field Arithmetic with PTX Instructions.}
The proposed digital signature, \poslo, is instantiated with elliptic curves. The basic operations are performed in $\Zq$, where $q$ is a 256-bit prime. A number $a \in \Zq$ can be decomposed into 4 64-bit words $\{a_i\}_{i=1}^4$. We use PTX-based arithmetic over 64-bit words, halving the number of PTX instructions per operation compared to the 32-bit approach in \cite{hu2023high}. Multi-precision addition and subtraction of two 256-bit numbers are performed using PTX 64-bit instructions, with carry and borrow propagated using the single carry flag bit, CC.CF. The implementation of $Add_q$ is illustrated as follows:

\begin{adjustbox}{max width=0.9\textwidth}
	\begin{lstlisting}
		__device__ uint64_t add(uint64_t a[4], uint64_t b[4], uint64_t r[4]) {
			uint64_t carry;
			asm volatile("add.cc.u64 %0,%1,%2;" :"=l"(r[0]):"l"(a[0]),"l"(b[0]));
			asm volatile("addc.cc.u64 %0,%1,%2;" :"=l"(r[1]):"l"(a[1]),"l"(b[1]));
			asm volatile("addc.cc.u64 %0,%1,%2;" :"=l"(r[2]):"l"(a[2]),"l"(b[2]));
			asm volatile("addc.cc.u64 %0,%1,%2;" :"=l"(r[3]):"l"(a[3]),"l"(b[3]));
			asm volatile("addc.u64 %0, 0, 0;" :"=l"(carry));
			return carry;
		}
	\end{lstlisting}
\end{adjustbox}
\vspace{2pt}

\noindent {\em AES with T-Table Approach.} The primary computational overhead of \poslo~signing and verification are hash function ($H$) and pseudo-random function ($\prf$) invocations. To optimize signing on resource-constrained devices, we instantiate $H$ and \prf~using AES-based constructions for \poslo+. It also provides an advantage on verification for distiller and \css~due to hardware-accelerated implementations on commodity hardware (e.g., x86/64) and GPU architectures. 
Numerous hardware-accelerated GPU-based techniques (e.g., T-table approach \cite{tezcan2021optimization}, bitsliced approach \cite{hajihassani2019fast}) have been proposed for AES block cipher on GPUs. 
We adopt the T-table approach by storing only a single T-table in shared memory. In \cite{tezcan2021optimization}, 32 copies of the T-table are used to avoid bank conflicts, but this creates warp divergence and consumes $32$KB of shared memory. The high shared memory usage and warp divergence hinder its deployment as a building block in digital signature primitives (e.g., \poslo), which rely on shared memory to store input messages, auxiliary data, and output results. 
Moreover, prior GPU-based AES implementations target encryption and decryption applications, where the user key is not updated, thus, the round key is precomputed before kernel launch. 
In contrast, the MMO and MDC-2 constructions using AES-128 execute the key expansion for each 128-bit chunk of the input data. 
Therefore, we implemented the AES key expansion algorithm on the device, using PTX instructions similar to the cipher computation. 
Copying 32 duplicates (as in \cite{tezcan2021optimization}) of T-tables of both key expansion and cipher algorithms would exceed the shared memory limit (74KB out of a maximum of 64KB). 
In our implementation, the total shared memory usage for AES computations in the CUDA kernel is only $2.3$KB per block, which includes two T-tables, the S-box table, and the round constant word array, while the round keys are stored in the register space. 

\vspace{2pt}
\noindent {\em Inline Macro and Loop Unrolling.} 
{\em (i)} Device functions called from the main kernel introduce additional overhead due to the need for the program to jump to the function code. To mitigate this, CUDA provides the {\_\_forceinline\_\_} macro, which forces the compiler to inline device functions. This optimization technique reduces the function call overhead, ultimately lowering the overall verification time. 
{\em (ii)} Loop unrolling enhances performance by increasing register usage and eliminating loop control overhead. However, when loops already consume a large number of registers, unrolling may lead to register saturation, reducing verification efficiency. To balance performance and resource constraints, we apply loop unrolling selectively for finite field operations but omit it in the implementation of AES-128 and SHA-256.


\begin{table*}[ht!]
	\centering
	\caption{Verification peak throughput and speedup of \aver~and \paver~for \poslo~and counterparts}
	\label{tab:gpu_comparison_extended}
	\footnotesize
	\vspace{-2mm}
	\begin{minipage}{\textwidth}
		\resizebox{\textwidth}{!}{
			\begin{tabular}{|l|ccc|ccc|ccc|}
				\hline
				\multirow{2}{*}{\textbf{Scheme}} & 
				\multicolumn{3}{c|}{\textbf{$\boldsymbol{|m| = 64}$-bit}} & 
				\multicolumn{3}{c|}{\textbf{$\boldsymbol{|m| = 128}$-bit}} & 
				\multicolumn{3}{c|}{\textbf{$\boldsymbol{|m| = 256}$-bit}} 
				\\ \cline{2-10}
				& \textbf{\aver (x86/64)} & \textbf{\paver (GPU)} & \textbf{\textbf{\aver} / \textbf{\paver} Ratio} 
				& \textbf{\aver (x86/64)} & \textbf{\paver (GPU)} & \textbf{\textbf{\aver} / \textbf{\paver} Ratio}
				& \textbf{\aver (x86/64)} & \textbf{\paver (GPU)} & \textbf{\textbf{\aver} / \textbf{\paver} Ratio}
				\\ \hline
				SchnorrQ~\cite{costello2016schnorrq}
				& $3.4 \times 10^{-5}$ & $0.0037$ & $108.82$
				& $3.4 \times 10^{-5}$ & $0.0037$ & $108.82$
				& $3.4 \times 10^{-5}$ & $0.0037$ & $108.82$
				\\ \hline
				BLS~\cite{BLS:2004:Boneh:JournalofCrypto}
				& $1.23 \times 10^{-5}$ & N/A & N/A
				& $1.23 \times 10^{-5}$ & N/A & N/A
				& $1.23 \times 10^{-5}$ & N/A & N/A
				\\ \hline \hline
				$\text{\poslo}$
				& $0.99$ & $89.83$ & $90.83$
				& $0.98$ & $91.33$ & $93.20$
				& $0.96$ & $85.76$ & $89.33$
				\\ \hline
				$\poslop$
				& $1.05$ & $1782.13$ & $1697.27$
				& $0.77$ & $1711.60$ & $2222.86$
				& $0.6$ & $1290.32$ & $2150.53$
				\\ \hline
				$\poslopp$
				& $1.75$ & $275.62$ & $157.50$
				& $1.76$ & $271.23$ & $154.21$
				& $1.1$ & $265.68$ & $241.53$
				\\ \hline
			\end{tabular}
		}
		\vspace{3pt}
		{
			\scriptsize
            \aver~and \paver~denote the throughput of $2^{20}$ log entry verifications per second. No GPU implementation is available for BLS.
		}
	\end{minipage}
\end{table*}

\begin{figure*}[t]
	\centering
	
	\begin{subfigure}[b]{0.32\textwidth}
		\centering
		\includegraphics[width=\textwidth]{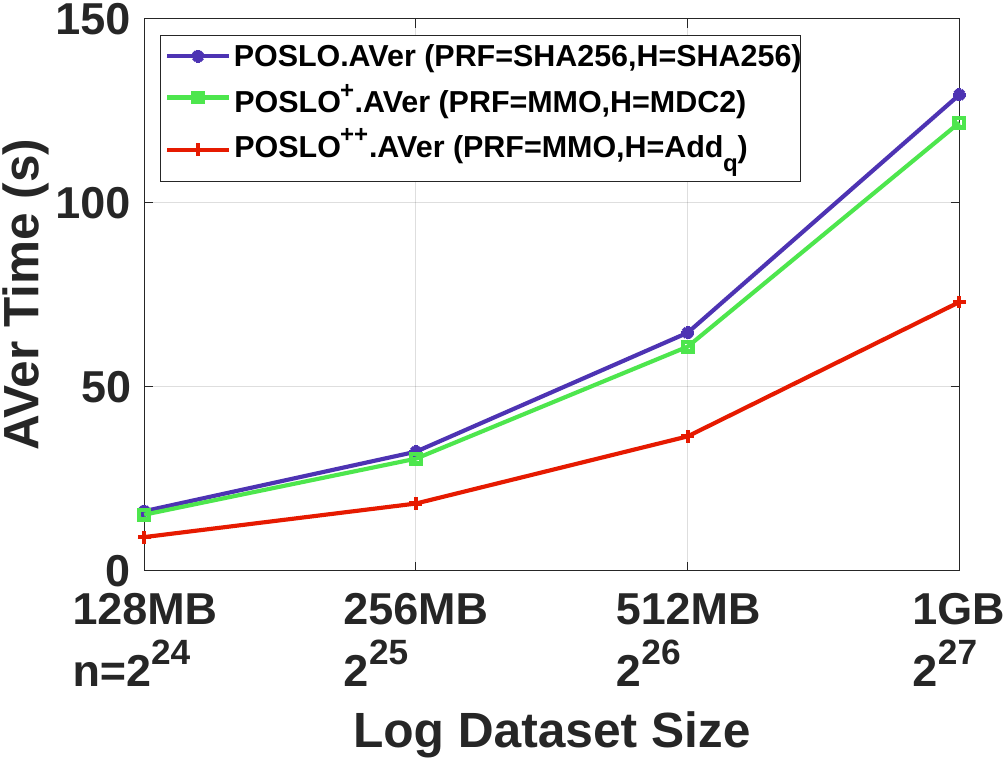}
	\end{subfigure} 
	\hfill
	\begin{subfigure}[b]{0.32\textwidth} 
		\centering
		\includegraphics[width=\textwidth]{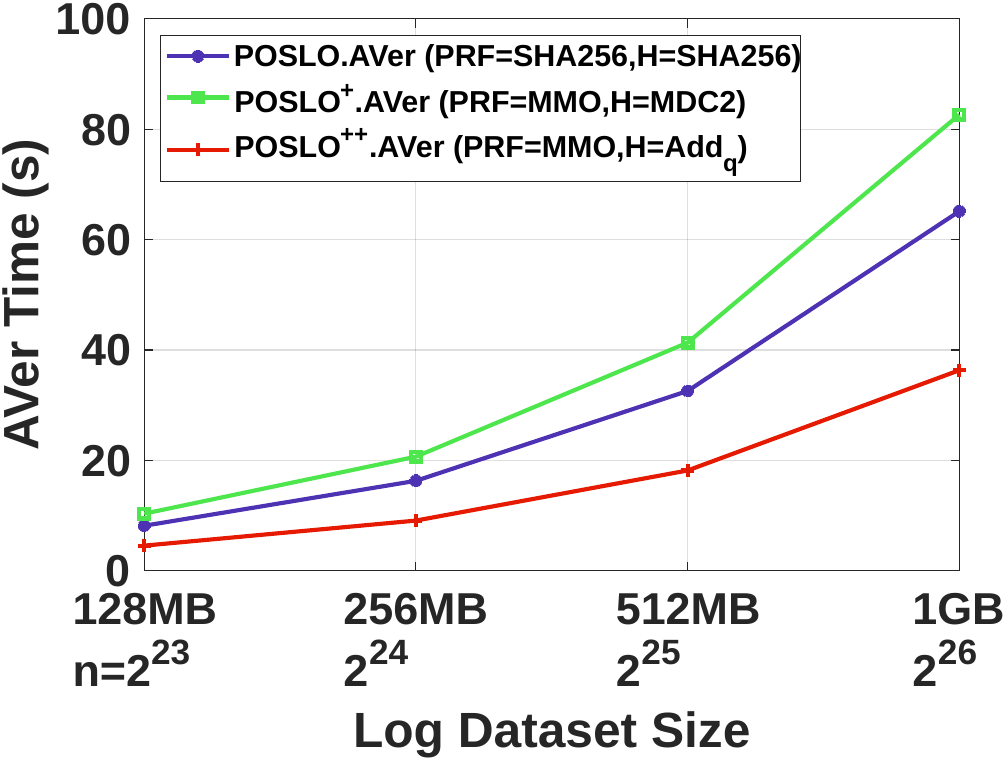}
	\end{subfigure}
	\hfill
	\begin{subfigure}[b]{0.32\textwidth}
		\centering
		\includegraphics[scale=0.265]{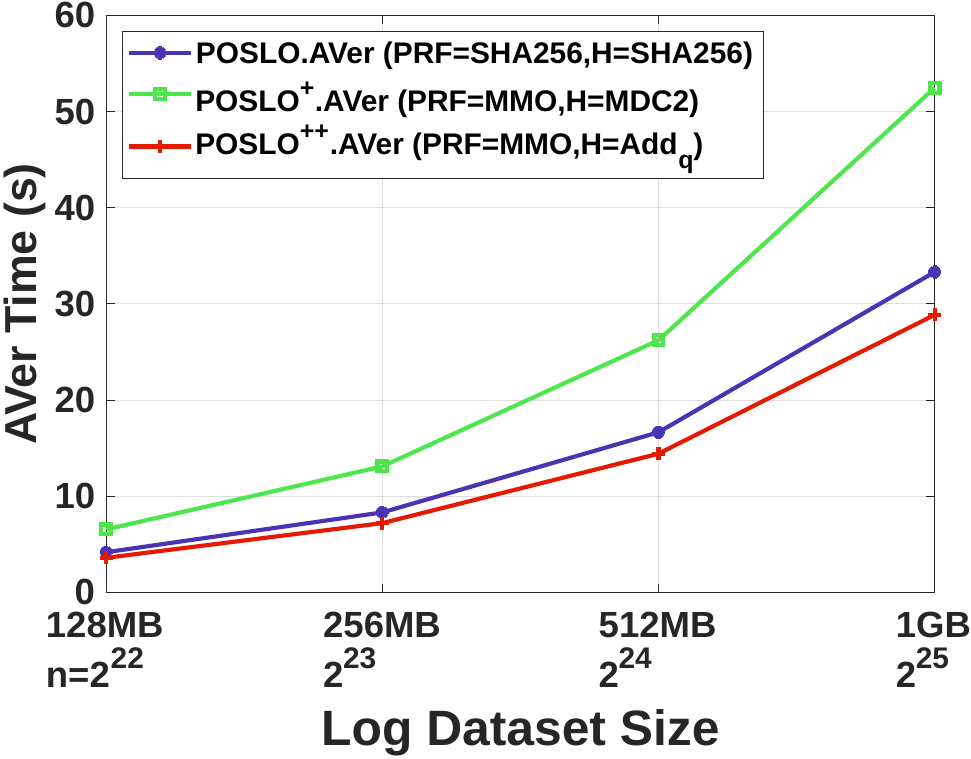}
	\end{subfigure}

	
	\begin{subfigure}[b]{0.32\textwidth}
		\centering
		\includegraphics[width=\textwidth]{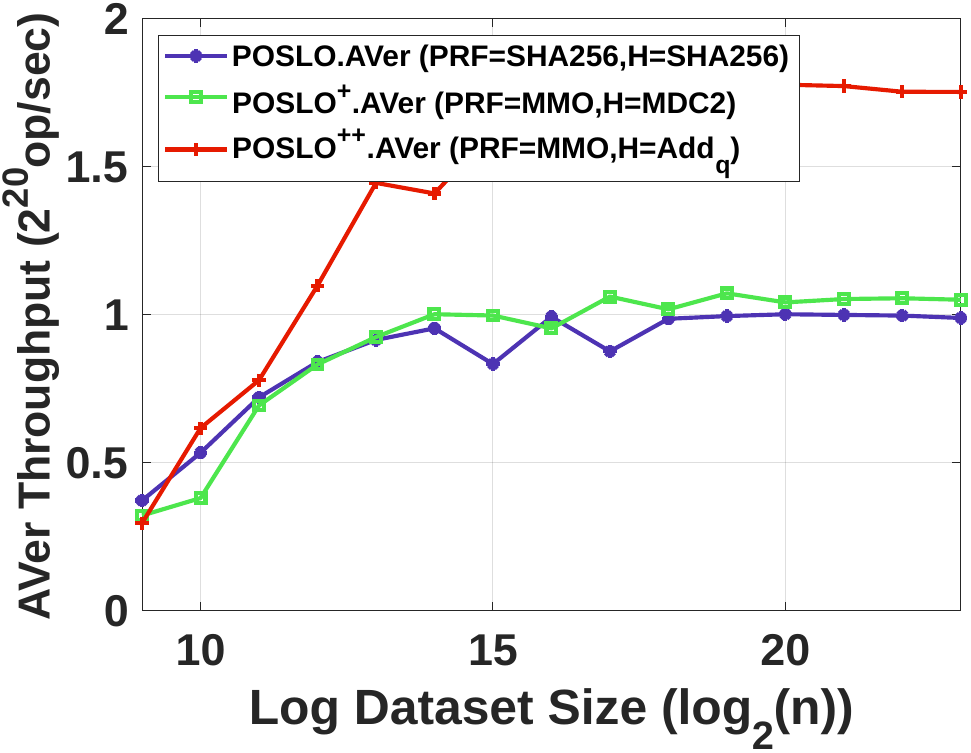}
	\end{subfigure} 
	\hfill
	\begin{subfigure}[b]{0.32\textwidth} 
		\centering
		\includegraphics[width=\textwidth]{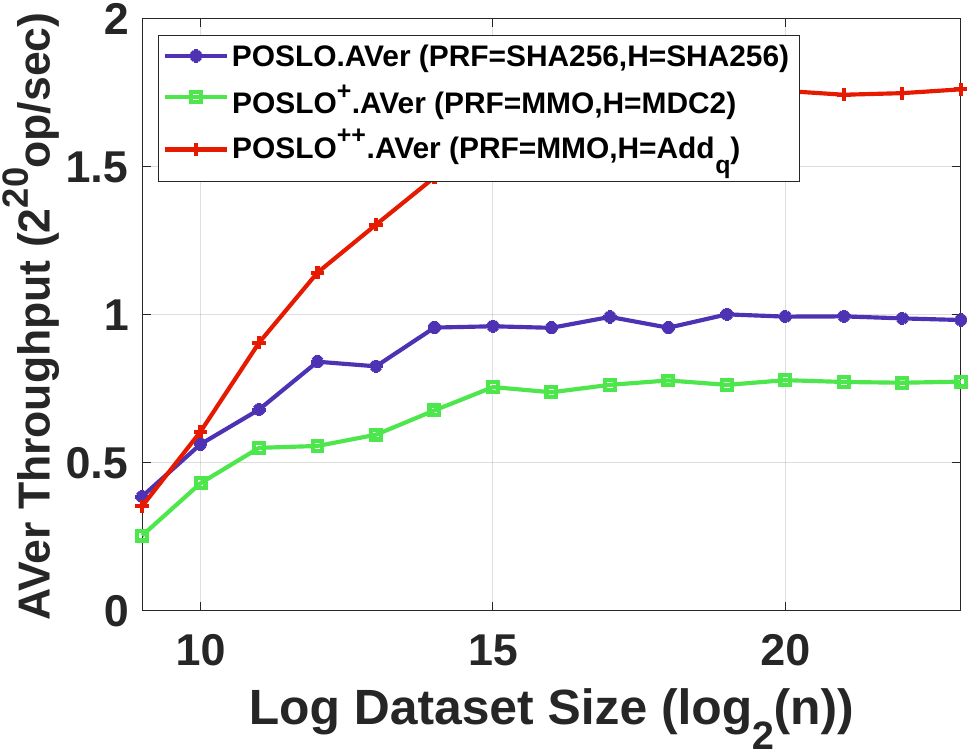}
	\end{subfigure}
	\hfill
	\begin{subfigure}[b]{0.32\textwidth}
		\centering
		\includegraphics[width=\textwidth]{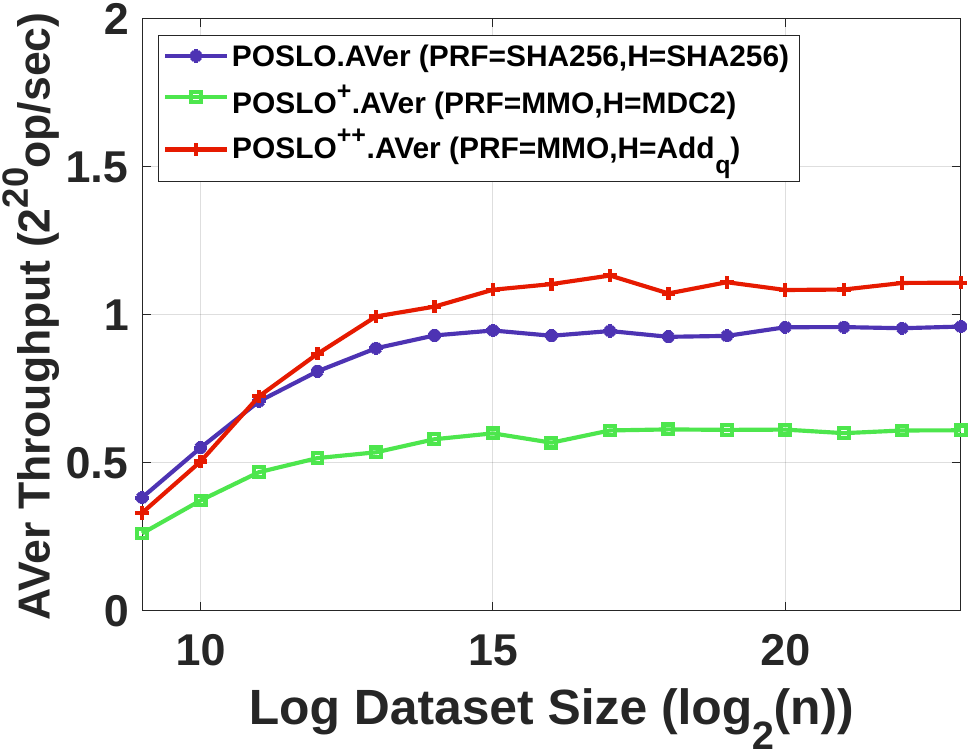}
	\end{subfigure}
    
	
	\begin{subfigure}[b]{0.32\textwidth}
		\centering
		\includegraphics[width=\textwidth]{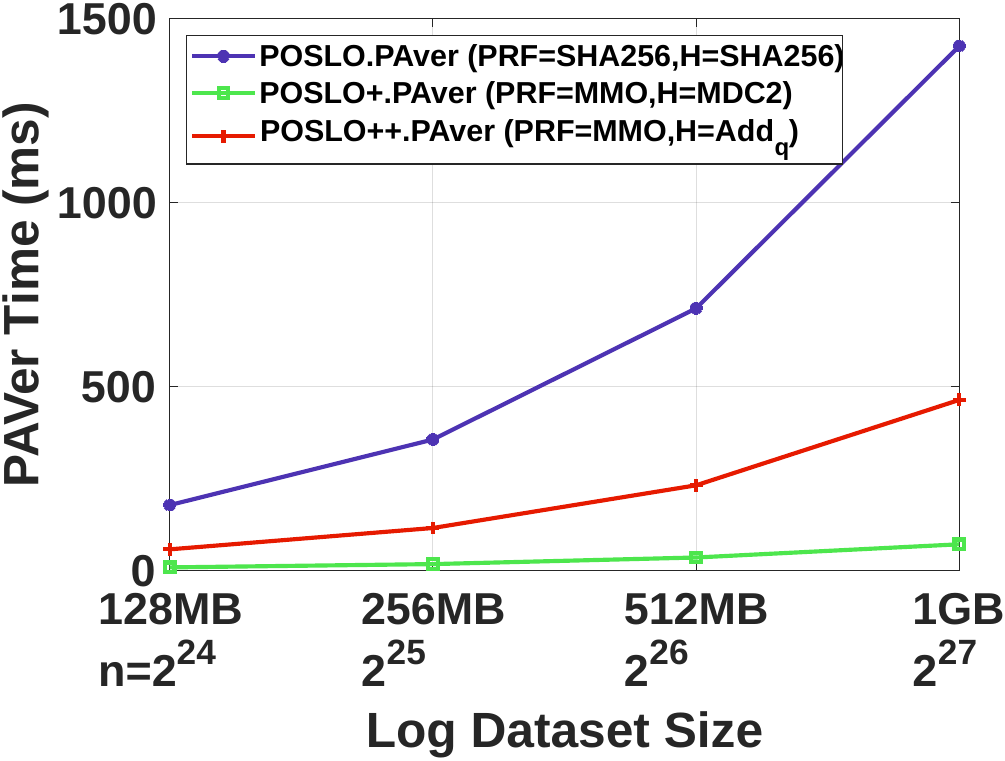}
	\end{subfigure} 
	\hfill
	\begin{subfigure}[b]{0.32\textwidth} 
		\centering
		\includegraphics[width=\textwidth]{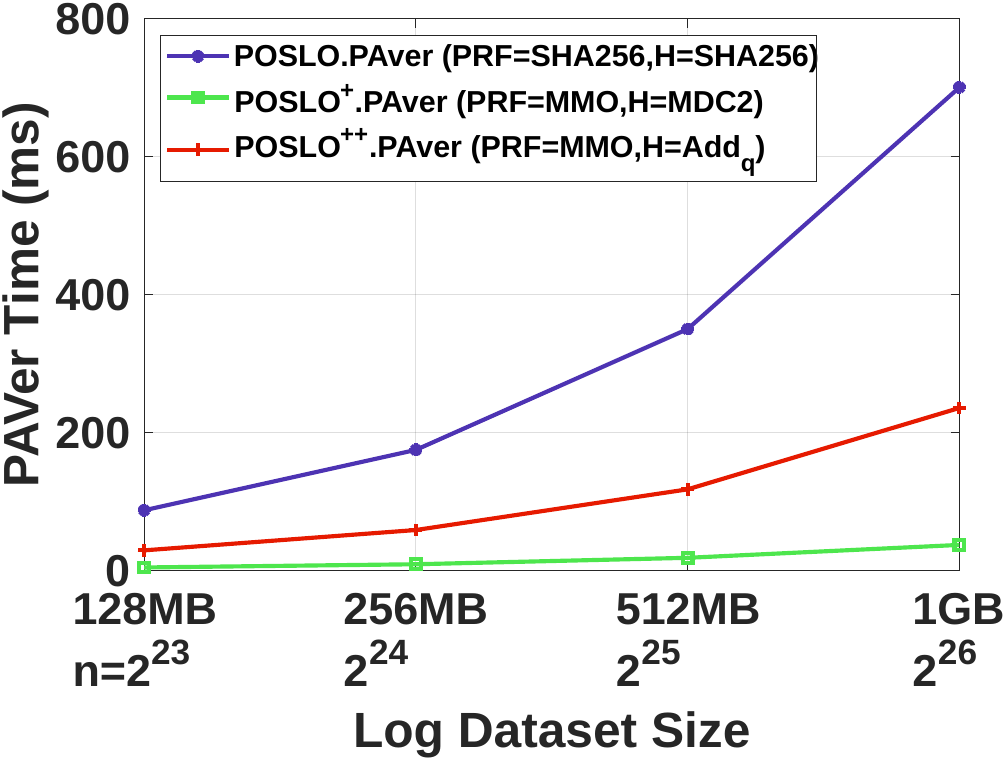}
	\end{subfigure}
	\hfill
	\begin{subfigure}[b]{0.32\textwidth}
		\centering
		\includegraphics[width=\textwidth]{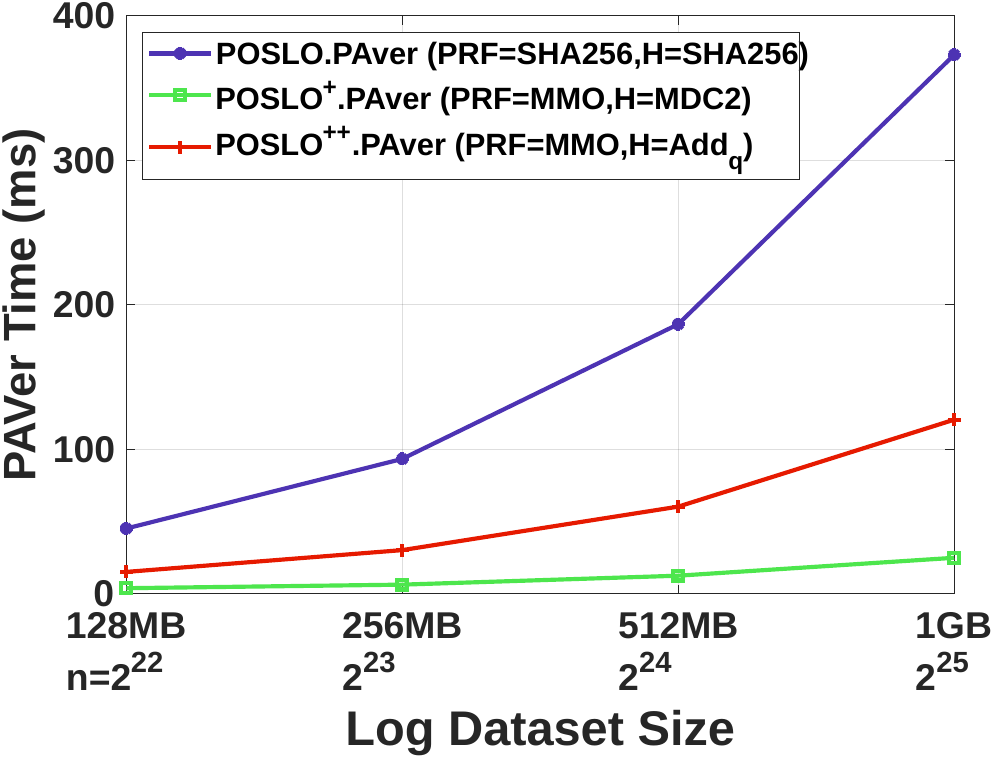}
	\end{subfigure}

	
	\begin{subfigure}[b]{0.32\textwidth}
		\centering
		\includegraphics[width=\textwidth]{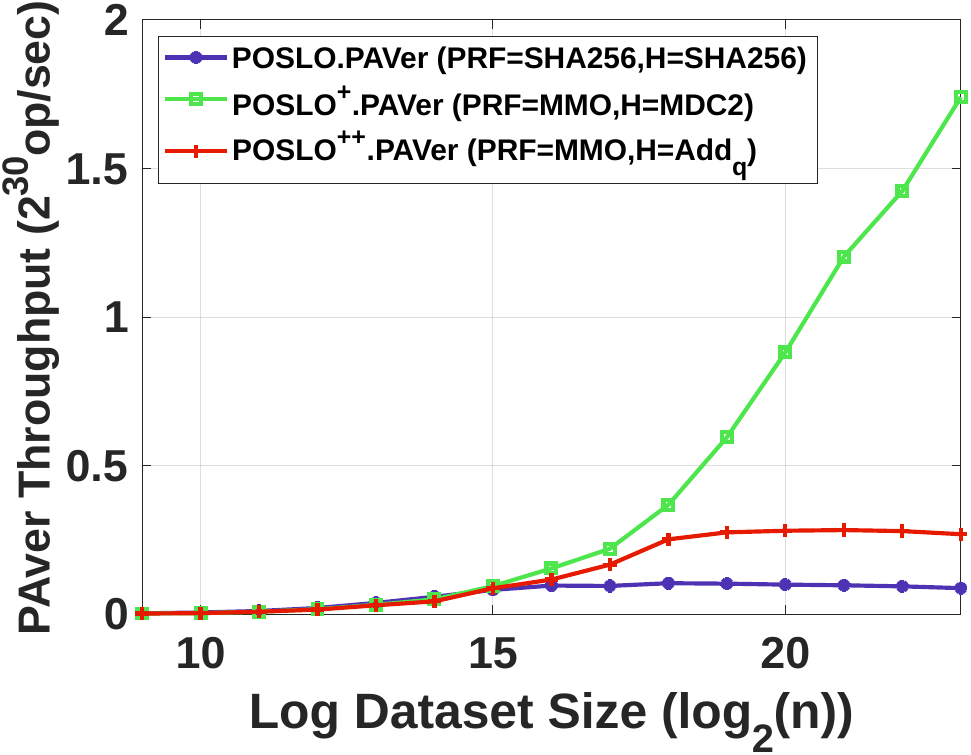}
		\caption{$|m|=64$-bit}
	\end{subfigure} 
	\hfill
	\begin{subfigure}[b]{0.32\textwidth} 
		\centering
		\includegraphics[width=\textwidth]{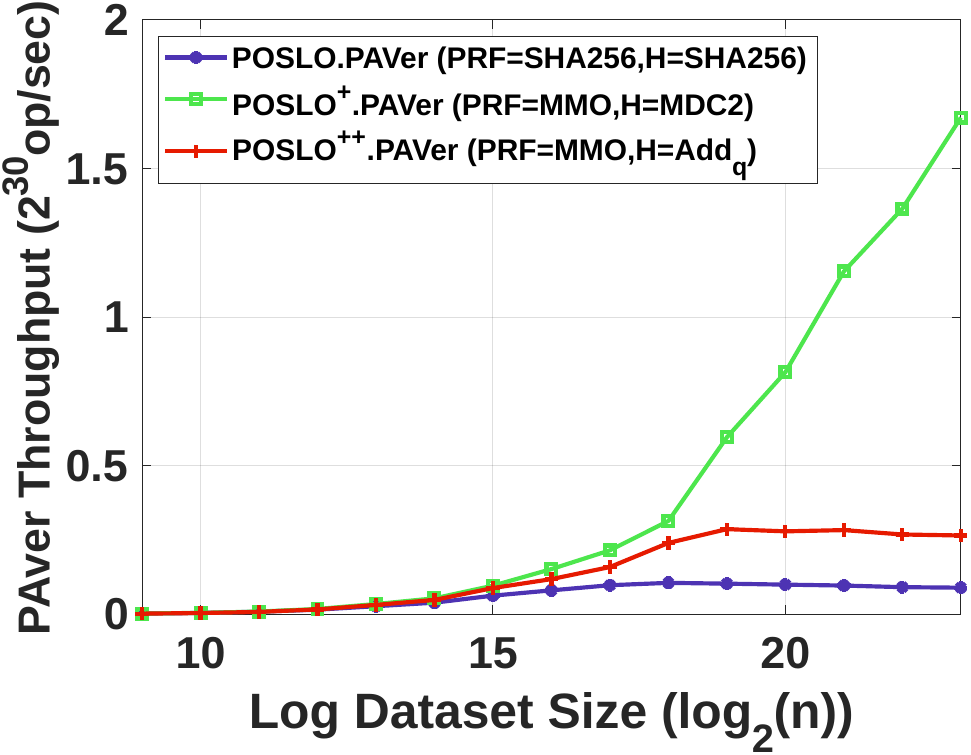}
		\caption{ $|m|=128$-bit}
	\end{subfigure}
	\hfill
	\begin{subfigure}[b]{0.32\textwidth}
		\centering
		\includegraphics[width=\textwidth]{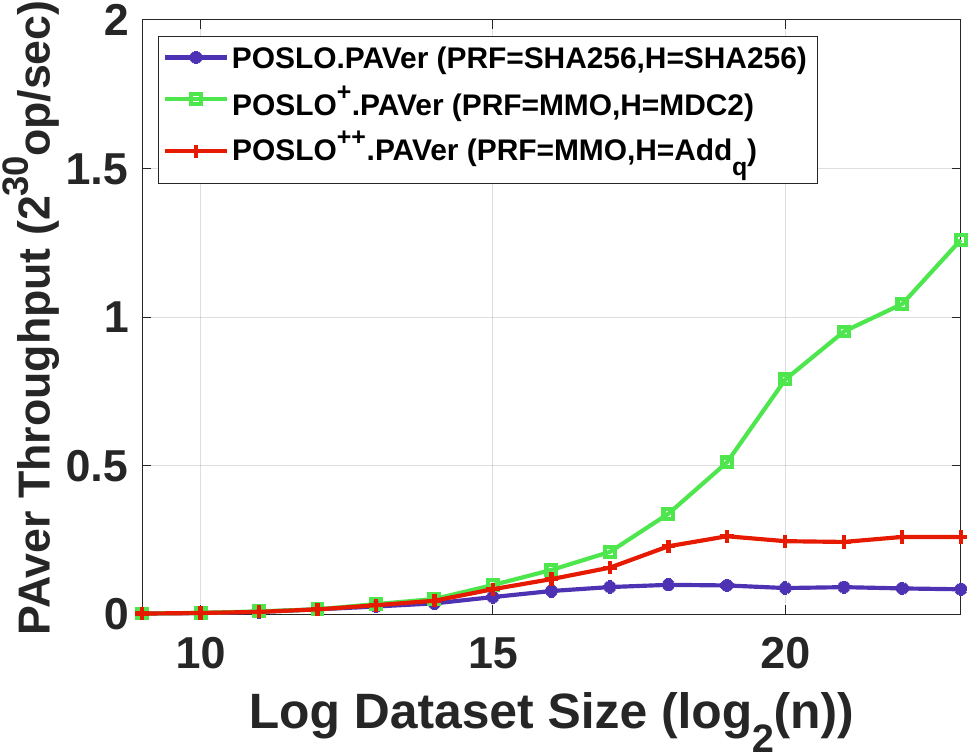}
		\caption{$|m|=256$-bit}
	\end{subfigure}
        \vspace*{-4mm}
	\caption{Performance of (parallel) batch verification (\texttt{P})\aver~of \poslo~instantiations on x86/64 (\posloaver: $1^\text{st}$-$2^{\text{nd}}$ rows) and GPU (\poslopaver: $3^\text{rd}$-$4^{\text{th}}$ rows)}
	\label{fig:gpu-perf-comparison}
	
		\vspace{-3mm}
\end{figure*}

\noindent $\bullet$ \underline{{\em Performance Results}:} 
Figure \ref{fig:gpu-perf-comparison} and Table \ref{tab:gpu_comparison_extended} highlight the verification efficiency on the GPU of \poslo~variants and counterparts, compared to the CPU-bounded x86/64 baseline implementation. Our evaluation is performed across different input and batch sizes, with three different combinations of $H$ and $\prf$ (as detailed in Table \ref{tab:instantations}). $\poslo+$, uses AES-based MMO and MDC-2 constructions, yields superior verification throughput up to $\approx 2^{31}$ log entry verification per second and consists of $2,222\times$ speedup over x86/64 for 128-bit messages. This is attributed to the GPU optimizations, specifically leveraging the T-Table AES optimization and PTX instructions. Considering a large log data (e.g., 1TB), $\poslo$+~batch verification time is only $24.8$ seconds, including the disk I/O overhead. In contrast, $\poslo$, instantiated with SHA256 for both $H$ and $PRF$, offers stronger security against collision attacks but achieves a $18\times$ lesser throughput compared to $\poslop$. $\poslopp$ variant, which uses modular arithmetic hashing for small inputs, achieves a throughput of $\approx 2^{28}$ log entry verification per second, which is $2.97\times$ and $6.3\times$ faster and slower than $\poslo$ and AES-based $\poslop$, respectively. However, \poslopp~outperforms both \poslo~and \poslop~in terms of verification time on x86/64 architecture due to the low overhead of modular addition compared to sequential SHA-256 and MDC-2 with AES-128. Therefore, \poslopp~is the best candidate when efficient distillation and batch verification on x86/64 is desirable but only for small input sizes, whereas \poslop~supports large input sizes and is the best candidate for both low-end devices and Cold-StaaS, especially when GPU acceleration is available at the Cold-StaaS. 

\poslo~instantiations performs several orders of magnitude better compared to the baseline non-aggregate SchnorrQ signature scheme, which uses identical EC-based operations.  Note that our peak performance is observed on $1$GB log data. The verification of larger log data (e.g., 1TB) can be performed via (1) sequential execution of $\posloaggekeys$ per $1$GB chunk to obtain the sub-aggregate of ephemeral keys (i.e., $\agg{e}$). (2) running in parallel $\posloaggekeys$ on $1$GB chunks across multiple GPUs, to further reduce the audit time. Moreover, the umbrella signature verification can be performed simultaneously during the final aggregation of sub-aggregate ephemeral keys. Overall, our performance results confirm the efficiency of our parallel \poslo~design that harnesses the epoch processing and mutual aggregation property, and further improves on the AES-based variant using T-Table implementation. This reaffirms that ${\poslop}$ is the most suitable candidate for the IoT-Cloud continuum by achieving the lowest energy consumption and the highest verification throughput on low-end IoT devices and cold storage servers, respectively.

\section{Conclusion} \label{sec:conclusion}
In this work, we introduced a new family of aggregate signatures, called \poslo, designed for secure logging in resource-constrained IoT networks. To the best of our knowledge, \poslo~achieves the strongest trade-off among provable security guarantees, computational efficiency, and cryptographic storage compactness, surpassing existing secure logging schemes across diverse performance metrics.

At its core, \poslo~incorporates a novel tree-based seed management strategy and a post-signature disclosure of one-time separated commitments. This combination enables a highly compact cryptographic state while supporting secure and efficient signing and verification. To preserve fine-grained verifiability, \poslo~introduces a tunable granularity parameter that allows selective retention of compact tags after signature distillation. Beyond the traditional CPU-bound implementations, we developed \poslopaver, a GPU-accelerated batch verification algorithm that exploits the homomorphic and decoupled-commitment structure of \poslo~to deliver ultra-efficient throughput. \poslopaver~achieves up to $2^{31}$ log verifications per second, and demonstrates multiple orders of magnitude speedup over GPU-accelerated non-aggregate EC-based SchnorrQ (as our baseline).

Our comprehensive experimental evaluation on resource-constrained devices, general-purpose CPUs, and GPU platforms confirms that \poslo~and its variants offer strong configurability, exceptional efficiency, and minimal cryptographic storage. We further implement three optimized instantiations—\poslo, \poslop, and \poslopp, each leveraging different cryptographic primitives (e.g., hash-based, AES-based, or modular arithmetic) to provide flexible performance-security trade-offs. We formally prove that \poslo~achieves \AEUCMA~security in the random oracle model. Our full-fledged implementation is open source and publicly available to support reproducibility, third-party evaluation, and future deployment.

        \section*{Acknowledgments}
        This research is partially supported by the National Science Foundation (NSF) grant CNS-2350213.

		\bibliographystyle{ACM-Reference-Format}
		\bibliography{crypto-etc, AttilaYavuz}


\begin{thebibliography}{55}


\ifx \showCODEN    \undefined \def \showCODEN     #1{\unskip}     \fi
\ifx \showDOI      \undefined \def \showDOI       #1{#1}\fi
\ifx \showISBNx    \undefined \def \showISBNx     #1{\unskip}     \fi
\ifx \showISBNxiii \undefined \def \showISBNxiii  #1{\unskip}     \fi
\ifx \showISSN     \undefined \def \showISSN      #1{\unskip}     \fi
\ifx \showLCCN     \undefined \def \showLCCN      #1{\unskip}     \fi
\ifx \shownote     \undefined \def \shownote      #1{#1}          \fi
\ifx \showarticletitle \undefined \def \showarticletitle #1{#1}   \fi
\ifx \showURL      \undefined \def \showURL       {\relax}        \fi
\providecommand\bibfield[2]{#2}
\providecommand\bibinfo[2]{#2}
\providecommand\natexlab[1]{#1}
\providecommand\showeprint[2][]{arXiv:#2}

\bibitem[Ahmad et~al\mbox{.}(2022)]%
        {ahmad2022hardlog}
\bibfield{author}{\bibinfo{person}{Adil Ahmad}, \bibinfo{person}{Sangho Lee}, {and} \bibinfo{person}{Marcus Peinado}.} \bibinfo{year}{2022}\natexlab{}.
\newblock \showarticletitle{Hardlog: Practical tamper-proof system auditing using a novel audit device}. In \bibinfo{booktitle}{\emph{2022 IEEE Symposium on Security and Privacy (SP)}}. IEEE, \bibinfo{pages}{1791--1807}.
\newblock


\bibitem[Anthoine et~al\mbox{.}(2021)]%
        {anthoine2021dynamic}
\bibfield{author}{\bibinfo{person}{Gaspard Anthoine}, \bibinfo{person}{Jean-Guillaume Dumas}, \bibinfo{person}{M{\'e}lanie de Jonghe}, \bibinfo{person}{Aude Maignan}, \bibinfo{person}{Cl{\'e}ment Pernet}, \bibinfo{person}{Michael Hanling}, {and} \bibinfo{person}{Daniel~S Roche}.} \bibinfo{year}{2021}\natexlab{}.
\newblock \showarticletitle{Dynamic proofs of retrievability with low server storage}. In \bibinfo{booktitle}{\emph{30th USENIX Sec. Symp.}} \bibinfo{pages}{537--554}.
\newblock


\bibitem[Ateniese et~al\mbox{.}(2008)]%
        {ateniese2008scalable}
\bibfield{author}{\bibinfo{person}{Giuseppe Ateniese}, \bibinfo{person}{Roberto Di~Pietro}, \bibinfo{person}{Luigi~V Mancini}, {and} \bibinfo{person}{Gene Tsudik}.} \bibinfo{year}{2008}\natexlab{}.
\newblock \showarticletitle{Scalable and efficient provable data possession}. In \bibinfo{booktitle}{\emph{Proc. of the 4th international conference on Security and privacy in communication netowrks}}. \bibinfo{pages}{1--10}.
\newblock


\bibitem[Bernstein et~al\mbox{.}(2012)]%
        {Ed25519}
\bibfield{author}{\bibinfo{person}{Daniel~J. Bernstein}, \bibinfo{person}{Niels Duif}, \bibinfo{person}{Tanja Lange}, \bibinfo{person}{Peter Schwabe}, {and} \bibinfo{person}{Bo-Yin Yang}.} \bibinfo{year}{2012}\natexlab{}.
\newblock \showarticletitle{High-speed high-security signatures}.
\newblock \bibinfo{journal}{\emph{Journal of Cryptographic Engineering}} \bibinfo{volume}{2}, \bibinfo{number}{2} (\bibinfo{date}{01 Sep} \bibinfo{year}{2012}), \bibinfo{pages}{77--89}.
\newblock
\showISSN{2190-8516}


\bibitem[Boneh et~al\mbox{.}(2004)]%
        {BLS:2004:Boneh:JournalofCrypto}
\bibfield{author}{\bibinfo{person}{Dan Boneh}, \bibinfo{person}{Ben Lynn}, {and} \bibinfo{person}{Hovav Shacham}.} \bibinfo{year}{2004}\natexlab{}.
\newblock \showarticletitle{Short Signatures from the Weil Pairing}.
\newblock \bibinfo{journal}{\emph{J. Cryptol.}} \bibinfo{volume}{17}, \bibinfo{number}{4} (\bibinfo{year}{2004}), \bibinfo{pages}{297–319}.
\newblock


\bibitem[Boyko et~al\mbox{.}(1998)]%
        {boyko1998speeding}
\bibfield{author}{\bibinfo{person}{Victor Boyko}, \bibinfo{person}{Marcus Peinado}, {and} \bibinfo{person}{Ramarathnam Venkatesan}.} \bibinfo{year}{1998}\natexlab{}.
\newblock \showarticletitle{Speeding up Discrete Log and Factoring Based Schemes via Precomputations}. In \bibinfo{booktitle}{\emph{EUROCRYPT '98} (\bibinfo{edition}{eurocrypt '98} ed.)}. \bibinfo{pages}{221--235}.
\newblock


\bibitem[Chandra(2001)]%
        {chandra2001parallel}
\bibfield{author}{\bibinfo{person}{Rohit Chandra}.} \bibinfo{year}{2001}\natexlab{}.
\newblock \bibinfo{booktitle}{\emph{Parallel programming in OpenMP}}.
\newblock \bibinfo{publisher}{Morgan kaufmann}.
\newblock


\bibitem[Chen et~al\mbox{.}(2024)]%
        {chen2024last}
\bibfield{author}{\bibinfo{person}{Changhua Chen}, \bibinfo{person}{Tingzhen Yan}, \bibinfo{person}{Chenxuan Shi}, \bibinfo{person}{Hao Xi}, \bibinfo{person}{Zhirui Fan}, \bibinfo{person}{Hai Wan}, {and} \bibinfo{person}{Xibin Zhao}.} \bibinfo{year}{2024}\natexlab{}.
\newblock \showarticletitle{The Last Mile of Attack Investigation: Audit Log Analysis towards Software Vulnerability Location}.
\newblock \bibinfo{journal}{\emph{IEEE Transactions on Information Forensics and Security}} (\bibinfo{year}{2024}).
\newblock


\bibitem[Chen et~al\mbox{.}(2021)]%
        {chen2021does}
\bibfield{author}{\bibinfo{person}{Yilei Chen}, \bibinfo{person}{Alex Lombardi}, \bibinfo{person}{Fermi Ma}, {and} \bibinfo{person}{Willy Quach}.} \bibinfo{year}{2021}\natexlab{}.
\newblock \showarticletitle{Does Fiat-Shamir require a cryptographic hash function?}. In \bibinfo{booktitle}{\emph{Annual International Cryptology Conference}}. Springer, \bibinfo{pages}{334--363}.
\newblock


\bibitem[Chi et~al\mbox{.}(2020)]%
        {chi2020coldstore}
\bibfield{author}{\bibinfo{person}{Mucong Chi}, \bibinfo{person}{Jun Liu}, {and} \bibinfo{person}{Jie Yang}.} \bibinfo{year}{2020}\natexlab{}.
\newblock \showarticletitle{ColdStore: a storage system for archival data}.
\newblock \bibinfo{journal}{\emph{Wireless Personal Communications}} \bibinfo{volume}{111}, \bibinfo{number}{4} (\bibinfo{year}{2020}), \bibinfo{pages}{2325--2351}.
\newblock


\bibitem[Costello and Longa(2016)]%
        {costello2016schnorrq}
\bibfield{author}{\bibinfo{person}{Craig Costello} {and} \bibinfo{person}{Patrick Longa}.} \bibinfo{year}{2016}\natexlab{}.
\newblock \showarticletitle{Schnorrq: Schnorr signatures on fourq}.
\newblock \bibinfo{journal}{\emph{MSR Tech Report}} (\bibinfo{year}{2016}).
\newblock


\bibitem[Dong et~al\mbox{.}(2018)]%
        {dong2018sdpf}
\bibfield{author}{\bibinfo{person}{Jiankuo Dong}, \bibinfo{person}{Fangyu Zheng}, \bibinfo{person}{Niall Emmart}, \bibinfo{person}{Jingqiang Lin}, {and} \bibinfo{person}{Charles Weems}.} \bibinfo{year}{2018}\natexlab{}.
\newblock \showarticletitle{sDPF-RSA: Utilizing floating-point computing power of GPUs for massive digital signature computations}. In \bibinfo{booktitle}{\emph{2018 IEEE International Parallel and Distributed Processing Symposium (IPDPS)}}. IEEE, \bibinfo{pages}{599--609}.
\newblock


\bibitem[Feng et~al\mbox{.}(2022)]%
        {feng2022accelerating}
\bibfield{author}{\bibinfo{person}{Zonghao Feng}, \bibinfo{person}{Qipeng Xie}, \bibinfo{person}{Qiong Luo}, \bibinfo{person}{Yujie Chen}, \bibinfo{person}{Haoxuan Li}, \bibinfo{person}{Huizhong Li}, {and} \bibinfo{person}{Qiang Yan}.} \bibinfo{year}{2022}\natexlab{}.
\newblock \showarticletitle{Accelerating elliptic curve digital signature algorithms on GPUs}. In \bibinfo{booktitle}{\emph{SC22: International Conference for High Performance Computing, Networking, Storage and Analysis}}. IEEE, \bibinfo{pages}{1--13}.
\newblock


\bibitem[Ferrara et~al\mbox{.}(2009)]%
        {ferrara2009practical}
\bibfield{author}{\bibinfo{person}{Anna~Lisa Ferrara}, \bibinfo{person}{Matthew Green}, \bibinfo{person}{Susan Hohenberger}, {and} \bibinfo{person}{Michael~{\O}stergaard Pedersen}.} \bibinfo{year}{2009}\natexlab{}.
\newblock \showarticletitle{Practical short signature batch verification}. In \bibinfo{booktitle}{\emph{Cryptographers’ Track at the RSA Conference}}. Springer, \bibinfo{pages}{309--324}.
\newblock


\bibitem[Glas et~al\mbox{.}(2012)]%
        {Yavuz:ESCAR:SecuritySafety}
\bibfield{author}{\bibinfo{person}{Benjamin Glas}, \bibinfo{person}{Jorge Guajardo}, \bibinfo{person}{Hamit Hacioglu}, \bibinfo{person}{Markus Ihle}, \bibinfo{person}{Karsten Wehefritz}, {and} \bibinfo{person}{Attila~A. Yavuz}.} \bibinfo{year}{2012}\natexlab{}.
\newblock \bibinfo{title}{Signal-based Automotive Communication Security and Its Interplay with Safety Requirements}.
\newblock \bibinfo{howpublished}{ESCAR, Embedded Security in Cars Conference, Germany, November 2012}.
\newblock


\bibitem[Goyal et~al\mbox{.}(2006)]%
        {goyal2006attribute}
\bibfield{author}{\bibinfo{person}{Vipul Goyal}, \bibinfo{person}{Omkant Pandey}, \bibinfo{person}{Amit Sahai}, {and} \bibinfo{person}{Brent Waters}.} \bibinfo{year}{2006}\natexlab{}.
\newblock \showarticletitle{Attribute-based encryption for fine-grained access control of encrypted data}. In \bibinfo{booktitle}{\emph{Proc of the 13th ACM conference on Computer and communications security}}. \bibinfo{pages}{89--98}.
\newblock


\bibitem[Grissa et~al\mbox{.}(2019)]%
        {grissa2019trustsas}
\bibfield{author}{\bibinfo{person}{Mohamed Grissa}, \bibinfo{person}{Attila~A Yavuz}, {and} \bibinfo{person}{Bechir Hamdaoui}.} \bibinfo{year}{2019}\natexlab{}.
\newblock \showarticletitle{TrustSAS: A trustworthy spectrum access system for the 3.5 GHz CBRS band}. In \bibinfo{booktitle}{\emph{IEEE INFOCOM 2019-IEEE Conference on Computer Communications}}. IEEE, \bibinfo{pages}{1495--1503}.
\newblock


\bibitem[Hajihassani et~al\mbox{.}(2019)]%
        {hajihassani2019fast}
\bibfield{author}{\bibinfo{person}{Omid Hajihassani}, \bibinfo{person}{Saleh~Khalaj Monfared}, \bibinfo{person}{Seyed~Hossein Khasteh}, {and} \bibinfo{person}{Saeid Gorgin}.} \bibinfo{year}{2019}\natexlab{}.
\newblock \showarticletitle{Fast AES implementation: A high-throughput bitsliced approach}.
\newblock \bibinfo{journal}{\emph{IEEE Transactions on parallel and distributed systems}} \bibinfo{volume}{30}, \bibinfo{number}{10} (\bibinfo{year}{2019}), \bibinfo{pages}{2211--2222}.
\newblock


\bibitem[Hartung(2016)]%
        {Logging_VerifiableExcerpt}
\bibfield{author}{\bibinfo{person}{Gunnar Hartung}.} \bibinfo{year}{2016}\natexlab{}.
\newblock \showarticletitle{Secure Audit Logs with Verifiable Excerpts}. In \bibinfo{booktitle}{\emph{Topics in Cryptology - CT-RSA 2016}}, \bibfield{editor}{\bibinfo{person}{Kazue Sako}} (Ed.). \bibinfo{publisher}{Springer International Publishing}, \bibinfo{address}{Cham}, \bibinfo{pages}{183--199}.
\newblock
\showISBNx{978-3-319-29485-8}


\bibitem[Hartung(2017)]%
        {SecureLogAttacks:2017:Hartung}
\bibfield{author}{\bibinfo{person}{Gunnar Hartung}.} \bibinfo{year}{2017}\natexlab{}.
\newblock \showarticletitle{Attacks on Secure Logging Schemes}. In \bibinfo{booktitle}{\emph{Financial Cryptography and Data Security}}. \bibinfo{publisher}{Springer International Publishing}, \bibinfo{address}{Cham}, \bibinfo{pages}{268--284}.
\newblock


\bibitem[Hofemeier and Chesebrough(2012)]%
        {hofemeier2012introduction}
\bibfield{author}{\bibinfo{person}{Gael Hofemeier} {and} \bibinfo{person}{Robert Chesebrough}.} \bibinfo{year}{2012}\natexlab{}.
\newblock \showarticletitle{Introduction to intel aes-ni and intel secure key instructions}.
\newblock \bibinfo{journal}{\emph{Intel, White Paper}}  \bibinfo{volume}{62} (\bibinfo{year}{2012}), \bibinfo{pages}{6}.
\newblock


\bibitem[Hopcroft et~al\mbox{.}(1983)]%
        {hopcroft1983data}
\bibfield{author}{\bibinfo{person}{John~E Hopcroft}, \bibinfo{person}{Jeffrey~D Ullman}, {and} \bibinfo{person}{Alfred~Vaino Aho}.} \bibinfo{year}{1983}\natexlab{}.
\newblock \bibinfo{booktitle}{\emph{Data structures and algorithms}}. Vol.~\bibinfo{volume}{175}.
\newblock \bibinfo{publisher}{Addison-wesley Boston, MA, USA:}.
\newblock


\bibitem[Hu et~al\mbox{.}(2023)]%
        {hu2023high}
\bibfield{author}{\bibinfo{person}{Xinyi Hu}, \bibinfo{person}{Debiao He}, \bibinfo{person}{Min Luo}, \bibinfo{person}{Cong Peng}, \bibinfo{person}{Qi Feng}, {and} \bibinfo{person}{Xinyi Huang}.} \bibinfo{year}{2023}\natexlab{}.
\newblock \showarticletitle{High-performance implementation of the identity-based signature scheme in IEEE P1363 on GPU}.
\newblock \bibinfo{journal}{\emph{ACM Transactions on Embedded Computing Systems}} \bibinfo{volume}{22}, \bibinfo{number}{2} (\bibinfo{year}{2023}), \bibinfo{pages}{1--35}.
\newblock


\bibitem[Kim et~al\mbox{.}(2024)]%
        {kim2024parallel}
\bibfield{author}{\bibinfo{person}{DongCheon Kim}, \bibinfo{person}{HoJin Choi}, {and} \bibinfo{person}{Seog~Chung Seo}.} \bibinfo{year}{2024}\natexlab{}.
\newblock \showarticletitle{Parallel Implementation of SPHINCS $+ $ With GPUs}.
\newblock \bibinfo{journal}{\emph{IEEE Transactions on Circuits and Systems I: Regular Papers}} (\bibinfo{year}{2024}).
\newblock


\bibitem[Kim and Oh(2019)]%
        {FAS_Asymptotic}
\bibfield{author}{\bibinfo{person}{Jihye Kim} {and} \bibinfo{person}{Hyunok Oh}.} \bibinfo{year}{2019}\natexlab{}.
\newblock \showarticletitle{FAS: Forward secure sequential aggregate signatures for secure logging}.
\newblock \bibinfo{journal}{\emph{Information Sciences}}  \bibinfo{volume}{471} (\bibinfo{year}{2019}), \bibinfo{pages}{115 -- 131}.
\newblock
\showISSN{0020-0255}


\bibitem[Kirk et~al\mbox{.}(2007)]%
        {kirk2007nvidia}
\bibfield{author}{\bibinfo{person}{David Kirk} {et~al\mbox{.}}} \bibinfo{year}{2007}\natexlab{}.
\newblock \showarticletitle{NVIDIA CUDA software and GPU parallel computing architecture}. In \bibinfo{booktitle}{\emph{ISMM}}, Vol.~\bibinfo{volume}{7}. \bibinfo{pages}{103--104}.
\newblock


\bibitem[Lee et~al\mbox{.}(2019)]%
        {lee2019hybrid}
\bibfield{author}{\bibinfo{person}{Sokjoon Lee}, \bibinfo{person}{Hwajeong Seo}, \bibinfo{person}{Hyeokchan Kwon}, {and} \bibinfo{person}{Hyunsoo Yoon}.} \bibinfo{year}{2019}\natexlab{}.
\newblock \showarticletitle{Hybrid approach of parallel implementation on CPU--GPU for high-speed ECDSA verification}.
\newblock \bibinfo{journal}{\emph{The Journal of Supercomputing}}  \bibinfo{volume}{75} (\bibinfo{year}{2019}), \bibinfo{pages}{4329--4349}.
\newblock


\bibitem[Li et~al\mbox{.}(2020)]%
        {li2020permissioned}
\bibfield{author}{\bibinfo{person}{Tian Li}, \bibinfo{person}{Huaqun Wang}, \bibinfo{person}{Debiao He}, {and} \bibinfo{person}{Jia Yu}.} \bibinfo{year}{2020}\natexlab{}.
\newblock \showarticletitle{Permissioned blockchain-based anonymous and traceable aggregate signature scheme for Industrial Internet of Things}.
\newblock \bibinfo{journal}{\emph{IEEE Internet of Things Journal}} \bibinfo{volume}{8}, \bibinfo{number}{10} (\bibinfo{year}{2020}), \bibinfo{pages}{8387--8398}.
\newblock


\bibitem[Li et~al\mbox{.}(2017)]%
        {li2017iot}
\bibfield{author}{\bibinfo{person}{Xin Li}, \bibinfo{person}{Huazhe Wang}, \bibinfo{person}{Ye Yu}, {and} \bibinfo{person}{Chen Qian}.} \bibinfo{year}{2017}\natexlab{}.
\newblock \showarticletitle{An IoT data communication framework for authenticity and integrity}. In \bibinfo{booktitle}{\emph{2017 IEEE/ACM 2nd International Conf. on Internet-of-Things Design and Implementation (IoTDI)}}. \bibinfo{pages}{159--170}.
\newblock


\bibitem[Li et~al\mbox{.}(2021)]%
        {li2021threat}
\bibfield{author}{\bibinfo{person}{Zhenyuan Li}, \bibinfo{person}{Qi~Alfred Chen}, \bibinfo{person}{Runqing Yang}, \bibinfo{person}{Yan Chen}, {and} \bibinfo{person}{Wei Ruan}.} \bibinfo{year}{2021}\natexlab{}.
\newblock \showarticletitle{Threat detection and investigation with system-level provenance graphs: A survey}.
\newblock \bibinfo{journal}{\emph{Computers \& Security}}  \bibinfo{volume}{106} (\bibinfo{year}{2021}), \bibinfo{pages}{102282}.
\newblock


\bibitem[Liao et~al\mbox{.}(2024)]%
        {liao2024semantic}
\bibfield{author}{\bibinfo{person}{Wenhao Liao}, \bibinfo{person}{Jia Sun}, \bibinfo{person}{Haiyan Wang}, \bibinfo{person}{Zhaoquan Gu}, {and} \bibinfo{person}{Jianye Yang}.} \bibinfo{year}{2024}\natexlab{}.
\newblock \showarticletitle{Semantic-Integrated Online Audit Log Reduction for Efficient Forensic Analysis}. In \bibinfo{booktitle}{\emph{International Conf. on Advanced Data Mining and Applications}}. \bibinfo{pages}{318--333}.
\newblock


\bibitem[Liu et~al\mbox{.}(2010)]%
        {liu2010efficient}
\bibfield{author}{\bibinfo{person}{Zhe Liu}, \bibinfo{person}{Johann Gro{\ss}sch{\"a}dl}, {and} \bibinfo{person}{Ilya Kizhvatov}.} \bibinfo{year}{2010}\natexlab{}.
\newblock \showarticletitle{Efficient and side-channel resistant RSA implementation for 8-bit AVR microcontrollers}. In \bibinfo{booktitle}{\emph{Workshop on the Security of the Internet of Things-SOCIOT}}, Vol.~\bibinfo{volume}{10}.
\newblock


\bibitem[Ma and Tsudik(2009)]%
        {SecureLogDiMaACMTrans09}
\bibfield{author}{\bibinfo{person}{Di Ma} {and} \bibinfo{person}{Gene Tsudik}.} \bibinfo{year}{2009}\natexlab{}.
\newblock \showarticletitle{A New Approach to Secure Logging}.
\newblock \bibinfo{journal}{\emph{Trans. Storage}} \bibinfo{volume}{5}, \bibinfo{number}{1}, Article \bibinfo{articleno}{2} (\bibinfo{year}{2009}), \bibinfo{numpages}{21}~pages.
\newblock
\showISSN{1553-3077}


\bibitem[Marson and Poettering(2014)]%
        {Logging_Seekable2}
\bibfield{author}{\bibinfo{person}{Giorgia~Azzurra Marson} {and} \bibinfo{person}{Bertram Poettering}.} \bibinfo{year}{2014}\natexlab{}.
\newblock \showarticletitle{Even More Practical Secure Logging: Tree-Based Seekable Sequential Key Generators}. In \bibinfo{booktitle}{\emph{Computer Security - ESORICS 2014}}. \bibinfo{address}{Cham}, \bibinfo{pages}{37--54}.
\newblock
\showISBNx{978-3-319-11212-1}


\bibitem[Menezes et~al\mbox{.}(1996)]%
        {CryptoHandBook}
\bibfield{author}{\bibinfo{person}{A.J. Menezes}, \bibinfo{person}{P.~C. {van Oorschot}}, {and} \bibinfo{person}{S.A. Vanstone}.} \bibinfo{year}{1996}\natexlab{}.
\newblock \bibinfo{booktitle}{\emph{Handbook of Applied Cryptography}}.
\newblock \bibinfo{publisher}{{CRC} Press}.
\newblock


\bibitem[Minerva et~al\mbox{.}(2020)]%
        {minerva2020digital}
\bibfield{author}{\bibinfo{person}{Roberto Minerva}, \bibinfo{person}{Gyu~Myoung Lee}, {and} \bibinfo{person}{Noel Crespi}.} \bibinfo{year}{2020}\natexlab{}.
\newblock \showarticletitle{Digital twin in the IoT context: A survey on technical features, scenarios, and architectural models}.
\newblock \bibinfo{journal}{\emph{Proc. IEEE}} \bibinfo{volume}{108}, \bibinfo{number}{10} (\bibinfo{year}{2020}), \bibinfo{pages}{1785--1824}.
\newblock


\bibitem[MITRE({[n.\,d.]})]%
        {mitrelog}
\bibfield{author}{\bibinfo{person}{MITRE}.} \bibinfo{year}{[n.\,d.]}\natexlab{}.
\newblock \bibinfo{title}{{Indicator Removal: Clear Linux or Mac System Logs }}.
\newblock \bibinfo{howpublished}{\url{https://attack.mitre.org/techniques/T1070/002/}}.
\newblock
\newblock
\shownote{Accessed: April 5, 2025}.


\bibitem[Mosenia and Jha(2016)]%
        {mosenia2016comprehensive}
\bibfield{author}{\bibinfo{person}{Arsalan Mosenia} {and} \bibinfo{person}{Niraj~K Jha}.} \bibinfo{year}{2016}\natexlab{}.
\newblock \showarticletitle{A comprehensive study of security of internet-of-things}.
\newblock \bibinfo{journal}{\emph{IEEE Transactions on emerging topics in computing}} \bibinfo{volume}{5}, \bibinfo{number}{4} (\bibinfo{year}{2016}), \bibinfo{pages}{586--602}.
\newblock


\bibitem[Nouma and Yavuz(2023)]%
        {nouma2023practical}
\bibfield{author}{\bibinfo{person}{Saif~E Nouma} {and} \bibinfo{person}{Attila~A Yavuz}.} \bibinfo{year}{2023}\natexlab{}.
\newblock \showarticletitle{Practical Cryptographic Forensic Tools for Lightweight Internet of Things and Cold Storage Systems}. In \bibinfo{booktitle}{\emph{Proc. of the 8th ACM/IEEE Conf. on Internet of Things Design and Implementation}}. \bibinfo{pages}{340--353}.
\newblock


\bibitem[Ozmen et~al\mbox{.}(2019)]%
        {Yavuz:CNS:2019}
\bibfield{author}{\bibinfo{person}{Muslum~Ozgur Ozmen}, \bibinfo{person}{Rouzbeh Behnia}, {and} \bibinfo{person}{Attila~A. Yavuz}.} \bibinfo{year}{2019}\natexlab{}.
\newblock \showarticletitle{Energy-Aware Digital Signatures for Embedded Medical Devices}. In \bibinfo{booktitle}{\emph{7th {IEEE} Conf. on Communications and Network Security ({CNS}), June}}.
\newblock


\bibitem[Pan et~al\mbox{.}(2016)]%
        {pan2016efficient}
\bibfield{author}{\bibinfo{person}{Wuqiong Pan}, \bibinfo{person}{Fangyu Zheng}, \bibinfo{person}{Yuan Zhao}, \bibinfo{person}{Wen-Tao Zhu}, {and} \bibinfo{person}{Jiwu Jing}.} \bibinfo{year}{2016}\natexlab{}.
\newblock \showarticletitle{An efficient elliptic curve cryptography signature server with GPU acceleration}.
\newblock \bibinfo{journal}{\emph{IEEE Trans. on Information Forensics and Security}} \bibinfo{volume}{12}, \bibinfo{number}{1} (\bibinfo{year}{2016}), \bibinfo{pages}{111--122}.
\newblock


\bibitem[Rohde et~al\mbox{.}(2008)]%
        {rohde2008fast}
\bibfield{author}{\bibinfo{person}{Sebastian Rohde}, \bibinfo{person}{Thomas Eisenbarth}, \bibinfo{person}{Erik Dahmen}, \bibinfo{person}{Johannes Buchmann}, {and} \bibinfo{person}{Christof Paar}.} \bibinfo{year}{2008}\natexlab{}.
\newblock \showarticletitle{Fast hash-based signatures on constrained devices}. In \bibinfo{booktitle}{\emph{Smart Card Research and Advanced Applications: 8th IFIP WG 8.8/11.2 International Conference, CARDIS 2008, London, UK, September 8-11, 2008. Proceedings 8}}. Springer, \bibinfo{pages}{104--117}.
\newblock


\bibitem[Sasi et~al\mbox{.}(2024)]%
        {sasi2024comprehensive}
\bibfield{author}{\bibinfo{person}{Tinshu Sasi}, \bibinfo{person}{Arash~Habibi Lashkari}, \bibinfo{person}{Rongxing Lu}, \bibinfo{person}{Pulei Xiong}, {and} \bibinfo{person}{Shahrear Iqbal}.} \bibinfo{year}{2024}\natexlab{}.
\newblock \showarticletitle{A comprehensive survey on IoT attacks: Taxonomy, detection mechanisms and challenges}.
\newblock \bibinfo{journal}{\emph{J. of Information and intelligence}} \bibinfo{volume}{2}, \bibinfo{number}{6} (\bibinfo{year}{2024}), \bibinfo{pages}{455--513}.
\newblock


\bibitem[Schneier and Kelsey(1999)]%
        {SecureLogBruceACMTrans99}
\bibfield{author}{\bibinfo{person}{B. Schneier} {and} \bibinfo{person}{J. Kelsey}.} \bibinfo{year}{1999}\natexlab{}.
\newblock \showarticletitle{Secure audit logs to support computer forensics}.
\newblock \bibinfo{journal}{\emph{ACM Transaction on Information System Security}} \bibinfo{volume}{2}, \bibinfo{number}{2} (\bibinfo{year}{1999}), \bibinfo{pages}{159--176}.
\newblock


\bibitem[Schnorr(1991)]%
        {schnorr1991efficient}
\bibfield{author}{\bibinfo{person}{Claus-Peter Schnorr}.} \bibinfo{year}{1991}\natexlab{}.
\newblock \showarticletitle{Efficient signature generation by smart cards}.
\newblock \bibinfo{journal}{\emph{Journal of cryptology}} \bibinfo{volume}{4}, \bibinfo{number}{3} (\bibinfo{year}{1991}), \bibinfo{pages}{161--174}.
\newblock


\bibitem[Shah et~al\mbox{.}(2019)]%
        {shah2019analyzing}
\bibfield{author}{\bibinfo{person}{Aashaka Shah}, \bibinfo{person}{Vinay Banakar}, \bibinfo{person}{Supreeth Shastri}, \bibinfo{person}{Melissa Wasserman}, {and} \bibinfo{person}{Vijay Chidambaram}.} \bibinfo{year}{2019}\natexlab{}.
\newblock \showarticletitle{Analyzing the impact of $\{$GDPR$\}$ on storage systems}. In \bibinfo{booktitle}{\emph{11th USENIX Workshop on Hot Topics in Storage and File Systems}}.
\newblock


\bibitem[Steinberger(2007)]%
        {steinberger2007collision}
\bibfield{author}{\bibinfo{person}{John~P Steinberger}.} \bibinfo{year}{2007}\natexlab{}.
\newblock \showarticletitle{The collision intractability of MDC-2 in the ideal-cipher model}. In \bibinfo{booktitle}{\emph{Advances in Cryptology-EUROCRYPT: 26th Annual International Conf. on the Theory and Applications of Cryptographic Techniques}}. \bibinfo{pages}{34--51}.
\newblock


\bibitem[Tezcan(2021)]%
        {tezcan2021optimization}
\bibfield{author}{\bibinfo{person}{Cihangir Tezcan}.} \bibinfo{year}{2021}\natexlab{}.
\newblock \showarticletitle{Optimization of advanced encryption standard on graphics processing units}.
\newblock \bibinfo{journal}{\emph{IEEE Access}}  \bibinfo{volume}{9} (\bibinfo{year}{2021}), \bibinfo{pages}{67315--67326}.
\newblock


\bibitem[Vallent et~al\mbox{.}(2021)]%
        {vallent2021efficient}
\bibfield{author}{\bibinfo{person}{Thokozani~F. Vallent}, \bibinfo{person}{Damien Hanyurwimfura}, {and} \bibinfo{person}{Chomora Mikeka}.} \bibinfo{year}{2021}\natexlab{}.
\newblock \showarticletitle{Efficient certificate-less aggregate signature scheme with conditional privacy-preservation for vehicular adhoc networks enhanced smart grid system}.
\newblock \bibinfo{journal}{\emph{Sensors}} \bibinfo{volume}{21}, \bibinfo{number}{9} (\bibinfo{year}{2021}).
\newblock


\bibitem[Verma et~al\mbox{.}(2021)]%
        {verma2021scbs}
\bibfield{author}{\bibinfo{person}{Girraj~Kumar Verma}, \bibinfo{person}{Neeraj Kumar}, \bibinfo{person}{Prosanta Gope}, \bibinfo{person}{BB Singh}, {and} \bibinfo{person}{Harendra Singh}.} \bibinfo{year}{2021}\natexlab{}.
\newblock \showarticletitle{SCBS: a short certificate-based signature scheme with efficient aggregation for industrial-internet-of-things environment}.
\newblock \bibinfo{journal}{\emph{IEEE Internet of Things Journal}} \bibinfo{volume}{8}, \bibinfo{number}{11} (\bibinfo{year}{2021}), \bibinfo{pages}{9305--9316}.
\newblock


\bibitem[Wang et~al\mbox{.}(2010)]%
        {wang2010secure}
\bibfield{author}{\bibinfo{person}{Cong Wang}, \bibinfo{person}{Ning Cao}, \bibinfo{person}{Jin Li}, \bibinfo{person}{Kui Ren}, {and} \bibinfo{person}{Wenjing Lou}.} \bibinfo{year}{2010}\natexlab{}.
\newblock \showarticletitle{Secure ranked keyword search over encrypted cloud data}. In \bibinfo{booktitle}{\emph{2010 IEEE 30th international conference on distributed computing systems}}. \bibinfo{pages}{253--262}.
\newblock


\bibitem[Wang et~al\mbox{.}(2011)]%
        {wang2011privacy}
\bibfield{author}{\bibinfo{person}{Cong Wang}, \bibinfo{person}{Sherman~SM Chow}, \bibinfo{person}{Qian Wang}, \bibinfo{person}{Kui Ren}, {and} \bibinfo{person}{Wenjing Lou}.} \bibinfo{year}{2011}\natexlab{}.
\newblock \showarticletitle{Privacy-preserving public auditing for secure cloud storage}.
\newblock \bibinfo{journal}{\emph{IEEE transactions on computers}} \bibinfo{volume}{62}, \bibinfo{number}{2} (\bibinfo{year}{2011}), \bibinfo{pages}{362--375}.
\newblock


\bibitem[Yavuz({[n.\,d.]})]%
        {DSSE:Yavuz:patent:ForwardSecureLog:2015}
\bibfield{author}{\bibinfo{person}{Attila~A. Yavuz}.} \bibinfo{year}{[n.\,d.]}\natexlab{}.
\newblock \bibinfo{title}{System and method for secure review of audit logs}.
\newblock \bibinfo{howpublished}{Robert Bosch, Provisional Application No. 62/006,476, Filing Date: June 2, 2014, PCT Application: June 2, 2015}.
\newblock


\bibitem[Yavuz(2018)]%
        {Yavuz:TDSC:OutsourcedDB}
\bibfield{author}{\bibinfo{person}{Attila~A. Yavuz}.} \bibinfo{year}{2018}\natexlab{}.
\newblock \showarticletitle{Immutable Authentication and Integrity Schemes for Outsourced Databases}.
\newblock \bibinfo{journal}{\emph{{IEEE} Trans. Dependable Sec. Comput.}} \bibinfo{volume}{15}, \bibinfo{number}{1} (\bibinfo{year}{2018}), \bibinfo{pages}{69--82}.
\newblock


\bibitem[Yavuz et~al\mbox{.}(2012)]%
        {Yavuz:2012:TISSEC:FIBAF}
\bibfield{author}{\bibinfo{person}{A.~A. Yavuz}, \bibinfo{person}{Peng Ning}, {and} \bibinfo{person}{Michael~K. Reiter}.} \bibinfo{year}{2012}\natexlab{}.
\newblock \showarticletitle{{BAF} and {FI-BAF}: Efficient and Publicly Verifiable Cryptographic Schemes for Secure Logging in Resource-Constrained Systems}.
\newblock \bibinfo{journal}{\emph{ACM Trans. on Inf. System Sec.}} \bibinfo{volume}{15}, \bibinfo{number}{2} (\bibinfo{year}{2012}), \bibinfo{numpages}{28}~pages.
\newblock


\end{thebibliography}

\section*{APPENDIX A}
\label{appendixA}

We provide the security proof of \posloc~scheme as below:

\newtheorem*{theorem1}{Theorem 5.1}
\begin{theorem1} \label{the:Theorem1}
	$\advsocosa \le \advdll$, where  $t'=O(t)+ O(n \cdot (\kappa^{3}+RNG))$.
\end{theorem1}

\noindent {\em Proof:} Let \A~be a \posloc~attacker. We construct a {\em DL-attacker} \F~that uses \A~as a sub-routine. That is, we set $(b\Rq,~B\as \alpha^{b} \bmod p)$ as defined in $\mathit{DL\mhyphen}$experiment (i.e., Definition~\ref{def.dlp}) and then run the simulator \F~by Definition~\ref{Def:AIEUCMA} (i.e., \AEUCMA~experiment) as follows:

\vspace{2mm} \noindent \underline{{\em Algorithm $F(B)$}} 
\begin{enumerate}[ ]
	\setlength{\itemsep}{2pt}
	\setlength{\parskip}{0pt}
	\setlength{\parsep}{0pt}
	
	\item \underline{{\em Setup:}} \F~maintains \lh, \lm, and \ls~to keep track of query results in the duration of the experiment. \lh~is a hash list in form of tuples $(M_l, h_l)$, where $M_l$ and $h_l$ denote the $l^{\text{th}}$ data item queried to \ro~and its corresponding \ro~answer, respectively. $\lh[l,0]$ and $\lh[l,1]$ denote the access to the element $M_l$, $h_l$ via the hash function $H$, respectively. \lm~is a list of messages, in which each of its elements $\lm[i]$ is a message set $\batch{m}_i$ (i.e., the $i^{\text{th}}$ batch query). \ls~is a signature list that is used to record answers given by $\poslocsig_{sk}$.

	\begin{itemize}
		\item \F~creates a simulated \posloc~public key \pk:
		
		\begin{enumerate}[a)]
			\item Set $(n_1,n_2)$ as in $\poslockg(.)$
			\item $Y \as B$ and $x_D[0] \Ra \{0,1\}^{\kappa}$
			\item \textbf{for} $l=1,\ldots,n$ \textbf{do}
			\\ ~~~~ \textit{i)} $R_l \as Y^{e_l} \cdot \alpha^{s_l} \bmod p$ where $(s_l, e_l) \Rq$
			
			\item \textbf{for} $i=1,\ldots,n_1$ \textbf{do}
			\\  ~~~~\textit{i)}  $\agg{R}_i \as \prod_{j=1}^{n_2}{R_{i \cdot n_1 + j}} \bmod p$
			\item $ \pk \as (Y, \batch{R})$, where $\batch{R} \as \{\agg{R}_i\}_{i=1}^{n_1}$ 
			\item $I \as (p,q,\alpha, n_1, n_2 )$ and initialize $l \as 0,~i\as 0$
		\end{enumerate}
	\end{itemize}

	\item \underline{Execute $\mathcal{A}^{\ro, \poslocsig_{sk}(.)}(\pk)$}: 
	
	\begin{list}{-}{}
		\item  \underline{Queries}: \A~queries $\poslocsig_{sk}(.)$~oracle on $n$ messages of her choice. It also queries \ro~oracle on up to $n'$ messages of her choice. These queries are as follows:\\
		
		\vspace{-2mm}
		$\bullet$~{\em How to Handle \ro~Queries}: \F~implements a function $\hsim(\delta, k)$ that works as \ro~as follows: If $\exists l' : \delta \in \lh[l']$ then return $\lh[l']$. Otherwise, return $h \Rq$ as the answer for $H$, insert new tuple $(\delta,h)$ to $\lh$ as $(\lh[l,0]\as \delta, \lh[l,1] \as h)$ and then update $l \as l+1$ . That is, the cryptographic hash function $H$ used in \posloc~is modeled as random oracle. 
		
		When \A~queries \ro~on a message $m_l$,  \F~returns $h_l \as \hsim(m_l)$ as described above. 
		
		$\bullet$~{\em How to respond to $\poslocsig_{sk}(.)$ queries:} 
		\begin{enumerate}[-]
			\itemsep 4pt
			\item For each batch query $\batch{m_i}$, \A~queries $\poslocsig(.)$~on $\batch{m_i}$ of her choice. If $i \geq n_1$, \F~rejects the query (i.e., the query limit is exceeded), else \F~continues as follows:
			\begin{enumerate}[a)]
				
				\item \F computes $x_0[i] \as \sconst(x_D[0], D,0,0,i)$
				\item Initialize $\agg{s}_i \as 0$
				\item \textbf{for} $j = 1, \ldots, n_2$  \textbf{do}
				\begin{enumerate}[i)]
					\item \F sets $x_{i}^{j} \as \prf_0(x_0[i] \| j)$. If $(m_i^j \| x_{i}^{j}) \in \lh$ then \F~{\em aborts}, else inserts $(m_i^j \| x_i^j)$ to \lh.
					\item \F computes $\agg{s}_{i} \as \poslocagg(\agg{s}_{i},  s_i^{j})$
				\end{enumerate}
				\item \F~sets $\agg{\sigma}_i \as ( \agg{s}_i ,\ds_i \as \sso(x_0[0],i) )$, inserts $(\batch{m_i},\agg{\sigma}_i)$ to (\lm,\ls) and $i \as i +1$.
			\end{enumerate}
			
			\item    \underline{Forgery of \A:} Eventually, \A~outputs a forgery on \pk~as $(\vec{\batch{m}}^{*} , \agg{\sigma}^{*})$, where $\vec{\batch{m}}^{*}=\{ \batch{m_i^*} \}_{i\in \batch{I}}$ and $\agg{\sigma}^{*}=(\agg{s}^{*},\ds^{*})$. By Definition \ref{Def:AIEUCMA}, \A~wins $\AEUCMA\mhyphen$experiment if $\poslocaver(\pk,\vec{\batch{m}}^{*},\agg{\sigma}^{*})=1$ and  $\vec{\batch{m}}^{*} \notin \lm$ hold. If these conditions hold, \A~returns $1$, else, returns $0$.
			
		\end{enumerate}
		
		\item \underline{Forgery of \F}: If \A~loses the \AEUCMA~experiment for \posloc, \F~also loses in the $\dl\mhyphen$experiment, and therefore \F~{\em aborts} and returns $0$. Else, if $\batch{m^{*}} \in \lh$ then \F~{\em aborts} and returns $0$ (i.e., \A~wins the experiment without querying \ro~oracle).  Otherwise, \F~continues:
		
		\vspace{2mm}
		$\agg{R} \equiv Y^{\agg{e}} \cdot \alpha^{\agg{s}} \bmod p$  holds for the aggregated variables $(\agg{R},\agg{e},\agg{s})$. That is, given the indices of corresponding previous messages $\batch{I}$, \F~retrieves  $(s_i,r_i)$ from $(\ls,\lh)$, and then computes 
		$\agg{e} \as \sum_{i \in \batch{I}} \sum_{j=1}^{n_2} e_{i \cdot n_1 + j} \bmod q$ and $\agg{s} \as \poslocagg(\{ \agg{s}_i \}_{i \in \batch{I}} )$.  Moreover, $\poslocaver(\pk,\batch{m^{*}},\sigma^{*})=1$ holds, and therefore $R \equiv Y^{{e}^{*}}\cdot \alpha^{s^{*}} \bmod p$ also holds.  Note that \A~queries \F~on $n_1$ batches  and $n$ messages in total. Hence, \F~disclosed the root of \oslot~tree, from which required seeds can be derived. \F~calls $x_0[i] \as \sret(\ds^*, i),~\forall i \in \batch{I}$. It then computes  $\agg{e}^* = \sum_{i \in \batch{I}} \sum_{j=1}^{n_2} \hsim(m_i^{j^*} \| x_i^{j^*}, 0 )  $ where $x_i^j \as \prf_0(x_0[i] \| j)$. Thus, the following equations hold:
		
		~~~~~~~~~~~~~~~~ $\agg{R} \equiv Y^{\agg{e}}\cdot \alpha^{\agg{s}} \bmod p, ~~
		\agg{R} \equiv Y^{\agg{e}^{*}}\cdot \alpha^{\agg{s}^{*}} \bmod p,$
		
		\F~then extracts $y'=b$ by solving the below modular linear equations (note that only unknowns
		are $y$ and $r$), where $Y=B$ as defined in the public key simulation:
		
		~~~~~~~~~~~~ $r  \equiv y'\cdot e+ s \bmod q, ~r \equiv y'\cdot e^{*} + s^{*} \bmod q$
		
		$B'\equiv \alpha^{b} \bmod p$ holds, since \A's forgery is valid and non-trivial on $B'=B$. By Definition \ref{def.dlp}, $\mathcal{F}$ wins the $\mathit{DL\mhyphen experiment}$.
		
	\end{list}
\end{enumerate}

\vspace{2mm}

\noindent The execution time and probability analysis are as follows:
\vspace{1pt}

\noindent \underline{{\em Execution Time Analysis}}: In this experiment, the runtime of \F~is that of \A~plus the time it takes to respond \ro~queries. 
\begin{itemize}
	\item {\em Setup phase:} \F~draws $2n+1$ random numbers, performs $2n$ modular exponentiations and multiplications. Hence, the total cost of this phase is $2n\cdot \mathcal{O}(\kappa^{3}+\kappa^{2})+(2n+1)\cdot \RNG$, where $\mathcal{O}(\kappa^{3})$ and $\mathcal{O}(\kappa^{2})$ denote the cost of modular exponentiation and modular multiplication, respectively. \RNG~denotes the cost of drawing a random number. 
	
	\item {\em Query phase:} \F draws $n_1 \cdot \log{n_1} \cdot \RNG$ to compute the epoch seeds and $n \cdot \RNG$ to derive one-time random keys. It also draws $n \cdot \RNG$  to handle \A's \ro~queries. The cost of query phase is bounded as $\mathcal{O}(T)\cdot \RNG$.
\end{itemize}
Therefore, the approximate total running time of \F~is $t'=O(t)+ O(n \cdot (\kappa^{3}+RNG))$.

\vspace{2mm}

\noindent \underline{{\em Success Probability Analysis}}: 
\F~succeeds if all below events occur.

\begin{enumerate}[-]
	\setlength{\itemsep}{0pt}
	\setlength{\parskip}{0pt}
	\setlength{\parsep}{0pt}
	\item \nab: \F does not abort during the query phase.
	
	\item \forge: \A~wins the \AEUCMA~experiment for \posloc.
	
	\item \nabb: \F~does not abort after \A's forgery.
	
	\item   \suc: \F~wins the \AEUCMA~experiment for \dl{\em-experiment}.
	
	\item  $Pr[\suc] = Pr[\nab]\cdot Pr[\forge|\nab]\cdot Pr[\nabb|\nab \wedge \forge]$
	
\end{enumerate}

$\bullet$ {\em The probability that event \nab~occurs}: During the query phase, \F~aborts if $(m_i^j||x_i^j)$ $\in\lh,~1\leq i \leq n_1,~1\leq j \leq n_2$ holds, {\em before} \F~inserts $(m_i^j\| x_i^j)$ into \lh. This occurs if \A~guesses $x_i^j$ (before it is released) and then queries $(m_i^j \| x_i^j)$ to \ro~{\em before} querying it to $\poslocsig(.)$. The probability that this occurs is $\frac{1}{2^{\kappa}}$, which is negligible in terms of $\kappa$. Hence, $Pr[\nab]=(1-\frac{1}{2^{\kappa}})\approx 1$.

$\bullet$ {\em The probability that event \forge~occurs}: If \F~does not abort, \A~also does not abort since the \A's simulated  view is {\em indistinguishable} from \A's real view (see the indistinguishability analysis). Thus, $Pr[\forge|\nab]=\advsocosa$.

$\bullet$ {\em The probability that event \nabb~occurs}: \F~does not abort if the following conditions are satisfied:
(i) \A~wins the \AEUCMA~experiment for \posloc~on a message $m^{*}$ by querying it to \ro. The probability that \A~wins without querying $m^{*}$ to \ro~is as difficult as a random guess.
(ii) After \F~extracts $y'=b$ by solving modular linear equations, the probability that $Y' \not\equiv \alpha^{y'} \bmod p$ is negligible in terms $\kappa$, since $(Y=B) \in \pk$ and $\poslocaver(\pk,m^{*},\sigma^{*})=1$. Hence, $Pr[\nabb|\nab \wedge \forge]=\advsocosa$.
Omitting the terms that are negligible in terms of $\kappa$, the upper bound on {\em \AEUCMA-advantage of \posloc}~is as follows:
\begin{eqnarray*}
	\advsocosa \le \advdll,
\end{eqnarray*}

\vspace{2mm}
\noindent {\em \underline{Indistinguishability Argument}}: The real-view of \Areal~is comprised of public key \pk, parameters $I$, the answers of $\poslocsig_{sk}(.)$~ and \ro, recorded by \F~in \ls~and \lm, respectively. These values are generated by \posloc~algorithms as in the real system, where $sk=(y,r, x_D[0])$ serves as initial randomness. The joint probability distribution of \Areal~is random uniform as that of \sk. 

The simulated view of \A~is as \Asim, and it is equivalent to \Areal~except that in the simulation, values $(s_l,e_l)$ for $l=1,\ldots,n$ are randomly selected from $\mathbb{Z}_{q}^{*}$. This then dictates the selection of $R_l$ for $l=1,\ldots,n$  as random via the public key simulation (step c)-ii). Note that the joint probability distribution of these variables is also random uniformly distributed and is identical to the original signature and hash outputs (since $H$ is modeled as \ro~via \hsim). $\poslocdistill(.)$~and $\poslocsebver(.)$~use $\poslocagg(.)$~and $\poslocaver(.)$, which are invoked in signature simulation and forgery/extraction phases. Since \ccd~only contains the values produced in the simulation, \Asim~for $\poslocdistill(.)$~and $\poslocsebver(.)$~are indistinguishable from that of \Areal. \hfill$\square$

We provide the security proof of \poslof~scheme as below:

\newtheorem*{theorem2}{Lemma 5.2}
\begin{theorem2}
	\poslof~is as secure as \posloc.
\end{theorem2}

\noindent {\em Proof:} In the sketch proof, we first show that \poslof~public key and signature simulations produce random uniformly distributed values as in \posloc. We then show that the forgery and extraction phases in \AEUCMA~experiment for both variants are identical. Finally, we provide an indistinguishability argument for the \AEUCMA~for \poslof.

$\bullet$~{\em Public Key Simulation}: $\poslofkg(.)$~Step 1-4 are identical to that of \posloc, except commitment values $R$ are generated via \bpv~generator. Therefore, \F~runs the public key simulation as in \posloc, expect $\batch{R}$ is not pre-stored as a part of the public key. All the values $\{s_l,e_l, R_l\}_{l=1}^{n}$ are as in the \posloc~simulation.

$\bullet$~{\em Signature Simulation}: \F~sets $\sigma_l = ( s_l,R_l,x_l)$, where $(s_l,R_l)$ are as defined above, and $(e_l,x_l)$ are obtained through \ro~as in \posloc~via \hsim~function. $\poslofsig(.)$~queries are individual, and therefore $\sigma_l$ is not aggregated via $\poslocagg(.)$. The abort conditions in both \posloc~and \poslof~are the same.

$\bullet$~{\em Forgery and Extraction}: \posloc~and \poslof~verifications are identical except for the first step, which identifies if the signature is on a single or batch of messages. If the forgery is an aggregate signature on a batch message, $\poslofaver(.)$ verifies it by performing aggregation as in $\poslocaver(.)$. Hence, the forgery and extraction are identical, wherein \A~might return a batch or individual forgery $(\sigma^*,M^*)$. \F~retrieves $(s,R,e)$ from \ls~since $R$ components are the part of signatures but not \pk~(unlike \posloc). 

$\bullet$~{\em Indistinguishability Argument}:  \Areal~of \poslof~is as in \posloc~except that $\{R_\ell \}_{l=1}^{n}$ (generated via \bpv) are not  part of \pk~but in individual signatures $\{\sigma_l=( s_l,R_l,\ds_l ) \}_{l=1}^{n}$. The joint probability distribution of the values in \Areal~are random uniformly distributed as all derived from \sk~(as in \posloc). 
Remark that  each $R_l$ is also random uniform because the distribution of \bpv~output $r_l$ is statistically close to the uniform random distribution with an appropriate choice of parameters $(v,k)$ \cite{boyko1998speeding}.
\Asim~is identical to \Areal~since public key and signature simulations produce random uniformly distributed values of equal size to \Areal. As in \posloc, $\poslofdistill(.)$~and $\poslofsebver(.)$~call $\poslofagg(.)$~and $\poslofaver(.)$, in which \ccd~values are produced by $\poslofsig(.)$~and \hsim. \hfill$\square$


\newtheorem*{theorem3}{Corollary 5.3} 
\begin{theorem3} \label{corollary:5.3}
	\poslop~and \poslopp~instantiations are as secure as \poslo.
\end{theorem3}

\begin{proof}
    In \poslop, the cryptographic hash function (SHA-256) used in \poslo~is replaced with block-cipher-based constructions: MMO for the \prf~primitive and MDC-2 for message hashing $H$. Their security is based on the pseudorandomness of the underlying block cipher (i.e., AES-128) \cite{CryptoHandBook}. MMO's output is indistinguishable from a random string provided that AES-128 is a random permutation. As a key derivation function $\prf_{0,1}$, AES-based MMO achieves preimage resistance with probability at most $\frac{1}{2^{128}}$ \cite{CryptoHandBook}. 
    As for MDC-2, Steinberger et al. \cite{steinberger2007collision} shows that the best theoretical collision attack in the ideal cipher model requires $2^{3\ell/5}$ queries, where $\ell$ is the block size. However, collision attacks still require close to $2^\ell$ queries ($\ell=128$ for AES-128), offering comparable security to SHA-256.
    In \poslopp, the MDC-2-based hash function $H$ is replaced by modular addition over a large prime field $Add_q(x,y) = x+y \mod q$. Chen el al. \cite{chen2021does} prove that such a non-cryptographic hash function (i.e., $Add_q$) is sufficient for Schnorr-based digital signatures despite missing full collision resistance, given that inputs are unpredictable and of size smaller than $q$ (i.e., lesser than 32-byte).
    Therefore, under standard assumtpions on AES-128 security and following the results of \cite{chen2021does}, \poslop~and \poslopp~achieve equivalent security to the original \poslo.
    %
\end{proof}

\end{document}